\documentclass{aa}  

\usepackage{graphicx}
\usepackage{xcolor}
\usepackage{soul}
\usepackage{gensymb}
\usepackage{upgreek}
\usepackage{amsmath}
\usepackage{textcomp}
\usepackage{txfonts}
\usepackage{hyperref}
\usepackage{hanging}
\newcommand{\hMsun}{h^{-1}\mathrm{M_\odot}}

\newcommand{\hkpc}{h^{-1}\mathrm{kpc}}
\newcommand{\hMpc}{h^{-1}\mathrm{Mpc}}
\newcommand{\hGpc}{h^{-1}\mathrm{Gpc}}

\newcommand{\kms}{\mathrm{km\,s^{-1}}}

\newcommand{\sqdeg}{\mathrm{deg}^2}

\begin{document} 


\title{The DESI One-Percent Survey: Modelling the clustering and halo occupation of all four DESI tracers with \textsc{Uchuu}}

\author{
F.~Prada \inst{1},
J.~Ereza \inst{1},
A.~Smith \inst{2},
J.~Lasker \inst{3},
R.~Vaisakh \inst{3},
R.~Kehoe \inst{3},
C.~A.~Dong-P\'aez \inst{1},
M.~Siudek \inst{4},
M.~S.~Wang \inst{5},
S.~Alam \inst{6},
F.~Beutler \inst{5},
D.~Bianchi \inst{7},
S.~Cole \inst{2},
Biprateep~Dey \inst{8},
D.~Kirkby \inst{9},
P.~Norberg \inst{2,10},
J.~Aguilar \inst{11},
S.~Ahlen \inst{12},
D.~Brooks \inst{13},
T.~Claybaugh \inst{11},
K.~Dawson \inst{14},
A.~de la Macorra \inst{15},
K.~Fanning \inst{16,17},
J.~E.~Forero-Romero \inst{18,19},
S.~Gontcho A Gontcho \inst{11},
C.~Hahn \inst{20},
K.~Honscheid \inst{21,22,23},
M.~Ishak \inst{24},
T.~Kisner \inst{11},
M.~Landriau \inst{11},
M.~Manera \inst{25,26},
A.~Meisner \inst{27},
R.~Miquel \inst{26,28},
J.~Moustakas \inst{29},
E.~Mueller \inst{30},
J.~Nie \inst{31},
W.~J.~Percival \inst{32,33,34},
C.~Poppett \inst{11,35,36},
M.~Rezaie \inst{37},
G.~Rossi \inst{38},
E.~Sanchez \inst{39},
M.~Schubnell \inst{40,41},
G.~Tarl\'{e} \inst{41},
M.~Vargas-Maga\~na \inst{15},
B.~A.~Weaver \inst{27},
S.~Yuan \inst{16,17},
and Z.~Zhou \inst{31},
\email{f.prada@csic.es}
}

\institute{Affiliations are in Appendix~\ref{sec:affil}}

 \date{...; ...}

\abstract{
We present results from a set of mock lightcones for the DESI One-Percent Survey, created from the \textsc{Uchuu} simulation. This $8~h^{-3}{\rm Gpc}^3$ $N$-body simulation comprises 2.1 trillion particles and provides high-resolution dark matter (sub)haloes in the framework of the Planck base-$\Lambda$CDM cosmology. Employing the subhalo abundance matching (SHAM) technique, we populate the \textsc{Uchuu} (sub)haloes with all four DESI tracers (BGS, LRG, ELG and QSO) to $z=2.1$. Our method accounts for redshift evolution as well as the clustering dependence on luminosity and stellar mass. The two-point clustering statistics of the DESI One-Percent Survey generally agree with predictions from \textsc{Uchuu} across scales ranging from $0.3\,\hMpc$ to $100\,\hMpc$ for the BGS and across scales ranging from $5\,\hMpc$ to $100\,\hMpc$ for the other tracers. 
We observe some differences in clustering statistics that can be attributed to incompleteness of the massive end of the stellar mass function of LRGs, our use of a simplified galaxy-halo connection model for ELGs and QSOs, and cosmic variance. We find that at the high precision of \textsc{Uchuu}, the shape of the halo occupation distribution (HOD) of the BGS and LRG samples are not fully captured by the standard 5-parameter HOD model. However, the ELGs and QSOs show agreement with an adopted Gaussian distribution for central haloes with a power law for satellites. We observe fair agreement in the large-scale bias measurements between data and mock samples, although the BGS data exhibits smaller bias values, likely due to cosmic variance. The bias dependence on absolute magnitude, stellar mass and redshift aligns with that of previous surveys. These results provide DESI with tools to generate high-fidelity lightcones for the remainder of the survey and enhance our understanding of the galaxy-halo connection.}


\authorrunning{F. Prada et al.}
\titlerunning{Modelling the clustering and HOD of DESI with Uchuu}

\maketitle



\section{Introduction}
\label{sec:intro}

Since the discovery of the accelerating expansion of the universe \citep{SupernovaSearchTeam:1998fmf, SupernovaCosmologyProject:1998vns}, there has been a significant emphasis within cosmology on ascertaining its underlying physical principles.  The initial measurement, facilitated by type Ia supernovae (SNe\,Ia)  as standardizable candles, has been substantially extended \citep[e.g.][]{JLA, Pantheon, DES3YR, Pantheon+}. Measurements of cosmic expansion have also been obtained using other methods, most notably data from the cosmic microwave background (CMB), such as that obtained from the Planck satellite \citep{Planck2020}. As these measurements have improved, a noticeable discrepancy has emerged when comparing the local Hubble constant value ($H_0$) from SNe Ia with that projected from CMB measurements \citep{Verde2019,Freedman:2021ahq,Mortsell:2021nzg}. This discrepancy, currently exceeding a $4\sigma$ significant level, continues to be investigated for potential hidden systematic effects or evidence of new physics \citep[see][and references therein]{Dainotti:2021pqg}. Although measurements of cosmic expansion and dark energy have been tested using additional methods, including the large-scale clustering of galaxies, the tension remains. Beyond addressing this question, understanding the specific behavior of dark energy, such as whether it manifests as a cosmological constant or arises from new physics, are key questions we must try to answer.

The large-scale structure of the universe becomes evident through the measurements of galaxy clustering obtained from large redshift surveys.  This structure naturally emerges from primordial fluctuations that originated in the early universe.  The propagation of baryon acoustic oscillations (BAO) generates matter density fluctuations that are frozen at the cosmic epoch of recombination. The characteristic BAO distance scale between galaxies provides a standard ruler, allowing us to investigate the expansion history of the universe. BAO analysis has emerged as a successful cosmological probe \citep{Cole2005, SDSS:2005xqv}, as demonstrated by its sensitivity to the BAO distance scale highlighted in the results obtained from the latest SDSS-III/BOSS \citep{Alam2017} and SDSS-IV/eBOSS \citep{Alam2021} large spectroscopic surveys. Combining BAO measurements with studies of redshift-space distortions \citep[RSD; see e.g.][for BOSS and eBOSS]{GilM2017, GilM2018} provides a critical complement to supernova and CMB results enabling us to measure the expansion of the universe and constrain cosmological models. 

Spectroscopic surveys greatly improve cosmological constraints in comparison to photometric surveys due their precise 3D measurements of galaxy clustering \citep{ParticleDataGroup:2016lqr}. The Dark Energy Spectroscopic Instrument \citep[DESI;][]{desifdr16a,desifdr16b} seeks the most precise measurements of the cosmic expansion history by using almost 40 million galaxy spectra to map the matter distribution across an unprecedented redshift range of $z<3.5$ \citep{levi2013}.  Four different galaxy tracers will be used to cover this entire range.  DESI is forecast to achieve sub-percent precision on the BAO distance scale, and an order of magnitude improvement in constraints on the dark energy equation of state compared to previous surveys \citep{desifdr16a}, and DESI is currently on target to meet these goals \citep{sv}. This advancement will enable DESI to meet its Dark Energy Task Force (DETF) Stage IV figure-of-merit performance goals \citep{Albrecht:2006um}. 

The cosmic expansion history and cosmological parameters inferred from BAO and RSD results reflect gravitational effects on dark matter and baryonic matter. Baryonic physics associated with astrophysical processes plays a crucial role in placing observable galaxies within dark matter halos.  However, it introduces a galaxy clustering bias relative to halos and determines which galaxies are detectable. This bias hinders our ability to establish a direct connection between our observations of galaxies and their haloes, thereby obscuring the underlying cosmology. Consequently,  removing this bias is a prerequisite of cosmological measurements. 
The challenging baryonic physics connecting the galaxies to haloes must be carefully modelled \citep[see][for a review]{Wechsler:2018pic}. Furthermore, it is essential that the uncertainties arising from these models on BAO measurements are controlled \citep[see][]{mattia2021}. Although hydrodynamical cosmological simulations encompass such physics \citep[e.g.][]{Pakmor2022}, their computational demands have made them prohibitive for larger volumes required to probe the BAO scale ($>100\, h^{-1}{\rm Mpc}$). However, the recent large volume runs in the FLAMINGO suite of \citealt{Flamingo2023} show a first attempt at a hydrodynamical simulation covering 1/8 the volume of Uchuu with 1/2 the mass resolution, a first step towards being able to cover larger volumes with sufficiently resolved hydrodynamical simulations. Instead, large volume $N$-body cosmological simulations are employed to populate galaxies within well-resolved dark matter halos and subhalos. Empirical methods that rely on the halo occupation distribution (HOD) statistics are popular in cosmological surveys \citep[see][and references therein]{Wechsler:2018pic}. This approach involves fitting the observed two-point clustering statistics of galaxies \citep[e.g.][]{Zehavi2011,White2011,Avila:2020rmp} and generating mock galaxies accordingly \citep[e.g.][]{White2014}. Another technique, known as subhalo abundance matching (SHAM), offers a more precise and extensively validated approach for populating simulated (sub)haloes with observed galaxies. SHAM matches the number density of observed galaxies selected by luminosity or stellar mass with the calculated density for haloes and subhaloes in the simulation, using a reliable proxy of halo mass \citep{Conroy06, trujillo-gomez11}. SHAM has been successfully used in numerous studies to accurately reproduce the clustering properties of observed
galaxies in large-scale surveys \citep[e.g.][]{Nuza13, Reddick13, RodriguezTorres16, Contreras2023}.
The depth and volume of the DESI survey poses computational challenges when generating the $N$-body cosmological simulations required for these empirical methods.  It is crucial for the mass resolution of the $N$-body simulations to be able to match the scales probed by the survey tracers.  To address this, the 2.1 trillion particle \textsc{Uchuu} simulation provides the necessary combination of a low particle mass and a large box size \citep{TIshiyama2021}. This simulation  allows us to account for dark matter haloes and subhaloes, including those on the scale of dwarf galaxies, across the entire volume covered by the DESI survey.

In this paper, we present clustering and halo occupancy results based on high-fidelity, simulated lightcones created from the \textsc{Uchuu} simulation in the Planck base-$\Lambda$CDM cosmology for the DESI One-Percent Survey~\citep{sv,dr}. These lightcones encompass all four DESI tracers (BGS, LRG, ELG and QSO) up to a redshift of 2.1.
We provide an overview of the DESI early data in Section~\ref{sec:desieda}. Section~\ref{sec:desiplanck} details the properties of each of the four tracer types and the specific SHAM prescriptions employed to generate the \textsc{Uchuu}-DESI lightcones for each tracer. In Section~\ref{sec:results}, we present the galaxy clustering measurements obtained from the \textsc{Uchuu}-DESI lightcones, along with the resulting halo occupation distributions and linear bias measurements for each tracer.  Finally, our conclusions are summarized in Section~\ref{sec:concl}. 

This work is part of a collection of papers released with the DESI Early Data Release, which examine the galaxy-halo connection for different tracers using various methodologies. This includes studies using the \textsc{AbacusSummit} \citep{maksimova2021abacussummit} simulations to obtain LRG and QSO HOD models \citep{abacusLRGQSO}, and ELG HOD models \citep{abacusELG}. \cite{inclusiveSHAM} presents a modified SHAM analysis for LRGs, ELGs and QSOs based on the UNIT simulation \citep{chuang2019unit}. Abundace matching was also employed in \citet{novelAM} to analyze the cross-correlations between LRGs and ELGs using the \textsc{CosmicGrowth} \citep{jing2019cosmicgrowth} simulation.

\section{DESI and Early Data Release}
\label{sec:desieda}

To achieve DETF Stage IV science goals, DESI is a highly multiplexed, robotically fibre-positioned spectroscopic array installed on the Mayall 4-meter telescope at Kitt Peak National Observatory \citep{desifdr16a}. DESI is capable of obtaining nearly 5000 simultaneous spectra in a $3~\deg$ field-of-view \citep{desifdr16b,silber22,DESIcorrector2022}. Each of the ten spectrographs covers the entire UV-to-near-IR spectral range with three arms: $3600-5930$~{\AA}, $5660-7720$~{\AA} and $7470-9800$~{\AA}.  Each arm is equipped with a $4k\times4k$ CCD. 
Galaxy tracers are identified using spectroscopic features, such as the [OII] doublet visible for emission-line galaxies \cite[ELGs;][]{Comparat:2012hz}. 
Currently, DESI is conducting a five-year survey spanning a 14,000 deg$^2$ sky footprint, designed to yield approximately 40 million galaxy and quasar spectra to measure cosmological parameters to sub-percent precision over $0<z<3.5$. 

The survey is supported by several software and data processing pipelines.  The imaging from the public DESI Legacy Imaging Surveys \citep{bass17,dey19,dr9} supports the target selection pipeline for spectroscopic follow-up.  Target selection employs quality cuts, as well as various colour selections and machine learning classification tools, tailored to provide a highly complete and low contamination sample for each tracer type. 
Identification and prioritization of all targets are described in \citet{myers23}.  Specific details are provided for the Bright Galaxy Survey \citep[BGS;][]{ruiz20}, luminous red galaxies \citep[LRGs;][]{zhou20,Zhou23}, emission line galaxies \citep[ELGs;][]{raichoor20, raichoor23}, and quasars \citep[QSOs;][]{yeche20,Chaussidon23}. A planning pipeline optimizes the tiling of observations throughout the survey \citep{ops}.  Fibres are assigned to targets for each pointing in another pipeline \citep{fba}.  Resultant spectra are processed with a `spectroperfectionist' \citep{spectroperf} data reduction pipeline \citep{spec2022} followed by a template fit yielding redshifts and a final classifications for each source \citep{redrock2022}.  

Since DESI will probe the galaxy distribution substantially deeper than prior large area surveys, a 4-month Survey Validation (SV) observing period was conducted to evaluate the science program~\citep{sv}.  A substantial effort was dedicated to the attainment of deeper spectra with substantially higher target densities than expected for the main survey.  Deeper spectra allowed the parent redshift distribution to be probed, and the determination of the number of tracers versus redshift that should be expected to be observed in the survey.  The many exposures required for the deeper spectroscopy were used to characterize the exposures and establish the statistical performance of the redshifts, including their statistical uncertainties, completeness after classification, and purity. The SV period was valuable for testing and finalizing calibration procedures for the observations.  A large number of sky fibers were used, which was crucial in testing the sky subtraction.  DESI aims to classify and measure redshifts for galaxies near the Poisson noise limit, which requires an excellent sky subtraction. SV allowed a determination of how many sky fibres will be required in the main survey.  As an essential element of observing, DESI employs a dynamic exposure time calculation \citep{kirkby22} that utilizes $in\,situ$ real-time measurements of observing conditions to optimize exposure times, minimizing overheads and inefficient data taking.  SV observing in a full range of atmospheric and Galactic extinction environments allows for calibration of the models used by the calculator. The exposure times were optimized to facilitate completion of the main survey in the designed 5 years. An extensive visual inspection regime was also carried out using the SV data to verify the performance of the instrument, as well as the data reduction pipeline and target selection algorithms.  For galaxies, more details can be obtained in \citet{Lan23}, while quasars are described in \citet{Alexander23}.  Overall, SV has been invaluable in providing inputs for subsequent modeling and analysis.

The final month of SV was dedicated to the One-Percent Survey, covering 140$\rm\,deg^2$ at the intended main survey spectroscopic depth.  A total of 239 (214) dark (bright) time tiles were observed over 33 (35) nights, with 375 (287) exposures and 88.2 (15.9) hours of effective exposure time.  The BGS sample was split into two samples: BGS-BRIGHT with $r\leq 19.5$ and BGS-FAINT with $19.5< r\leq 20.175$, with the bulk of the sample in BGS-BRIGHT.  Target selection was optimized for high completeness and low background contamination. Stars and galaxies were distinguished by  comparing $G_{Gaia}$ from $Gaia$ \citep{gaia18} with $r$ from the DR9 Legacy Imaging Survey \citep{dr9} and removing sources that were bright in $Gaia$.  In order to isolate galaxies, colour cuts using $z, g, r$ and $W1$ from WISE \citep{wright10,lang16b} were used.  Further details can be obtained from \citet{Hahn22}.
LRG selection was optimized to yield uniform comoving number density in the redshift range of $0.4<z<0.8$, while WISE photometry was used to effectively veto stars.  $z, g, r$ and $W1$ filters were utilized to remove lower redshift and bluer galaxies.  LRGs have a lower priority than QSOs but a higher priority than ELGs, ensuring high completeness. Further details on LRG final selection can be found in \citet{Zhou23}.
Our final ELG selection was divided into two redshift bins due to their potential overlap with LRGs. A higher priority `ELG\_LOP' sample covers the range of $1.1<z<1.6$, which is inaccessible to LRGs, while a low priority `ELG\_VLO' sample covers the entire range of the DESI ELG targets $0.6<z<1.6$. Quality cuts were applied to reject bright stars, and $g, z, r$ and $g_\mathrm{fib}$ filters were used to select star-forming instead of passive galaxies. Further details on ELG selection can be found in \citet{raichoor23}.
Selection of quasars relies on colours from $z, g, r$ as well as $W1, W2$ bands.  The near-infrared fluxes are valuable to separate bluer stars from redder QSOs.  Ten colours from these 5 bands are used and employ a Random Forest technique to improve efficiency.  Quasars are observed at highest priority, and DESI achieves 99\% efficiency. Further details on the selection of all tracers can be found in \citet{myers23}. From the One-Percent Survey footprint, DESI projects number densities of $988~\rm deg^{-2}$, $533~\rm deg^{-2}$, $\rm 1121~deg^{-2}$ and $\rm 205~deg^{-2}$ for BGS, LRG, ELG, and QSO, respectively.  

DESI has demonstrated excellent performance in the One-Percent Survey. 
Redshift comparisons between the DESI tracers and data from DEEP2, SDSS, BOSS and eBOSS indicate offsets of $6.5~\kms$, $<10~\kms$ and $1~\kms$ for the BGS, LRGs, and ELGs, respectively.
The One-Percent Survey results have enabled DESI to meet and, in many cases, substantially surpass its prior projections and requirements. DESI  will obtain nearly 40 million unique galaxy and QSO redshifts over a 5 year survey. Catastrophic redshift failures, redshifts which are incorrect by more than $1000 \kms$, must be less than 5\% for LRGs and ELGs, and DESI has achieved 0.2\%. Similarly, the fractional redshift error must be less than $2 \times10^{-4}(1+z)$ for ELGs and $4\times10^{-4}(1+z)$ for QSOs.  The One-Percent Survey has achieved redshift errors of $3.3\times 10^{-6}(1+z)$ and $8.7\times10^{-5}(1+z)$ for these populations, respectively. SV results also allow us to better project the main survey's performance on cosmological parameter precision.  DESI forecasts a statistical precision of $\delta H(z)\sim0.28\%$, of RSD figure of merit $\delta R(z)\sim0.24\%$, and of $\delta f\sigma_8\sim1.56\%$ for $z<1.1$.  At higher redshift, $\delta H(z)$ will reach a precision of 0.39\% and 0.46\% in $1.1<z<1.9$ and $z>1.9$ ranges, respectively.  For further details about Survey Validation, the One-Percent Survey, and the projected DESI Survey results, refer to \citet{sv}.

\section{Cosmological Modelling of the DESI One-Percent Survey}
\label{sec:desiplanck}

\begin{figure*}
\centering
\includegraphics[width=\linewidth]{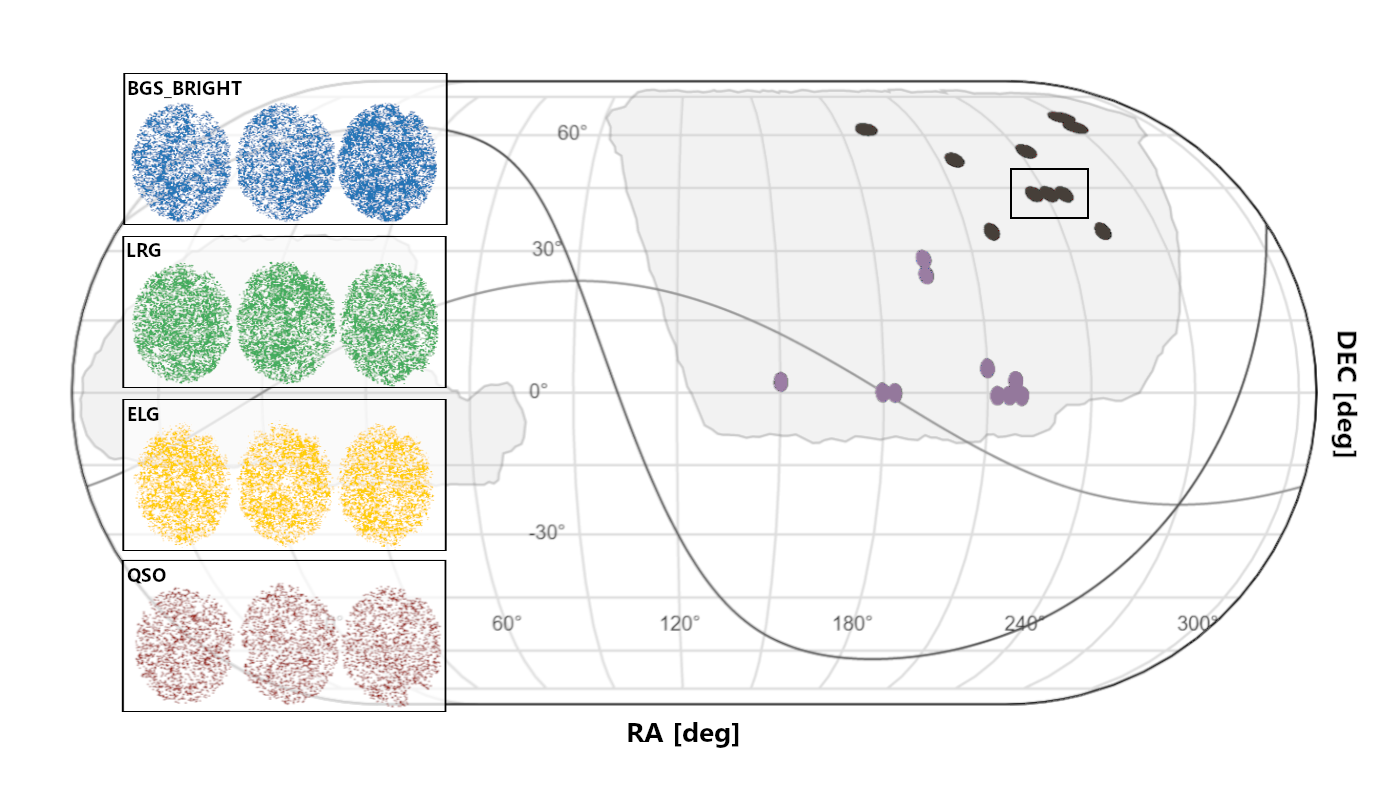}
\caption[]{The sky coverage of the DESI One-Percent Survey for the BGS-BRIGHT, LRG, ELG, and QSO cosmological tracers used in this analysis. The 20 rosettes that make up the One-Percent footprint are split into `North' (in black) and `South' (in purple). The grey-shaded regions indicate the expected DESI Year-5 sky coverage. The four small panels are zoomed in on a section of the footprint covered by 3 rosettes, for each tracer. The colour-coding represents the angular weighted number density, where darker colours indicate a higher density.
} 
    \label{fig:figure1}
\end{figure*}

\begin{table}
\setlength{\tabcolsep}{4pt}
	\centering
	\begin{tabular}{lcccccc} 
		\hline
		Sample     & Redshift range       & $z_\mathrm{med}$  & $A_\mathrm{eff}$ & $N_{\rm eff}$ & $10^2\times V_{\rm eff}$  \\
                   &         &       & ($\deg^2$)          &               & ($h^{-3}\mathrm{Gpc}^{3}$)\\
		\hline
        BGS- & $0.05<z<0.5$  &  0.21 & 173.5  & 142341 &  3.72\\
        BRIGHT & & & & & \\
        \hline
        LRG        & $0.45<z<0.85$  & 0.76 & 166.9 & 58764 &  7.97 \\
        \hline
        ELG        & $0.88<z<1.34$ & 1.07  & 168.6   & 156891 & 12.95 \\
        \hline
        QSO        & $0.9<z<2.1$ & 1.53  & 174.6   & 23085 &  1.87\\
		\hline
	\end{tabular}
 \caption{Basic properties of the DESI One-Percent Survey samples used in this work: the redshift interval, median redshift ($z_\mathrm{med}$), effective area of the sky footprint weighted by completeness ($A_\mathrm{eff}$), number of galaxies ($N_\mathrm{eff}$), and effective volume ($V_\mathrm{eff}$).}
\label{tab:all-basic}
\end{table}

This section details the process by which the \textsc{Uchuu} One-Percent mock lightcones were generated for each tracer to closely match the clustering properties of the DESI One-Percent Survey. The basic properties of the DESI One-Percent data, such as sky coverage and number densities of each tracer, are presented in Section~\ref{sec:one_percent_properties}. Section~\ref{sec:uchuu_simulation} describes the high-resolution \textsc{Uchuu} $N$-body simulation that was used to create our simulated lightcones. An overview of the SHAM methods adopted to populate galaxies and quasars into the \textsc{Uchuu} halo catalogues to build lightcones for each of the DESI tracers is provided in Section~\ref{sec:uchuu_one_percent_lightcones}.

\subsection{Properties of the One-Percent Survey}
\label{sec:one_percent_properties}

Figure~\ref{fig:figure1} shows the sky coverage of the DESI One-Percent BGS-BRIGHT, LRG, ELG, and QSO samples.\footnote{This figure was created with the help of the web interface provided by D. Kirkby (\url{https://observablehq.com/@dkirkby/desi-tutorial})} We list the basic properties of these samples as used in our analysis in Table~\ref{tab:all-basic}, which includes information such as the redshift ranges, sky area, total number of galaxies, and effective volume as a measure of constraining power. The effective volume \citep[equation~1 of][]{Wang2013_veff} is calculated as
\begin{equation}
    V_{\rm eff} = \sum_{i}\left(\frac{n(z_i)P_0}{1+n(z_i)P_0}\right)^2~\Delta V(z_i),
    \label{eq:eff_vol}
\end{equation}
where $n(z_i)$ is the weighted number density of galaxies and $\Delta V(z_i)$ is the comoving survey volume at redshift bin $z_i$. We adopt $P_0$, the power spectrum value at a scale of $k = 0.15~h\mathrm{Mpc}^{-1}$ where we desire to minimize power spectrum variance, to be $7000~h^{-3}\mathrm{Mpc}^3$ for BGS, $P_0 = 10000~h^{-3}\mathrm{Mpc}^3$ for LRG, $P_0 = 4000~h^{-3}\mathrm{Mpc}^3$ for ELG, and $P_0 = 6000~h^{-3}\mathrm{Mpc}^3$ for QSO \citep{dr}.

\begin{figure}
	\includegraphics[width=\columnwidth]{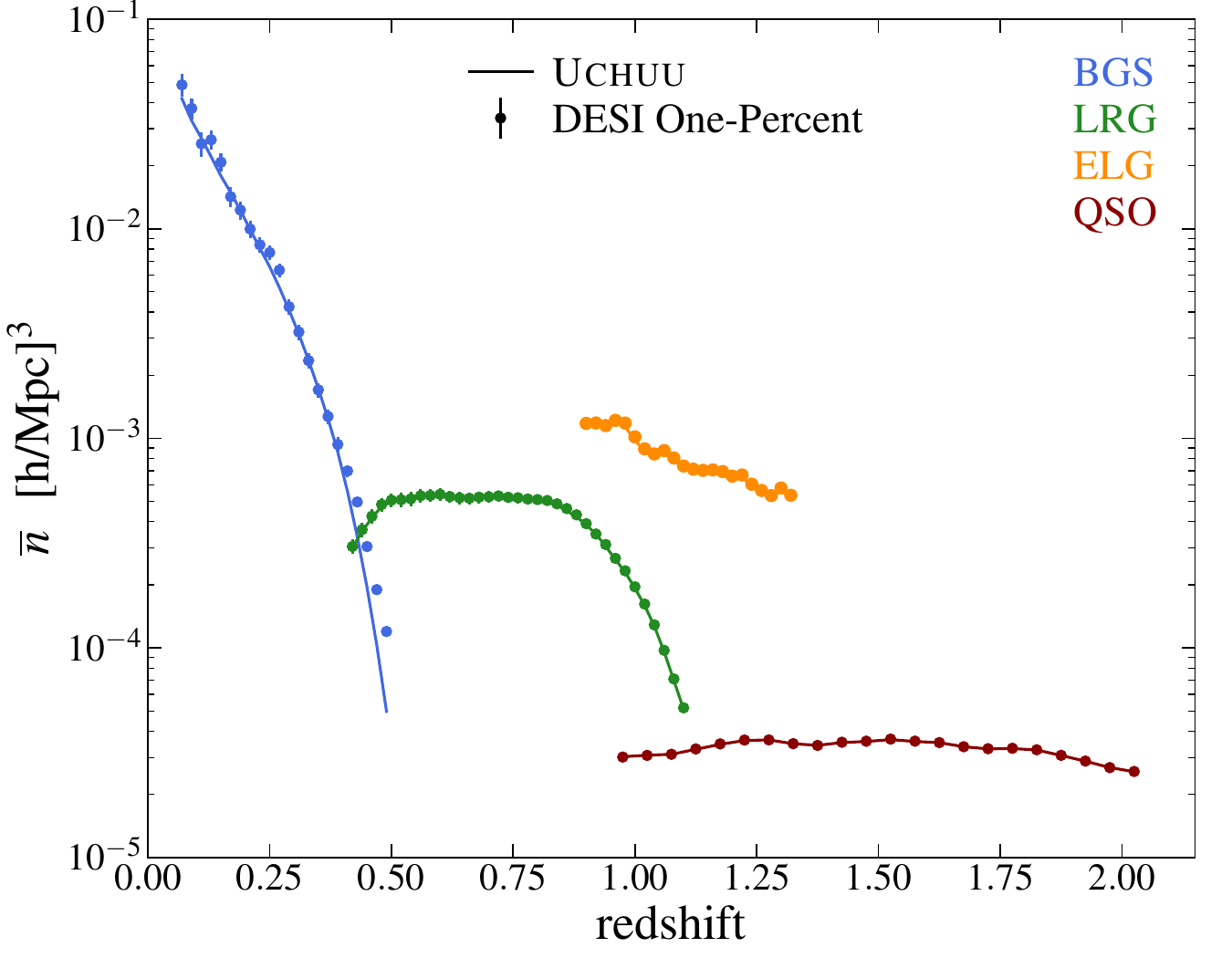}
    \caption{The comoving number density of the four DESI One-Percent tracer samples (points) and the average of the corresponding \textsc{Uchuu}-DESI mock lighcones (solid line) over the entire redshift range $0.1 < z < 2.1$. Data error bars are obtained from the ensemble of \textsc{Uchuu} One-Percent lighcones built in this work. 
    }
    \label{fig:all-ndens}
\end{figure}

The four cosmological tracers cover a wide redshift range, extending from $z=0.05$ to $z=2.1$, with the BGS-BRIGHT and QSO samples having the highest and lowest densities, respectively. Figure \ref{fig:all-ndens} displays the comoving number density for the BGS-BRIGHT, LRG, ELG, and QSO samples taken from the DESI One-Percent Survey. We utilize the entire data sample taken during the One-Percent Survey, including areas only tiled to partial completeness, leading to a larger effective area than the 140 sq. degrees reported in \cite{sv}. This necessitates the inclusion of weighting schemes to correct for the incompleteness which we discuss in Section~\ref{sec:results-clustering}. We converted the redshifts of the One-Percent galaxies and quasars to comoving distances using a fiducial flat $\Lambda$CDM cosmological model with the Planck-15 parameters $h = 0.6774$, $\Omega_{\rm m} = 0.3089$, $\Omega_{\rm b} = 0.0486$, $n_{\rm s} = 0.9667$, $\Omega_\Lambda=0.6911$, and $\sigma_8 = 0.8159$ \citep{Planck16}.

\subsection{The \textsc{Uchuu} simulation}
\label{sec:uchuu_simulation}

In order to model the clustering signal of the DESI One-Percent Survey in the flat $\Lambda$CDM Planck cosmology, we utilized the \textsc{Uchuu} $N$-body simulation \citep{TIshiyama2021}. Designed specifically to model the DESI survey, \textsc{Uchuu} boasts high numerical resolution which enables the resolution of dark matter haloes and subhaloes down to small masses on a very large volume. This resolution, in turn, allows us to apply the SHAM technique to populate the \textsc{Uchuu} haloes and subhaloes with DESI galaxies and quasars, generating mock lightcones to reproduce the number density and predict the clustering of each DESI tracer. Section~\ref{sec:uchuu_one_percent_lightcones} provides a more detailed description of the construction of these \textsc{Uchuu}-DESI lightcones. In Section~\ref{sec:results}, we provide a thorough comparison of the predicted clustering signal of these lightcones in the Planck cosmology to that observed in the DESI One-Percent survey. This comparison allows us to further investigate the halo occupation distribution and large-scale bias of all four tracers.

The \textsc{Uchuu} simulation was run using the TreePM code \textsc{GreeM} \citep{Ishiyama09,Ishiyama12}. The box has a comoving side length of $2~\hGpc$, with $12,800^3$ dark matter particles. The mass resolution and gravitational softening length are $3.27\times 10^8~\hMsun$ and $4.27~\hkpc$, respectively. The initial conditions were generated using the second-order Lagrangian Perturbation Theory (2LPT) approximation at $z_\mathrm{init} = 127$, and the simulation followed the growth of cosmic structures in the Planck-15 flat $\Lambda$CDM cosmology. We saved 50 snapshots of the particle distribution from $z=14$ to $z=0$, and identified bound structures using the \textsc{Rockstar} phase-space halo/subhalo finder \citep{Behroozi13}. We constructed merger trees for these structures using a parallel version of the \textsc{ConsistentTrees} algorithm \citep{Behroozi2013b}. Additionally, we obtained the peak value of the maximum circular velocity, $V_\mathrm{max} = \rm{max}(\sqrt{\frac{GM(r)}{r}})$,  over the history of each (sub)halo, denoted as $V_\mathrm{peak}$. We measure the maximum circular velocity at each of the 50 redshift outputs, and take the maximum value as $V_\mathrm{peak}$.  We used this to implement the SHAM method for populating \textsc{Uchuu} haloes with DESI galaxies and quasars. For more information on the simulation methodology and performance, we refer the reader to \citet{TIshiyama2021}. All \textsc{Uchuu} data products are publicly available through \textsc{Skies \& Universes}.\footnote{\url{https://www.skiesanduniverses.org/Simulations/Uchuu/}}

\subsection{\textsc{Uchuu} One-Percent Lightcones}
\label{sec:uchuu_one_percent_lightcones}

In the following, we provide a brief overview of the creation of  \textsc{Uchuu} mock lightcones for DESI One-Percent galaxies and quasars in the Planck cosmology. Further details will be provided in forthcoming papers on each tracer type utilizing first year of the observations of the DESI main survey to conduct a comprehensive study of the clustering signal, including the BAO scale.  This dataset covers a much larger sky area than the One-Percent survey, yielding over an order of magnitude more galaxies.

In the DESI One-Percent survey, the BGS-BRIGHT sample is magnitude limited and is thus well suited to a traditional SHAM method using the absolute magnitude. Of the three dark time tracers in DESI (LRGs, ELGs, and QSOs), LRGs are the only ones which are believed to be complete in halo mass occupation \citep{Alam_2020}, so they are the only tracers which can use a traditional abundance matching technique. ELGs and QSOs are not complete in any known parameter space \citep{Favole16,rodrigueztorres17}, so the traditional SHAM method must be modified. We describe both the traditional (Section~\ref{sec:uchuu_sham}) and modified SHAM (Section~\ref{sec:uchuu_mod_sham}) techniques used below. 
\subsubsection{Subhalo abundance matching: BGS-BRIGHT and LRG}
\label{sec:uchuu_sham}

\begin{figure}
    \includegraphics[width=\columnwidth]{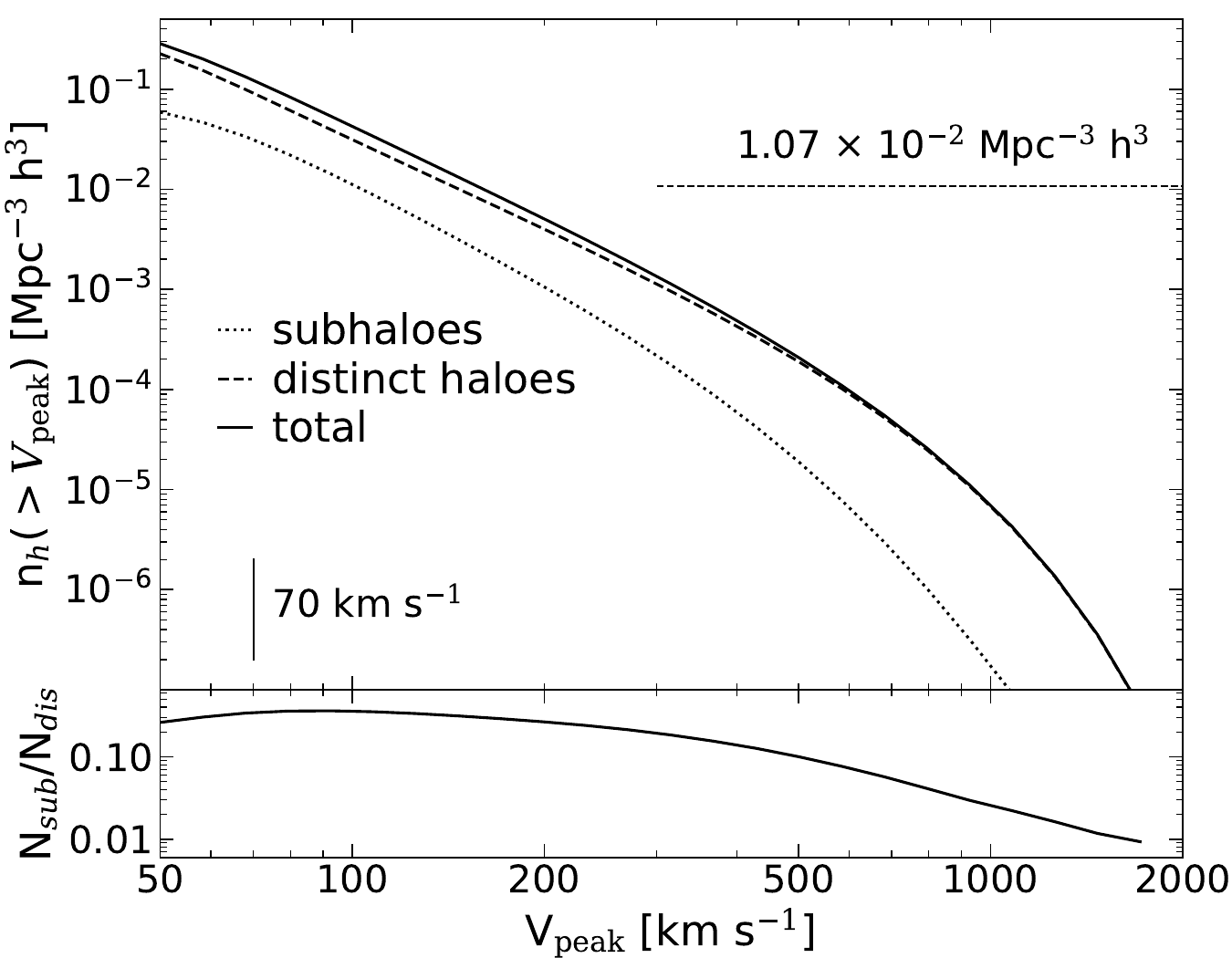}
    \caption{
    \textit{Top panel:} Cumulative number density of distinct haloes ($N_\mathrm{dis}$) (dashed curve), subhaloes ($N_\mathrm{sub}$) (dotted curve), and all haloes (solid curve) as a function $V_\mathrm{peak}$ in the \textsc{Uchuu} simulation box at $z=0.2$, the median redshift of the BGS-BRIGHT sample in the DESI One-Percent Survey.  The horizontal line indicates the mean number density of the BGS-BRIGHT sample.  The vertical line indicates the completeness threshold for \textsc{Uchuu}.
    \textit{Bottom panel:} Cumulative subhalo fraction measured as a function of $V_{\rm peak}$.
    }
    \label{fig:uchuu-vpeak}
\end{figure}

We employ the (traditional) SHAM algorithm to construct lightcones for 
BGS-BRIGHT and LRG DESI tracers from the \textsc{Uchuu} simulation boxes. 
For these samples, the SHAM method assigns luminosities or stellar masses to all \textsc{Uchuu} (sub)haloes in the simulation boxes. Specifically, we match their $V_\mathrm{peak}$ cumulative distribution function to our chosen BGS-BRIGHT luminosity and LRG stellar mass functions, with a certain level of intrinsic scatter $\sigma$ in the $V_\mathrm{peak}$-luminosity/stellar mass relationship as the only free parameter when creating our \textsc{Uchuu} One-Percent BGS-BRIGHT and LRG lightcones in the Planck cosmology. We present the methodology in 
Sections~3.3.1.1~and~3.3.1.2.

In this work, we adopt the peak maximum circular velocity $V_\mathrm{peak}$ as a proxy for (sub)halo mass. $V_\mathrm{peak}$ has been extensively used in numerous studies to accurately reproduce the properties of observed galaxies in large-scale surveys \citep[e.g.][]{Conroy06, trujillo-gomez11, Nuza13, Reddick13, chaves-montero16, RodriguezTorres16, Safonova21}.

We are able to reach the lowest luminosities and smallest stellar masses in the BGS-BRIGHT and LRG DESI galaxy samples, respectively, thanks to the high completeness level of subhaloes and distinct haloes in \textsc{Uchuu}.  As estimated previously~\citep{TIshiyama2021, DongPaez22} in comparisons with the much higher-mass resolution \textsc{Shin-Uchuu} simulation, satellite subhaloes in \textsc{Uchuu} have a completeness of $90\%$ down to $V_\mathrm{peak} \sim 70~\kms$.  Haloes have a completeness level of $90\%$ down to $V_\mathrm{peak} \sim 50~\kms$. 

Figure~\ref{fig:uchuu-vpeak} illustrates in the top panel the cumulative number density of (sub)haloes vs. $V_\mathrm{peak}$ in the \textsc{Uchuu} box at $z=0.2$, which is the median redshift of BGS-BRIGHT galaxies.  At this redshift, we use $V_\mathrm{peak} > 170~\kms$ for abundance matching with the BGS-BRIGHT sample, which is well above the completeness limit of \textsc{Uchuu}.  
The mean number density of the BGS-BRIGHT sample at its median redshift is indicated by the horizontal dotted line. This indicates that (sub)haloes hosting BGS-BRIGHT galaxies are well-resolved with the \textsc{Uchuu} simulation. The cumulative subhalo fraction, measured from the cumulative number densities in Figure~\ref{fig:uchuu-vpeak}, is shown as a function of $V_\mathrm{peak}$ in the bottom panel. For values of $V_\mathrm{peak}$ greater than $170~\kms$, the cumulative subhalo fraction is approximately $25\%$. Note that at redshifts lower than $z=0.2$, the faintest galaxies live in (sub)haloes with smaller $V_\mathrm{peak}$ than this, but only a small fraction have $V_\mathrm{peak} < 100~\kms$, which are still well resolved. For more insight, Figure 4 in \citet{Nuza13} presents the number density of (sub)haloes in the MultiDark simulation at $z=0.53$.  This is similar to the BOSS-CMASS LRG sample, corresponding to (sub)haloes with $V_\mathrm{peak}$ above $370 \, \kms$, consistent with typical DESI LRGs living in much more massive haloes than BGS-BRIGHT galaxies.  In this case, the subhalo fraction is typically about 10\%.  In the following sections, we will present the results obtained from our analysis of the halo occupation distribution of all four DESI tracers which we obtained from the (modified) SHAM \textsc{Uchuu} lightcones.
\\

\noindent \textit{3.3.1.1 \textsc{Uchuu} BGS BRIGHT}
\vspace{0.2cm}\\ 
\noindent We follow the SHAM methodology as outlined in Section~3.2 of \citet{DongPaez22}, which has been successfully applied to SDSS, to construct our \textsc{Uchuu} One-Percent BGS lightcones. SHAM assumes that the most massive (sub)haloes host the most luminous galaxies. To generate our simulated flux-limited BGS sample, we utilize a parametrised luminosity function from both SDSS and GAMA \citep[see][for the creation of a DESI-BGS lightcone from the Millennium-XXL simulation]{Smith17,Smith22b}.

We assigned galaxy magnitudes as a function of halo $V_\mathrm{peak}$ using a SHAM algorithm with intrinsic scatter, based on \citet{McCullagh2017} and \citet{Safonova21}. For simplicity, we adopt a constant scatter parameter of $\sigma=0.5 \, \mathrm{mag}$. This value is calibrated to match the observed SDSS clustering \citep{DongPaez22}.

We apply SHAM to the (sub)halo catalogues from \textsc{Uchuu} boxes at redshifts 0, 0.093, 0.19, 0.3, 0.43, and 0.49. The implementation details of the SHAM algorithm, and adopted scatter, are described in \citet{DongPaez22}. Finally, we combine the snapshots to create the \textsc{Uchuu} lightcone (see Section~\ref{sec:uchuu_lightcones} for more details).
\\

\noindent \textit{3.3.1.2 \textsc{Uchuu} LRG}
\vspace{0.2cm}\\ 
\noindent To construct the \textsc{Uchuu}-LRG lightcones, we follow the SHAM approach introduced in section~4.1 of \citet{RodriguezTorres16}, which has been previously applied to the BOSS survey. This method assumes the most massive galaxies are hosted by the most massive (sub)haloes.

\begin{figure}
	\includegraphics[width=\columnwidth]{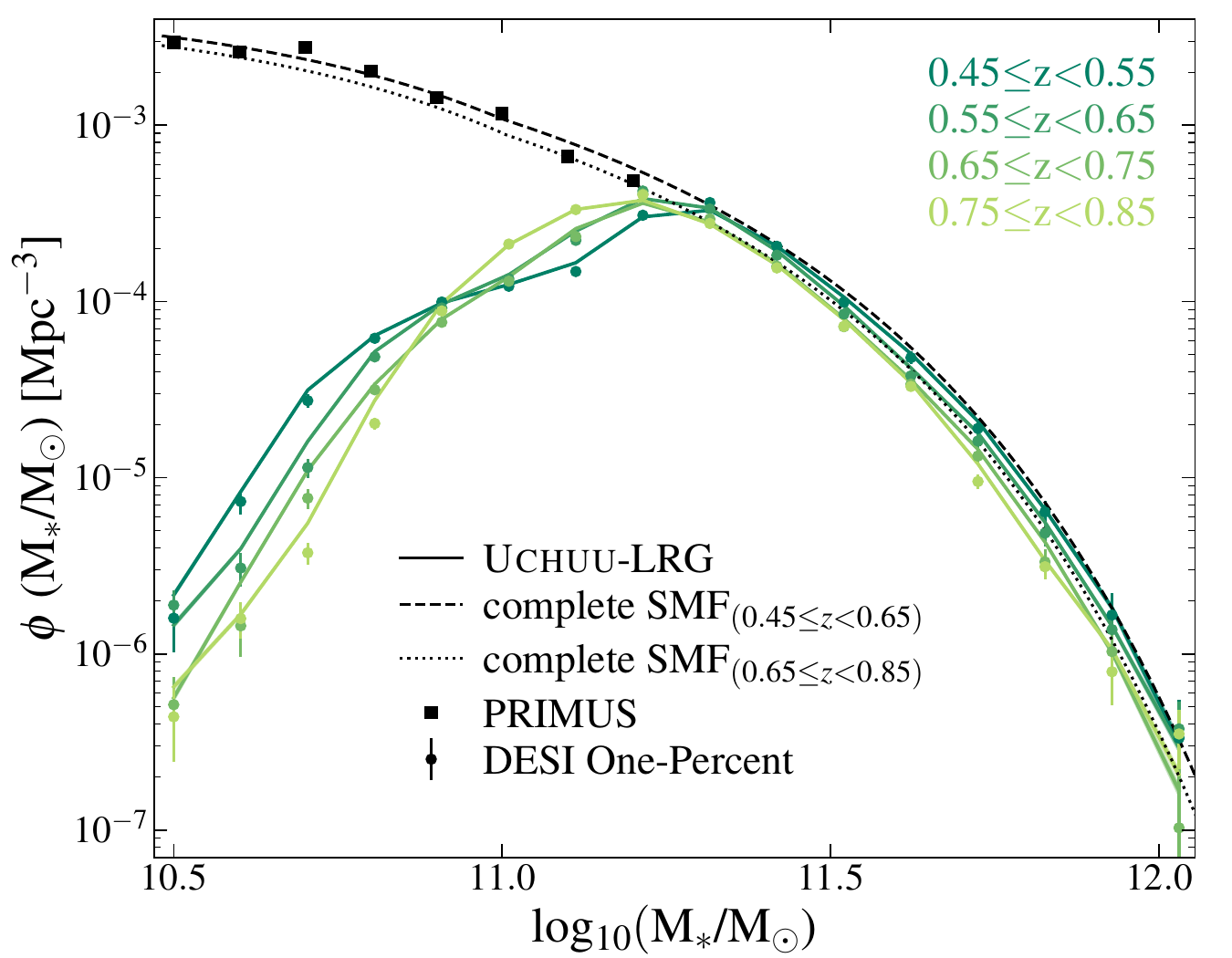}
    \caption{Stellar mass functions for LRG in the DESI One-Percent Survey (points) and the mean of our \textsc{Uchuu}-LRG lightcones (solid curves) are shown for several redshift bins within the range $0.45 < z < 0.85$. The dashed curves represent the complete SMF adopted in each redshift range, indicated in the legend.  Data error bars and the model shaded area represent the standard deviation of our set of 102 \textsc{Uchuu} lightcones.}
    \label{fig:lrg_stellarmass}
\end{figure}

In Figure \ref{fig:lrg_stellarmass}, we present the stellar mass function of LRG obtained from the DESI One-Percent survey using the CIGALE tool \citet{Boquien19} to estimate individual stellar masses \citep{siudek_tbd}. To fit the spectral energy distribution (SED), we used three optical photometry bands ($g,r,z$) from the DESI Legacy Survey \citep[DECaLS;][]{dey19}, complemented by four WISE mid-infrared bands (W1, W2, W3, and W4) from the NEOWISE-Reactivation project \citep[NEOWISER;][]{Mainzer14}. The CIGALE SED-fitting tool is based on the principles of the energetic balance between the absorbed stellar emission in the ultraviolet and optical bands and its re-emission in the infrared by dust. We adopted a grid of stellar population models with a delayed star formation history (SFH), including an optional exponential burst, a \citet{Chabrier03} initial mass function (IMF), and solar metallicity.  To model the effect of dust extinction, we used the reddening law of \citet{Calzetti2000}, and we adopted the updated dust templates from \citet{Draine14} to model the IR emission from dust reprocessed from the absorbed UV/optical stellar emission. We also incorporated the standard nebular emission model from \citet{Inoue11} and the AGN emission models from \citet{Fritz2006}. We performed a Bayesian-like analysis to fit the SEDs of these models to the DESI galaxy SEDs. The quality of the fit is expressed by the reduced $\chi^2$, and 
in this paper, we decided to limit ourselves between the threshold recommended by \citet{Siudek2017}, $\chi^2_r < 17$ and the most restrictive $\chi^2_r$ < 5 to ensure reliable stellar mass estimates. \citet{siudek_tbd} will provide a detailed description of the SED fitting procedure.

Although we were able to estimate the LRG stellar mass function (SMF), we lack information on the shape of the SMF at low masses due to the selection function. To supplement our analysis, we incorporate the SMF measurements obtained from PRIMUS presented by \citet{Moustakas_Primus}, which is shown by the square symbols in Figure \ref{fig:lrg_stellarmass}. It is worth noting that we do not consider the redshift evolution of the PRIMUS SMF in our analysis, as it has been shown to have a negligible impact on our results and is consistent with the findings of the PRIMUS survey.

To account for the observed evolution in the shape of the SMF with redshift (as shown in Figure \ref{fig:lrg_stellarmass}), we have employed two different complete SMF fits in our SHAM method: one that characterizes a complete galaxy population at $0.45\leq z<0.65$ (low-$z$, represented by the dashed line), and another that characterizes a complete population at $0.65\leq z<0.85$ (high-$z$, represented by the dotted line).

We apply SHAM to the (sub)halo catalogues from \textsc{Uchuu} boxes at redshifts 0.49, 0.63, 0.78, and 0.86 to cover the interval $0.45 \leq z < 0.85$. To ensure a consistent approach, we generate LRGs in the first two boxes using the low-$z$ complete SMF, and the remaining two boxes with the high-$z$ complete SMF. 
However, as the LRG DESI One-Percent stellar mass distribution is incomplete, we account for this by randomly down-sampling galaxies from the complete SMF in each \textsc{Uchuu} box to match the observed DESI One-Percent SMF. We then combine the resulting snapshots to generate the \textsc{Uchuu} lightcone (see Section~\ref{sec:uchuu_lightcones}).
Through this method, we do not intend for our lightcones to fit the data at all costs, but rather, we aim to analyze the observed sample. In searching for the complete SMF fit, we have assumed that the high-mass end of the observed SMF is no longer affected by the survey's selection criteria, that is, that the high-mas observed galaxy population is complete. Whether our lightcones, produced under this assumption, are able to replicate the data across all stellar mass cuts and redshift ranges will illustrate the validity of our assumption.

\subsubsection{Modified subhalo abundance matching: ELG and QSO}
\label{sec:uchuu_mod_sham}

To model the DESI ELGs and QSOs, we adopt a modified SHAM approach as described in section~3.1 of \citet{rodrigueztorres17}, which was previously applied to the BigMultiDark Planck simulations to build eBOSS-QSO lightcones. This modification allows the use of SHAM-like methods to model incomplete tracers  such as ELGs and QSOs, for which traditional SHAM would not typically be possible such as both ELGs and QSOs. However, we use $V_\mathrm{peak}$ instead of $V_\mathrm{max}$, the maximum circular velocity, adopted in \citet{rodrigueztorres17}. 
The modification from \citet{rodrigueztorres17} accounts for the incompleteness of the tracer population in terms of halo mass or luminosity/stellar mass \citep[see also][for the same method applied to {[O\textsc{ii}]} galaxy emitters]{Favole2017}.

This methodology is implemented by selecting (sub)halo samples from \textsc{Uchuu} adopting independent Gaussian distributions for central and satellite galaxies with the same mean $V_\mathrm{peak}$ ($V_\mathrm{mean}$) and standard deviation $\sigma_V$. This is performed separately for ELGs and QSOs using different $V_\mathrm{mean}$ and $\sigma_V$. The Gaussian distributions are normalized to match the observed ELG/QSO number densities, with the satellite fraction ($f_\mathrm{sat}$) treated as a free parameter in each case. This can be expressed in the form as described in \citet{rodrigueztorres17}.

Equation~\ref{eq:modsham1} describes the modified SHAM model in terms of its final distribution of ELG/QSO $V_{\rm peak}$ ($\phi_{\rm ELG/QSO} (V_{\rm peak})$)  as a combination of two Gaussians, one for the satellites, and one for the centrals ($\mathcal{G}_{\rm s/c}$) with model parameters $V_{\rm mean}$ and $\sigma_V$ controlling the center and width of the Gaussians. It is further broken down into the selection probabilities for satellites and centrals ($P_{\rm s/c}$) from the simulated main and satellite halo $V_{\rm peak}$ function ($\phi_{\rm sim}^{s/c}$),

\begin{equation}
\label{eq:modsham1}
\begin{split}
     \phi_{\rm ELG/QSO}(V_{\rm peak}) = &  \hspace{3pt} \phi_{ELG/QSO}^s + \phi_{ELG/QSO}^c \\ 
     = &  \hspace{3pt} P_{\rm s}(V_{\rm peak}; V_{\rm mean}, \sigma_V)*\phi_{\rm sim}^s \\
    &  +  P_{\rm c}(V_{\rm peak}; V_{\rm mean}, \sigma_V)*\phi_{\rm sim}^c \\
     = & \hspace{3pt} \mathcal{G}_{\rm s}(V_{\rm peak}; V_{\rm mean}, \sigma_V) \\
    &  + \mathcal{G}_{\rm c}(V_{\rm peak}; V_{\rm mean}, \sigma_V).
\end{split}
\end{equation}

The Gaussians are then normalized so that they exactly match the number density of the data ($\rho(z)$) in bins of $\Delta z= 0.02$  for the comoving volume ($V_{\rm c}(z)$) over the redshift ranges of each lightcone shell as shown in Equation~\ref{eq:modsham2} below. The relative normalization of the satellite and central Gaussians is controlled by the satellite fraction model parameter ($f_{\rm sat}$).

\begin{equation}
\label{eq:modsham2}
\begin{split}
     \int_0^\infty  \mathcal{G}_{\rm s}(V_{\rm peak}, z; V_{\rm mean}, \sigma_V) =&  \hspace{3pt} V_{\rm c}(z)*\rho(z)*f_{\rm sat} \\ 
    \int_0^\infty  \mathcal{G}_{\rm c}(V_{\rm peak}, z; V_{\rm mean}, \sigma_V) =&  \hspace{3pt} V_{\rm c}(z)*\rho(z)*( 1 - f_{\rm sat})
\end{split}
\end{equation}

We generate a grid of full sky ELG/QSO lightcone mocks in this parameter space, compute the monopole of the two-point correlation function (2PCF) for each of these mocks as described in Section~\ref{sec:results-clustering}, and compute a $\chi^2$ statistic for each 2PCF monopole with respect to that of the DESI One-Percent data, in the separation range $\sim 5~\hMpc$ to $\sim 30~\hMpc$, using the square root of the diagonal of the (N=60) jackknife covariance matrix of the data 2PCF as the uncertainty. The best fit mock parameters were determined by first finding the minimum $\chi^2$ value over the grid of parameters and then fitting a 1 dimensional parabola to the $\chi^2$ values vs. $V_\mathrm{mean}$ and $f_{\rm sat}$ independently while holding the other parameter fixed at the grid value where the $\chi^2$ is minimized.  The best fit parameter values were determined to be the location of the minimum of the parabola fit to each parameter's $\chi^2$ values.
\\

\noindent \textit{3.3.2.1 \textsc{Uchuu} ELG}
\vspace{0.2cm}\\ 
\noindent The best-fit $V_\mathrm{mean}$ and $f_\mathrm{sat}$ parameters used to generate our \textsc{Uchuu} One-Percent mock lightcones for ELGs are listed in the first row of Table~\ref{tab:elg-shamparm}. The subsequent rows show the best fit parameters for each box used in the construction of the mocks fit separately. $\sigma_{V}$ was fixed at $30~\kms$ as in \citet{rodrigueztorres17}. This was due to the relative lack of effect of $\sigma_{V}$ on the clustering in the mocks. The above scheme will be explained in more detail in a paper on Year 1 data.  We apply the modified SHAM method to the (sub)halo catalogues from \textsc{Uchuu} boxes at redshifts 0.94, 1.03, and 1.22 to cover the interval $0.88 < z < 1.34$. These boxes cover an irregularly spaced set of redshifts designed to equipartition the ELG data sample. 
\\

\begin{table*}
\centering
    \begin{tabular}{ccccccccccc}
        \hline
        $z_\mathrm{min}$ & $z_\mathrm{max}$ & $N_{\rm eff}$ & $V_{\rm eff}$ & $n_{\rm g}^{\rm ELG}$ & $\chi^2/{\rm d.o.f}$ & $V_\mathrm{mean}$ & $f_{\rm sat}$ & $f_{\rm sat}^{\rm Uchuu}$ & $b^{\rm ELG}$ & $b^{\rm Uchuu}$ \\
        \hline
        0.88 & 1.34 & 200997 & 0.259  & $0.776\times10^{-3}$ & 2.33 & 156.5$\pm$4.2 & 0.131$\pm$0.012 & 0.131 & 1.34$\pm$0.03 & 1.34$\pm$0.01 \\
        0.88 & 1.00 &  66699 & 0.059  & $1.13\times10^{-3}$  & 4.88 & 133.2$\pm$6.1 & 0.179$\pm$0.029 & --- & 1.35$\pm$0.04 & 1.27$\pm$0.01  \\
        1.00 & 1.16 & 69110  & 0.089  & $0.78\times10^{-3}$  & 3.62 & 156.5$\pm$6.9 & 0.131$\pm$0.023 & --- & 1.25$\pm$0.06 & 1.35$\pm$0.01  \\
        1.16 & 1.34 &  65187 & 0.111  & $0.60\times10^{-3}$  & 2.04 & 127.8$\pm$5.4 & 0.246$\pm$0.034 & --- & 1.43$\pm$0.01 & 1.42$\pm$0.01  \\
	\hline
	\end{tabular}
\caption{Best-fit modified SHAM parameters for the ELGs for each redshift bin and for the entire mock. The first two columns show the minimum and maximum redshift used to define the different ELG samples. The following columns give, for each redshift bin, the number of galaxies, the effective volume (in $h^{-3}\mathrm{Gpc}^{3}$) and the galaxy number density (in $h^{3}\mathrm{Mpc}^{-3}$).
The $\chi^2/{\rm d.o.f}$ is obtained by computing the monopole of the two-point correlation functions of the Uchuu ELG mocks for the best-fit model parameters and comparing to the observed monopole of the DESI One-Percent sample two-point correlation function, in the separation range $\sim 5~\hMpc$ to $\sim 30~\hMpc$. This was the same method used to obtain the best-fit $V_\mathrm{mean}$ (in $\kms$), and $f_\mathrm{sat}$ parameters used to generate our \textsc{Uchuu} One-Percent mock lightcones. 
The last two columns show the bias calculated from the data and the mocks respectively ($b^{\rm ELG}$ and $b^{\rm Uchuu}$). The errors reported for the data are statistical only from the fitting code, the errors reported for the mocks are the standard error in the mean of the 102 independent footprint mocks.}
\label{tab:elg-shamparm}
\end{table*}


\noindent \textit{3.3.2.2 \textsc{Uchuu} QSO} 
\vspace{0.2cm}\\ 
\noindent The best-fit $V_\mathrm{mean}$ and $f_\mathrm{sat}$ parameters
are listed in Table~\ref{tab:qso-shamparm}, used to generate our \textsc{Uchuu} One-Percent mock lightcones for QSO. For the same reasons as in Section~3.3.2.1, we fixed $\sigma_{V}$ at $30~\kms$. We work with the (sub)halo catalogues from \textsc{Uchuu} boxes at four different redshifts, namely $z =$ 1.03, 1.32, 1.65, and 1.9 to cover the redshift range of the QSO sample (0.9 to 2.1). Estimates of quasar redshift have large uncertainties \citep{Chaussidon23} of a few hundred $\kms$ due to the broadness of the emission lines and the intrinsic shifts from other emission lines 
\linebreak 
\citep{youles2022effect}. Hence we introduce Gaussian redshift errors such that
\begin{equation}
    z_\mathrm{final} = z + \mathcal{G}(0, \sigma).
\end{equation}
Here, $z_\mathrm{final}$ is the final redshift distribution for the mock quasar catalogs, and $\mathcal{G}(0, \sigma)$ is the Gaussian random error added to the initial redshift distribution $z$. The dispersion $\sigma$ was set as a constant $500~\kms$ for the One-Percent sample over all redshifts. This value was determined by comparing the power spectrum of QSO mocks to the power spectrum of the One-Percent Survey QSO sample with varying values of dispersion.

\begin{table*}
\centering
    \begin{tabular}{ccccccccccc}
        \hline
        $z_\mathrm{min}$ & $z_\mathrm{max}$ & $N_\mathrm{eff}$ & $V_{\rm eff}$ & $n_{\rm g}^{\rm QSO}$ & $\chi^2/{\rm d.o.f}$ & $V_\mathrm{mean}$ & $f_{\rm sat}$ & $f_{\rm sat}^{\rm Uchuu}$ & $b^{\rm QSO}$ & $b^{\rm Uchuu}$ \\
        \hline
        0.9 & 2.1 & 23085 & 0.66  & $3.52 \times 10^{-5}$ & 1.49 & 309$\pm$16 & 0.14$\pm$0.05 & 0.14 & 2.13$\pm$0.20  & 2.29$\pm$0.02  \\
        0.9 & 1.2 &  4605 & 0.12  & $3.84 \times 10^{-5}$ & 0.47 & 280$\pm$44 & 0.15$\pm$0.13 & --- & 1.74$\pm$0.11  & 1.86$\pm$0.01  \\
        1.2 & 1.5 &  6241 & 0.17  & $3.67 \times 10^{-5}$ & 0.94 & 307$\pm$29 & 0.14$\pm$0.09 & --- & 2.19$\pm$0.22  & 2.11$\pm$0.02 \\
        1.5 & 1.8 &  6583 & 0.19  & $3.46 \times 10^{-5}$ & 1.70 & 305$\pm$26 & 0.13$\pm$0.09 & --- & 2.44$\pm$0.18  &  2.44$\pm$0.02  \\
        1.8 & 2.1 &  5656 & 0.17  & $3.33 \times 10^{-5}$ & 1.30 & 333$\pm$39 & 0.12$\pm$0.11 & --- & 2.95$\pm$0.18 & 2.74$\pm$0.02 \\
		\hline
	\end{tabular}
\caption{Same as Table \ref{tab:elg-shamparm} but for QSOs. The $\chi^2/{\rm d.o.f}$ is obtained by computing the monopole of the two-point correlation functions of the Uchuu QSO mocks for the best-fit model parameters and comparing to the observed monopole of the DESI One-Percent sample two-point correlation function, in the separation range $\sim 5~\hMpc$ to $\sim 30~\hMpc$. This was the same method used to obtain the best-fit $V_\mathrm{mean}$ (in $\kms$), and $f_\mathrm{sat}$ parameters used to generate our \textsc{Uchuu} One-Percent mock lightcones. }
\label{tab:qso-shamparm}
\end{table*}

\subsubsection{Constructing the \textsc{Uchuu} lightcones}
\label{sec:uchuu_lightcones}

\begin{figure}
	\includegraphics[width=\columnwidth]{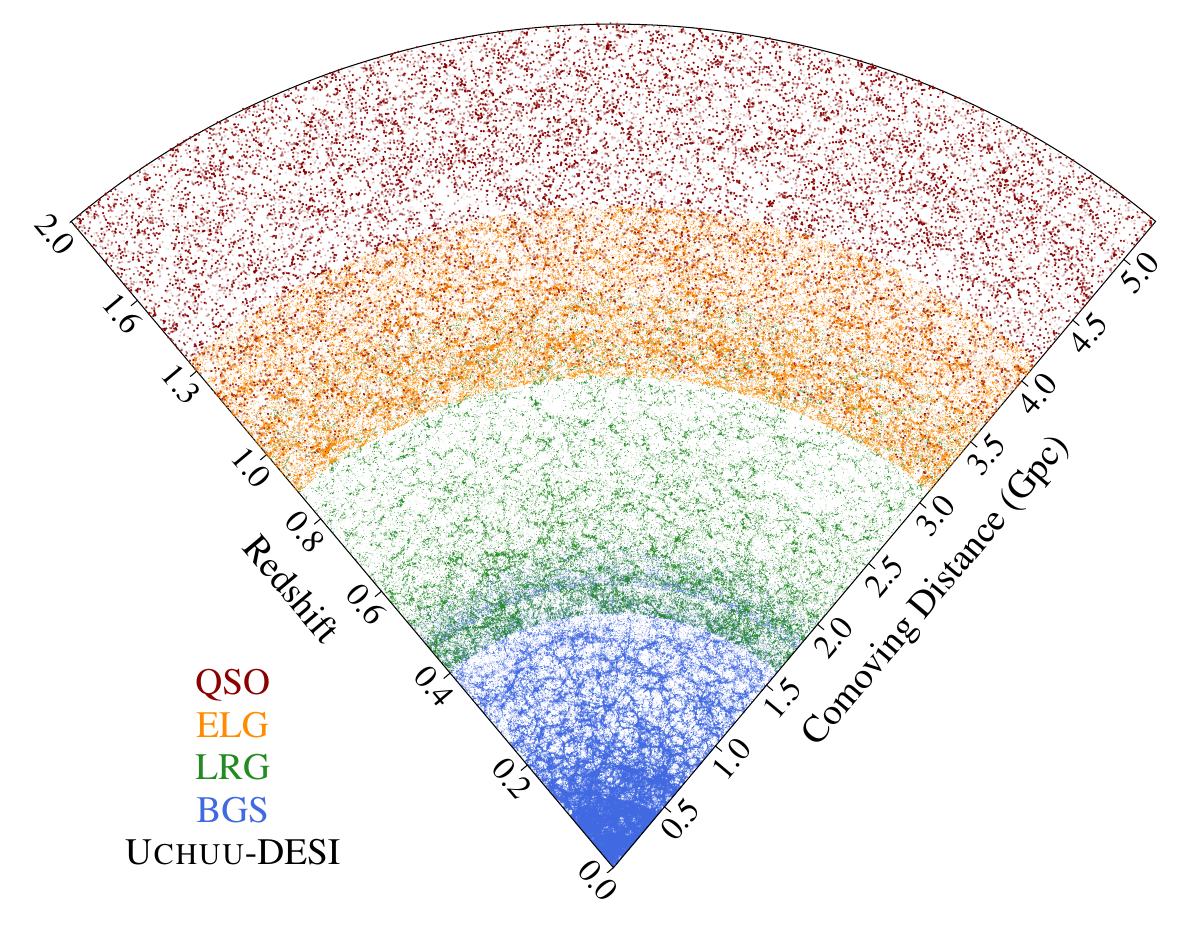}
    \caption{
    Slice through the \textsc{Uchuu}-DESI catalogues with objects coloured by tracer type: BGS (blue), LRG (green), ELG (orange), QSO (red). The slice shows an $80~\deg$ wedge projected into comoving coordinates using a Planck background cosmology, extending out to a maximum redshift of 2. The projected thickness is $1~\deg$ for QSOs and $0.5~\deg$ for ELGs.  The thickness for the BGS and LRG samples is adjusted with redshift to achieve a constant average projected number density.  The transparency of points out of the projection plane falls off with Gaussian weighting.
    }
    \label{fig:uchuu-desi}
\end{figure}

After applying the SHAM method to populate the simulation catalogues with galaxies and quasars, we generate the \textsc{Uchuu}-DESI lightcones for each tracer by joining together the cubic boxes in spherical shells \citep[see][for a detailed explanation of this method]{Smith22b}. The following steps are involved in creating the lightcone:

\begin{enumerate}
    \item The $(x,y,z)$ Cartesian coordinates of each galaxy/quasar cubic-box snapshot are transformed so that for any chosen observer position, the observer is at the origin. For the BGS mocks, periodic wrapping is applied to the galaxy coordinates so that the origin is the centre of the box and the corner of the original un-transformed box is used as the observer position. For the LRG, ELG, and QSO mocks, the center of the box is chosen as the observer position. 
    \item The cubic box is cut into a spherical shell, centred on the origin. The comoving distance between the observer and the inner/outer edges of the shell corresponds to the redshift halfway between this snapshot and the next/previous snapshot. In cases where the outer edge of the shell is bigger than the cubic box, periodic replications are applied. If this shell is at a high enough redshift, the inner edge will also be bigger than the central box. 
    \item The spherical shells from each snapshot are joined together to make the lightcone.
    \item The Cartesian coordinates are converted to (RA, Dec, $z$). When computing the redshift, we also include the effect of peculiar velocities of galaxies along the line of sight. 
    \item Depending on the tracer, an extra step is applied to make sure the correct number density of galaxies/quasars is achieved in the lightcone, and to avoid discontinuities at the interfaces between shells:
    \begin{enumerate}
        \item For the BGS lightcone, to obtain an $r$-band luminosity function that evolves smoothly with redshift, 
        we apply a rescaling to the magnitudes in the final step. This rescaling is described in \citet{DongPaez22} and ensures that the correct number density of galaxies is achieved in the lightcone, while avoiding any discontinuities between shells. Next, we assign a $g-r$ colour to each galaxy using the method and colour distributions presented in \citet{Smith22a}. To convert the absolute $r$-band magnitudes to the observed apparent magnitude, we use a set of colour-dependent $k$-corrections from the GAMA survey\footnote{The colour-dependent GAMA $k$-corrections we use are polynomial functions that depend on the galaxy properties that exist in the mock (redshift, absolute $r$-band magnitude and rest-frame $g-r$ colour). These are the same $k$-corrections that have been applied to the One-Percent survey BGS data \citep{dr}.} \citep[see][]{Smith22a}. Finally, we apply a magnitude cut of $r<19.5$ to match the faint apparent magnitude limit of the BGS-BRIGHT survey.
        \item For the LRG lightcone, we randomly downsample the galaxy population in those redshift ranges where the $n(z)$ of the lightcone is above the observed one. Additionally, in this step, we extend our LRG lightcone up to redshifts 0.4 and 1.1 for the sole purpose of producing Figures~\ref{fig:all-ndens}~and~\ref{fig:uchuu-desi}.
    \end{enumerate}
    
\end{enumerate}

We applied the aforementioned steps to each of the four DESI tracers to create their respective full-sky lightcones. The lightcones were then cut to match the northern and southern areas of the DESI One-Percent Survey footprint, as shown in Figure \ref{fig:figure1}. In this study, we retained all objects within the survey footprint, regardless of completeness, for all tracers. Since our methods emulate an observed catalogue rather than a parent catalogue, applying any correction for fibre collisions or the effects of applying fibre assignment on the mock catalogue would be incorrect and lead to an underselection of tracers. 

Figure~\ref{fig:uchuu-desi} presents a visual representation of all four DESI tracers within a thin slice of an \textsc{Uchuu}-DESI lightcone. The slice shows an $80~\deg$ wedge projected into comoving coordinates using a Planck background cosmology, extending out to a maximum redshift of 2. Since the effective volume of the DESI One-Percent Survey is small, its footprint can be replicated over the full sky to generate a significant number of \textsc{Uchuu} One-Percent lightcones for each tracer, enabling us to compute covariance errors for the clustering measurements. 

This is achieved by first moving the position of rosettes into a small rectangular region, where the separation of the rosettes is only conserved for closely separated rosettes (i.e. the triplet of rosettes highlighted in Figure~\ref{fig:figure1}, the three close pairs of rosettes, and the cluster of five rosettes at RA $\sim 0$). This rectangular region is then replicated across the sky, and for each mock, the rosettes (or clusters of closely separated rosettes) are taken from different copies of the rectangular region. Finally, the positions of the rosettes are transformed back to match the One-Percent footprint. This enables us to make 102 One-Percent lightcones for each tracer. However, since the relative positions of the rosettes were not fixed, we only trust clustering measurements on scales smaller than the clusters of rosettes. For the BGS tracer at $z \sim 0.2$, this corresponds to a comoving separation of $\sim 100~\hMpc$, and is larger for the other tracers at higher redshifts. It is worth noting that the mocks are not fully independent above $z = 0.36$, due to periodic replications of the box. Nevertheless, we checked the degree of overlap by tracking repeated halo IDs within the footprint of the mocks. We found that, for all four tracers and for the effective volume we are considering, the overlap is less than 5$\%$. Therefore, we treat the 102 lightcones as if they were completely independent. The area on the sky of each of our 102 lightcones is $191.4~\sqdeg$. This is larger than the area of the data catalogues, since the Uchuu lightcones are complete, and no regions are masked.

Figure \ref{fig:all-ndens} shows a comparison between the comoving number density of the DESI One-Percent Survey (points) and the mean comoving number density of the \textsc{Uchuu} One-Percent mock lightcones (solid lines) constructed for each of the four tracer samples. 
Overall, the agreement between \textsc{Uchuu} and DESI is good, with the differences in the number density being within the error bars for the dark time tracers and with the difference from BGS being explained by the redshift dependence of the GAMA-derived E-corrections not being consistent with the DESI data as described in Section~\ref{sec:clus_bymass}. 

\begin{figure}
	\includegraphics[width=0.95\columnwidth]{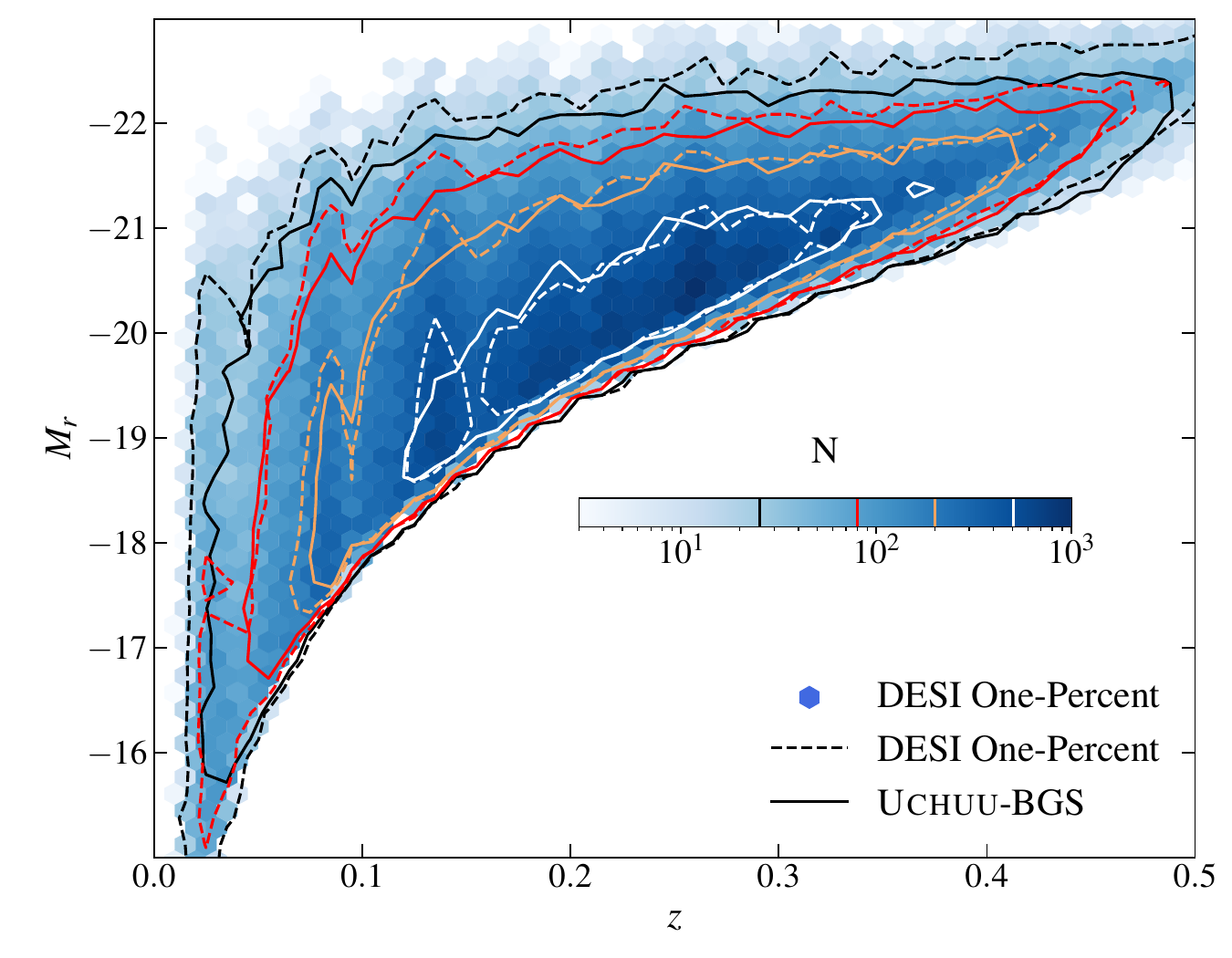}
    \includegraphics[width=0.95\columnwidth]{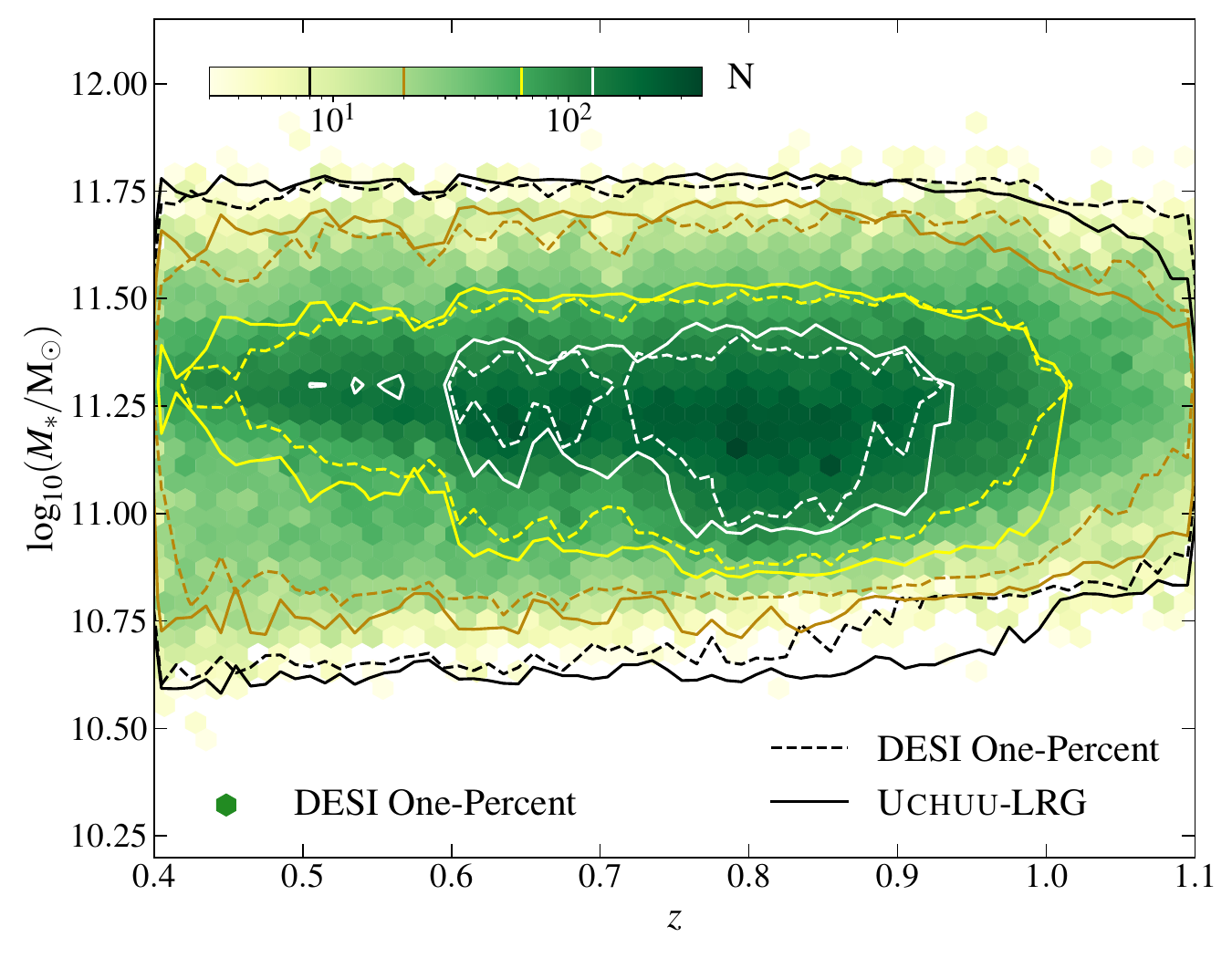}
    \caption{\textit{Top panel}: Absolute magnitude vs. redshift for the DESI One-Percent BGS sample (hexagonal bins), compared to one of the \textsc{Uchuu}-BGS lightcones (contours). Absolute magnitudes have been $k$- and $E$-corrected.  
    \textit{Bottom panel}: Logarithm of the stellar mass vs. redshift for the DESI One-Percent LRG sample (hexagonal bins), compared to one of the \textsc{Uchuu}-LRG lightcones (contours).}
    \label{fig:bgs_magnitude_vs_redshift}
\end{figure}

In Figure \ref{fig:bgs_magnitude_vs_redshift}, the top panel displays the distribution of $r$-band absolute magnitude for the BGS sample in the DESI One-Percent Survey (hexagonal bins) and one of the \textsc{Uchuu}-BGS lightcones (contours). The same colour-dependent $k$-correction has been applied to the galaxies in the data and mock \citep[see][]{Smith17}, as well as the same E-correction, $E(z) = Q_0 (z-z_0)$, where $Q_0=0.97$ and $z_0=0.1$. 
The bottom panel of Figure \ref{fig:bgs_magnitude_vs_redshift} shows the logarithm of stellar mass versus redshift for the DESI One-Percent LRG sample (hexagonal bins) and for the \textsc{Uchuu}-LRG lightcones (contours). This figure illustrates how the stellar mass distribution varies with redshift. This trend is also evident in Figure \ref{fig:lrg_stellarmass}, which presents the stellar mass function (SMF) for LRGs in the data and \textsc{Uchuu} One-Percent lightcones across several redshift bins. We account for incompleteness at the low-mass end of the stellar mass function in the \textsc{Uchuu}-LRG lightcones by randomly downsampling galaxies from the complete SMF adopted from \citet{RodriguezTorres16} (represented by a dashed line in Figure \ref{fig:lrg_stellarmass}). The completeness of the SMF measured in each redshift bin is further analyzed in Section~\ref{sec:results-clustering}. Both the luminosity and stellar mass included in our \textsc{Uchuu} galaxy catalogues allow us to study the dependence of the two-point correlation function on these properties for BGS and LRG in the DESI One-Percent survey.

\section{Results}
\label{sec:results}

In this section, we compare the clustering signal measured for each of the galaxy and quasar samples in the DESI One-Percent Survey with that predicted by the Planck cosmology using our \textsc{Uchuu} One-Percent mock lightcones, as described in the previous section. Additionally, we explore the dependence of the galaxy clustering on luminosity and stellar mass for BGS and LRG galaxies, respectively, and the dependence of the galaxy clustering on redshift for ELGs and QSOs. We estimate the halo occupancy and large-scale bias for all four targets.

\subsection{Clustering statistics: DESI vs. \textsc{Uchuu}}
\label{sec:results-clustering}

While a more detailed description of the calculation of the clustering statistics will be given in \citet{lasker24a}, we provide a brief overview specifically tailored to our analysis.

We use the Landy-Szalay \citep{LS93} estimator to measure the two-dimensional correlation function, $\xi(s,\mu)$, where $s$ represents the separation between a pair of objects in units of $\hMpc$, and $\mu$ is the cosine of the angle between the pair separation vector and the line-of-sight. We used logarithmically spaced bins of separation and linearly spaced bins in $\mu$ between -1 and 1. The number and range of separation bins and the number of $\mu$ bins were chosen based on the needs of each tracer. BGS used 30 separation bins between 0.03 and 100 $\hMpc$ and 201 $\mu$ bins. LRGs used 17 separation bins between 5 and 100 $\hMpc$ and 61 $\mu$ bins. ELGs used 16 separation bins between 5 and 100 $\hMpc$ and 201 $\mu$ bins. QSOs used 15 separation bins between 5 and 100 $\hMpc$ and 51 $\mu$ bins.
The Landy-Szalay estimator is given by the following equation:
\begin{equation}
    \xi(s,\mu) = \frac{DD(s,\mu) - 2DR(s,\mu) + RR(s,\mu)}{RR(s,\mu)}
\end{equation}
\noindent where the normalized pair counts in the correlation function estimate with $DD$ providing counts of data galaxies at each ($s,\mu$) with respect to other data galaxies, $DR$ providing counts of data galaxies with random points, and $RR$ providing the counts of random points with other random points. 

We then decompose $\xi(s,\mu)$ into Legendre polynomials,
\begin{equation}
\xi_\ell(s) = \frac{2\ell+1}{2} \int^1_{-1} \xi(s,\mu)P_\ell(\mu)d\mu .
\end{equation}
We measure the monopole and quadrupole ($\ell=0,2$), which are the first non-zero Legendre multipoles of the redshift-space two-point correlation function. To account for the selection function, we generate random samples that are 50 times larger than the One-Percent data and use them to estimate the data-random and random-random pair counts for each tracer. We estimate the two-point correlation functions using the Python package \textsc{pycorr}\footnote{\url{https://github.com/cosmodesi/pycorr}}, which is a wrapper for correlation function estimation wrapping a modified version of \textsc{Corrfunc} \citep{Corrfunc2020}. For the Uchuu lightcones, we create a different set of uniform randoms for each tracer, which match the same footprint as the mocks.

The data sample is primarily weighted using the Pairwise Inverse Probability (PIP) weights \citep{BianchiAndPercival}, which reflect whether pairs of galaxies would have been observed in 128 alternate realizations of DESI, in addition to which pairs are observed in the actual survey \citep{lasker24a}. The PIP weights are combined with FKP weights \citep{Feldman1994}, $w_\mathrm{FKP}$, to account for the inhomogeneous sampling density of the data sample, which are defined as
\begin{equation}
w_\mathrm{FKP} = \frac{1}{1+P_0 n(z)},
\end{equation}
where $n(z)$ is the weighted number density, and $P_0$ is the same power spectrum value chosen in Section~\ref{sec:one_percent_properties}, equation \ref{eq:eff_vol}. 


For ELG and QSO data clustering measurements, we apply angular upweighting based on the angular clustering of the parent and data catalogues \citep{PercivalAndBianchi,angup}. Random points, on the other hand, are only weighted by FKP weights.

For the BGS, we correct for incompleteness using Individual Inverse Probability weights (IIP) weights instead of PIP weights. These weights were computed from the same 128 alternate realizations of DESI as the PIP weights \citep{lasker24a}. However, where PIP weights up-weight galaxy \textit{pairs} by the inverse of the fraction of realizations in which the \textit{pairs} are observed, IIP weights up-weight \textit{individual galaxies} by the inverse of the fractions of realizations in which the \textit{individual galaxies} were observed. The difference in the 2PCF monopole over the separation range used in the fit is $\sim$2\%.

We use only FKP weights to measure the clustering in our \textsc{Uchuu} lightcones, with the same $P_0$ values as for the data measurements. The $n(z)$ is estimated from the mock.

\begin{figure*}
    \centering
    \includegraphics[width=0.95\columnwidth]{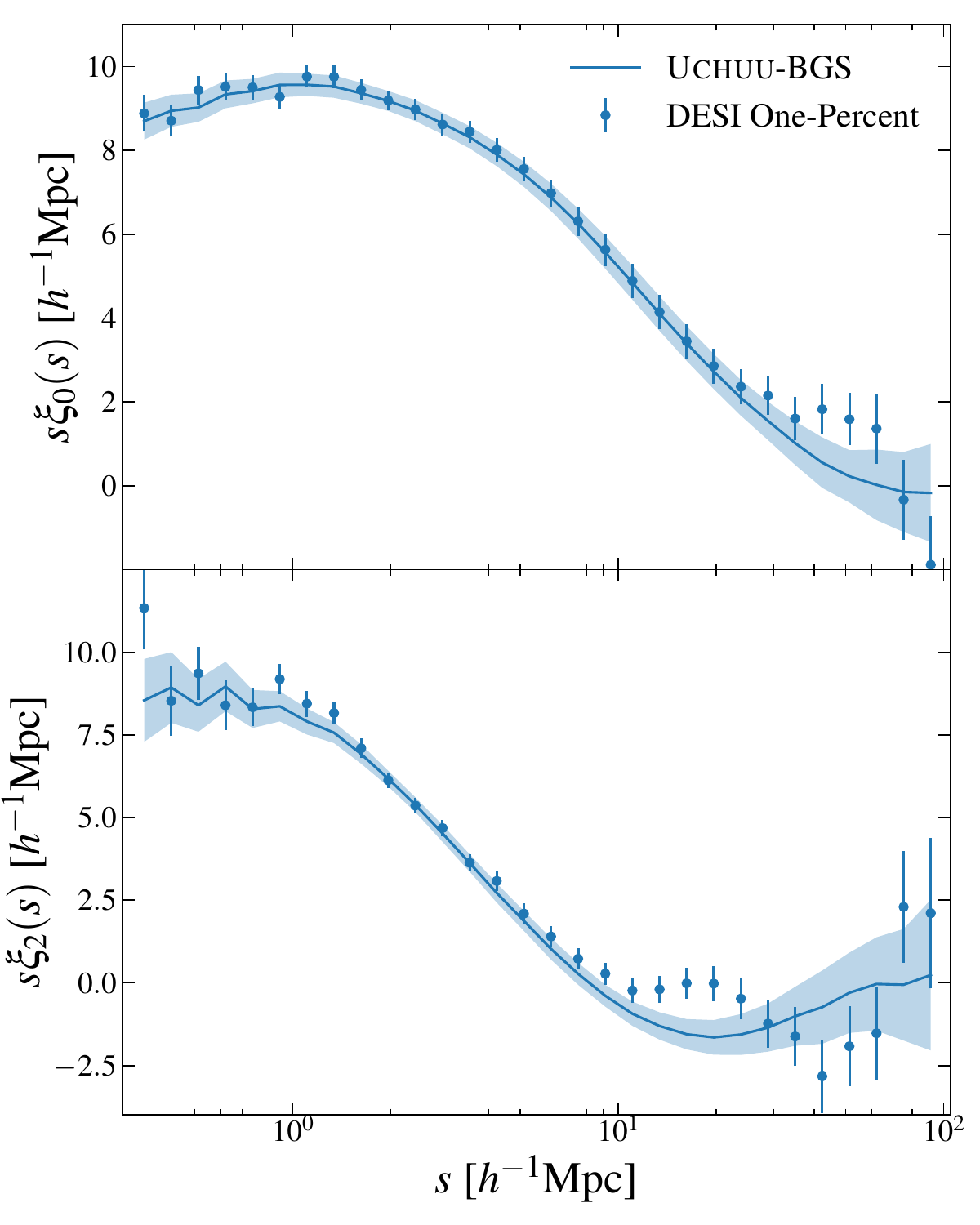}
    \includegraphics[width=0.95\columnwidth]{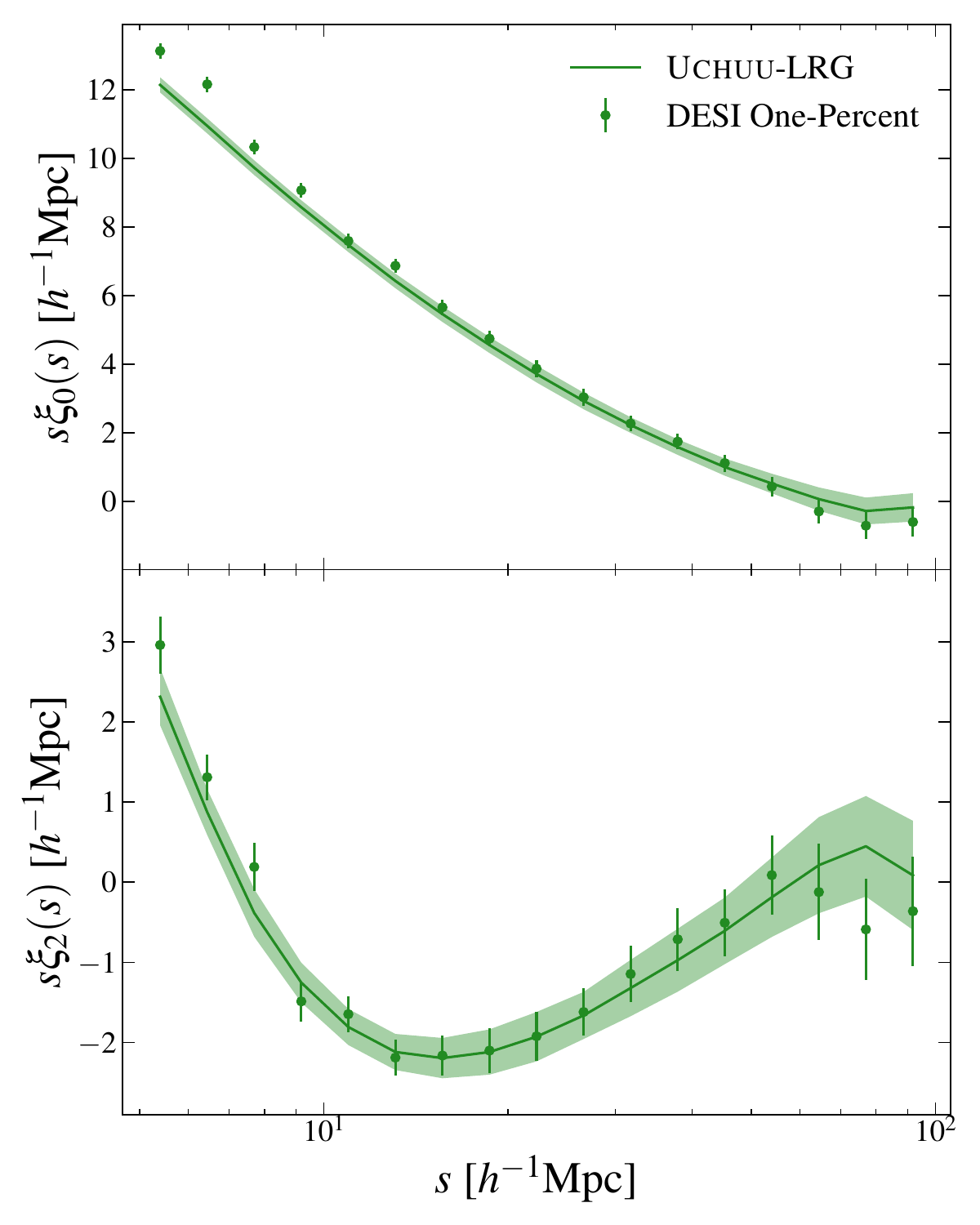}
    \includegraphics[width=0.95\columnwidth]{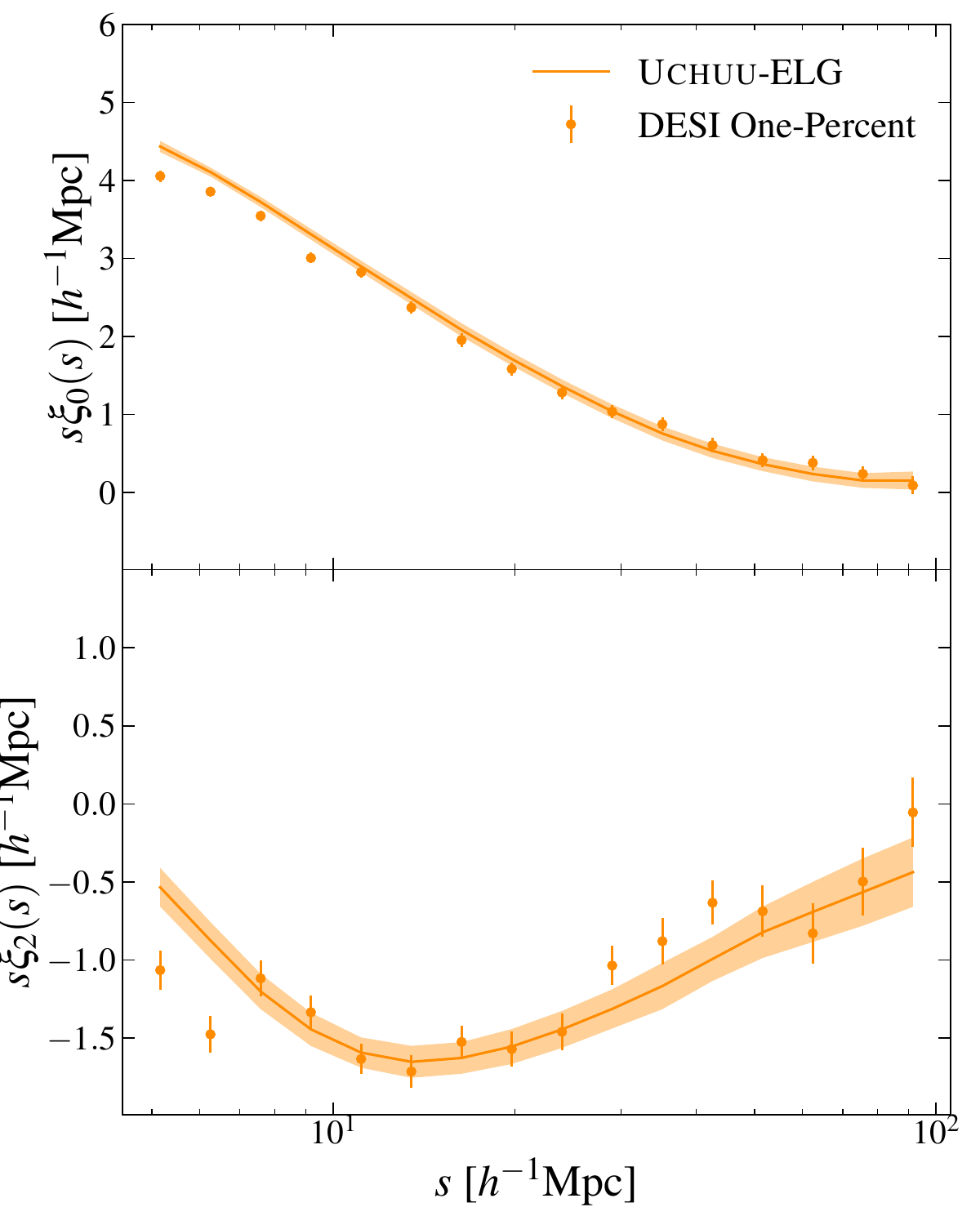}
    \includegraphics[width=0.95\columnwidth]{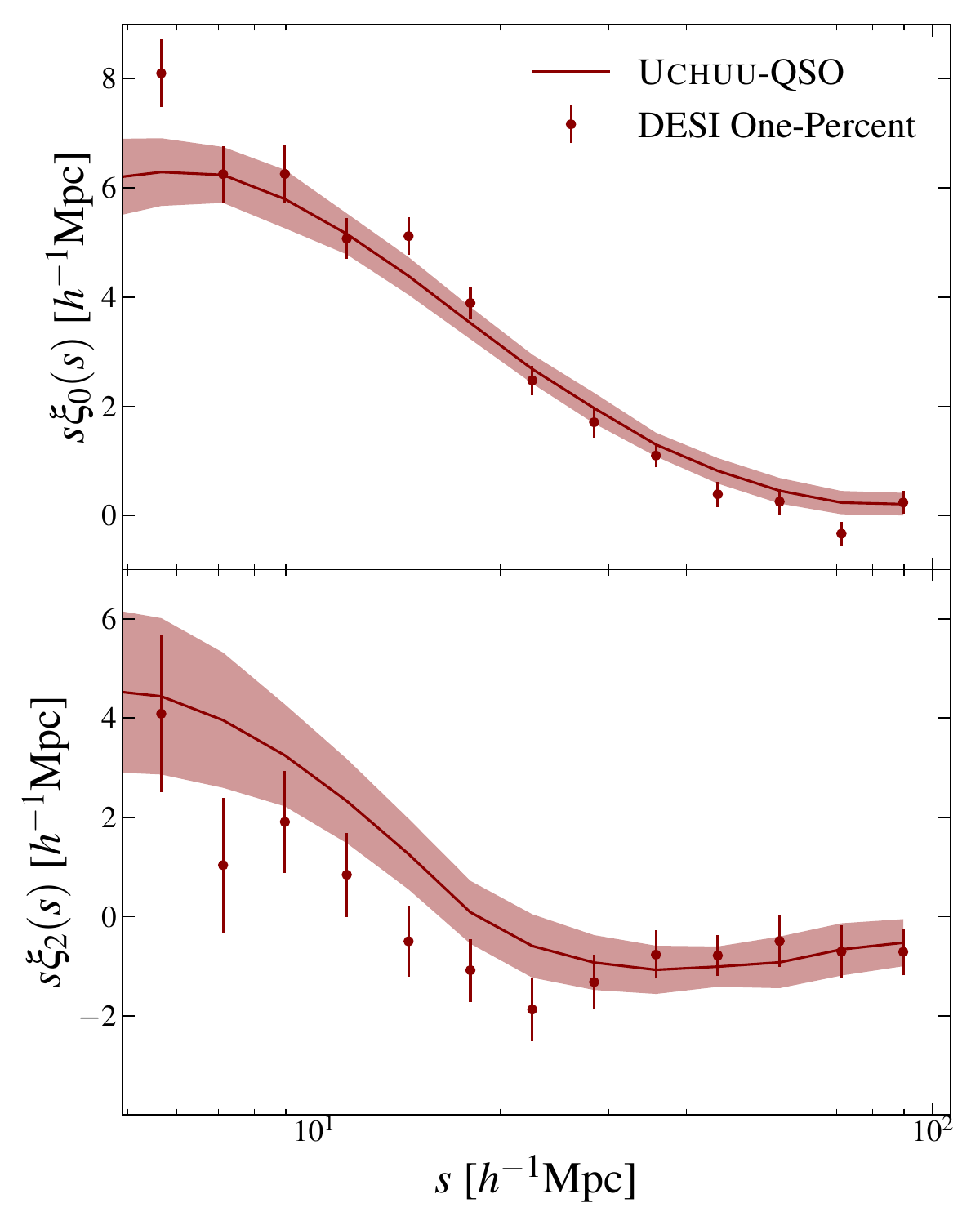}
    \caption{Measurements of the monopole and quadrupole of the redshift-space correlation function for all four tracers from the DESI One-Percent samples, in the redshift intervals $0.1 < z < 0.3$ (BGS), $0.45 < z < 0.85$ (LRG), $0.88 < z < 1.34$ (ELG) and $0.9 < z < 2.1$ (QSO). The theoretical predictions from the mean of the independent \textsc{Uchuu}-DESI lightcones generated for each tracer are shown as solid curves, while the shaded areas correspond to the error from the RMS of the 102 mocks. The clustering measurements for the BGS, LRG, ELG and QSO  are shown in blue, green, orange and red, respectively, with the monopole and quadrupole shown in the upper and lower panels. The points with error bars represent the measurements from the DESI One-Percent Survey.}
    \label{fig:sv3-2pcf}
\end{figure*}

\subsubsection{Redshift-space correlation function}

Figure~\ref{fig:sv3-2pcf} presents the measurements of the redshift-space correlation function for the four tracers drawn from the DESI One-Percent parent samples. The monopole, $\xi_0(s)$, and quadrupole, $\xi_2(s)$, are shown for different redshift intervals indicated in the legends. The points with errors indicate the DESI One-Percent clustering measurements, and the solid curves represent the theoretical predictions based on Planck cosmology, determined from the mean of the \textsc{Uchuu} One-Percent lightcones, in the redshift intervals indicated in the figure caption (see also Table~\ref{tab:all-basic}). The $1\sigma$ errors are estimated from the diagonal component of the covariance matrix obtained from our sample of \textsc{Uchuu} One-Percent lightcones generated for each tracer.

For the BGS, we find agreement between \textsc{Uchuu} lightcone and DESI One-Percent monopole measurements on small scales. On larger scales cosmic variance becomes a factor, as the BGS One-Percent sample has a small volume. The quadrupole agrees within $10\%$ at separations below $\sim 6~\hMpc$, but as with the monopole, it is affected by cosmic variance on large scales.

For the LRG sample, we observe agreement between \textsc{Uchuu} lightcones and DESI One-Percent monopole and quadrupole measurements at scales larger than $5\,\hMpc$. However,
on scales below $\sim 5~\hMpc$, we find a systematically low prediction of \textsc{Uchuu} compared to DESI. This finding is consistent with the results obtained from our study of the dependence of clustering on stellar mass and redshift that we discuss in Section~\ref{sec:clus_bymass} (see Figure \ref{fig:sv3-2pcfmass}).
Residuals from the 2PCF monopole, denoted as $r=\xi_0^\mathrm{Uchuu}/\xi_0^\mathrm{data} -1$, maintain values under $10\%$ ranging from $5\hMpc$ up to $50\hMpc$. 

For the ELG monopole, we find a systematically $10\%$ low prediction from Uchuu compared to the data. The quadrupole of the mock shows agreement with the data in the fit window ($5~\hMpc < s < 30~\hMpc$). The bump in the data outside of the fit range ($30 \leq s \leq 45 \hMpc$) is explained by a correlated fluctuation due to the small volume covered by the DESI One-Percent survey and the limited number of pairs at separations comparable to the size of the rosettes (s~ 70 $\hMpc$). 

For the QSO sample, the monopole shows general agreement between the data and mock.  The quadrupole agrees at large scales, but the mock is consistently below the data for smaller scales. 
We will comment further on the consistent over-prediction of the quadrupole at scales below 25 $\hMpc$ when we discuss the clustering results in redshift bins.

\begin{figure}
    \includegraphics[width=\linewidth]{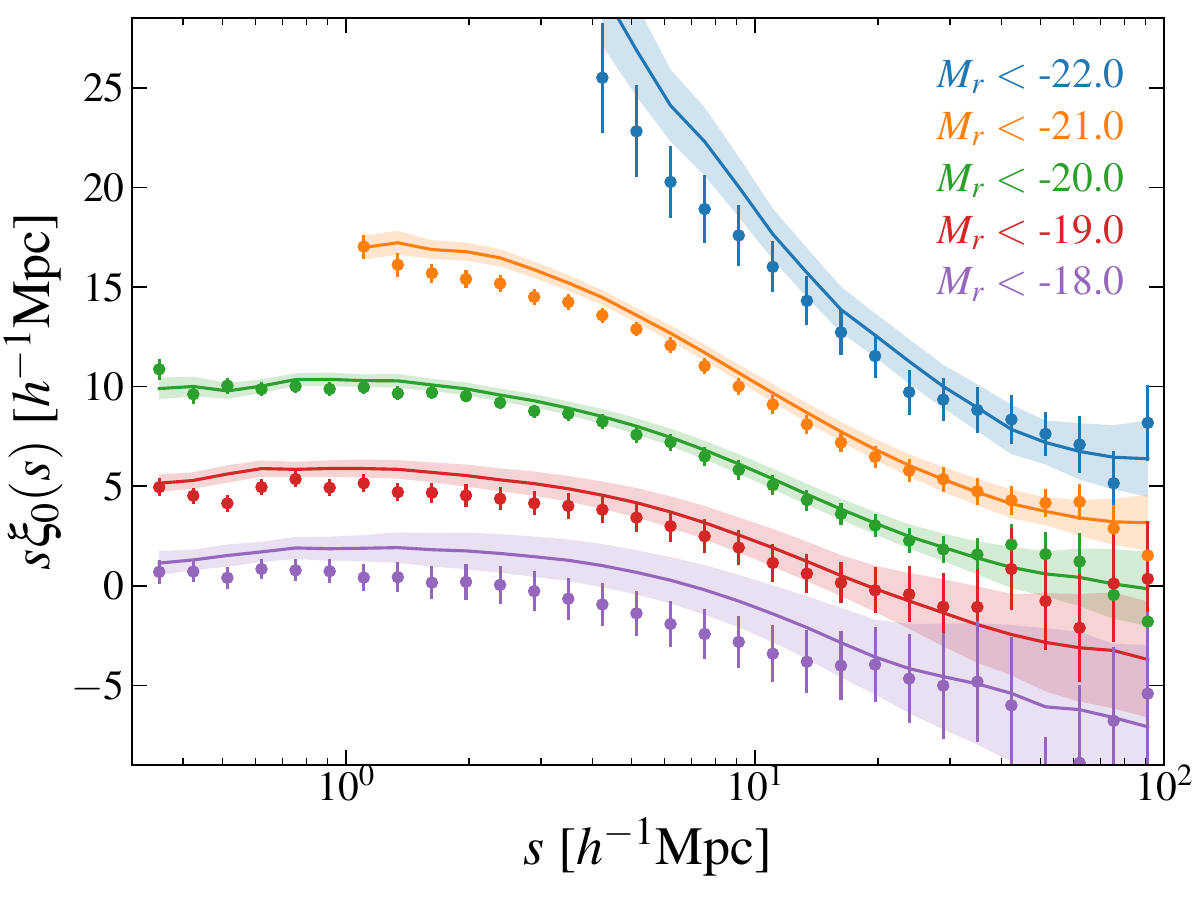}
    \caption{The monopole of the correlation function for BGS galaxies in volume-limited samples. The legend shows different magnitude thresholds with corresponding colours. The solid curves represent the mean of 102 independent \textsc{Uchuu} One-Percent mocks, while the shaded region denotes the error from the RMS of the mocks. DESI One-Percent clustering measurements are indicated by the points with error bars, where the errors are the 1$\sigma$ scatter between the mocks. Each magnitude threshold sample is vertically offset relative to the $M_r<-20$ sample. }
    \label{fig:sv3-2pcfmag}
\end{figure}

\begin{figure*}
    \centering
    \includegraphics[width=\linewidth]{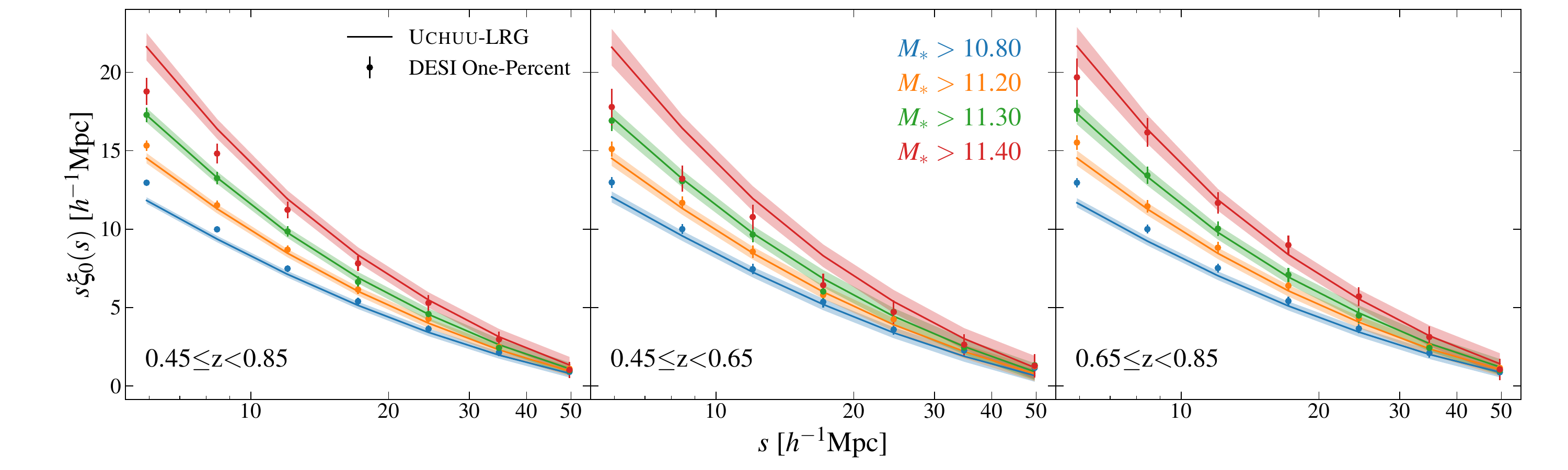}
    \caption{The monopole of the correlation function for LRG galaxies, in stellar mass threshold samples and for different redshift bins. The legend shows different stellar mass thresholds with corresponding colours, and the redshift ranges are $0.45\leq z<0.85$, $0.45\leq z<0.65$, and $0.65\leq z<0.85$ from left to right.}
    \label{fig:sv3-2pcfmass}
\end{figure*}

\subsubsection{Clustering dependence on luminosity, stellar mass and redshift}
\label{sec:clus_bymass}

We have calculated the monopole for different BGS volume-limited samples corresponding to distinct magnitude thresholds as provided in Table~\ref{tab:bgs-hod}.  Figure~\ref{fig:sv3-2pcfmag} illustrates the monopole distribution for five of these thresholds with a vertical offset applied between samples. Although we have calculated the clustering measurements for nine different volume-limited samples, we only display five in the figure to enhance clarity, with a vertical offset applied between samples. We see agreement at a level of 4\% between the mock and BGS data for the intermediate $M_r < -20$ sample. For the brighter $M_r < -21$ and -22 samples, the monopole of the mock shows stronger clustering than the data, with residuals within 10\% and 15\%, respectively. The magnitude threshold used to define these samples is bright, where the luminosity function drops rapidly, so any small changes in the magnitudes due to e.g. errors in the E-corrections will have a large effect on the number density and clustering. Currently, we apply E-corrections to the data which come from luminosity function measurements from the GAMA survey \citep{McNaught-Roberts2014}. We leave it for future work to improve the E-corrections in the data by measuring how the BGS luminosity function evolves with redshift, and apply a consistent evolution to the luminosity function of the mock. We have found that adjusting the data's magnitude thresholds by 0.1 mag improves the agreement of the clustering with the mock. For the fainter samples, which are less affected by uncertainties in the E-corrections, the clustering in the mock is also stronger than in the One-Percent data. For example, the residual for the faintest $M_r$ < -18 sample is at a level of $\sim 20\%$ at $1~\hMpc$. This difference is due to cosmic variance in this small volume, and the One-Percent data happens to have a low clustering amplitude by chance. We have verified this by comparing the clustering of the \textsc{Uchuu}-BGS lightcones with a larger dataset consisting of the first 2 months of DESI observations, and we find agreement with the \textsc{Uchuu} mock. 

For the LRG sample, Figure \ref{fig:sv3-2pcfmass} illustrates agreement between the \textsc{Uchuu} lightcones and DESI One-Percent for stellar masses $\log M_{\ast}>11.4$ (in red) in the redshift range $0.65<z<0.85$ (right panel), indicating that the LRG population in DESI One-Percent is complete in that range. 
However, at $0.45<z<0.65$ (middle panel), we note that \textsc{Uchuu} predictions remains above the data clustering signal. Our lightcones are produced assuming an observed SMF that is complete to its high mass end.  The lack of agreement with data for small separations implies that our assumption is not correct. We do observe agreement from $5~\hMpc$ to $45~\hMpc$ in all redshift bins for $log M_{\ast}>11.2$ and $11.3$ (orange and green lines respectively). Residuals are within $7\%$ and $10\%$, respectively. For $log M_{\ast}>10.8$ the \textsc{Uchuu} monopole is slightly below the data. This result suggested that the complete SMF we have defined in the stellar mass range between $10.8$ and $11.2$, yields a $n(z)$ that surpasses that of the actual complete LRG population. We verified this by the clustering of galaxies with stellar masses in the mocks and data. The results confirm that the \textsc{Uchuu} signal is lower than that from the data.
To estimate the complete LRG population in the $10.8<\log_{10}M_{\ast}<11.2$ range, we make use of the complete PRIMUS galaxy sample presented in \citet{Moustakas_Primus} (squares in Figure \ref{fig:lrg_stellarmass}). Nonetheless, some discrepancies between DESI and PRIMUS samples may account for this discrepancy, such as the use of spectroscopic and photometric redshifts, respectively.

\begin{figure*}
    \centering
    \includegraphics[width=0.95\columnwidth]{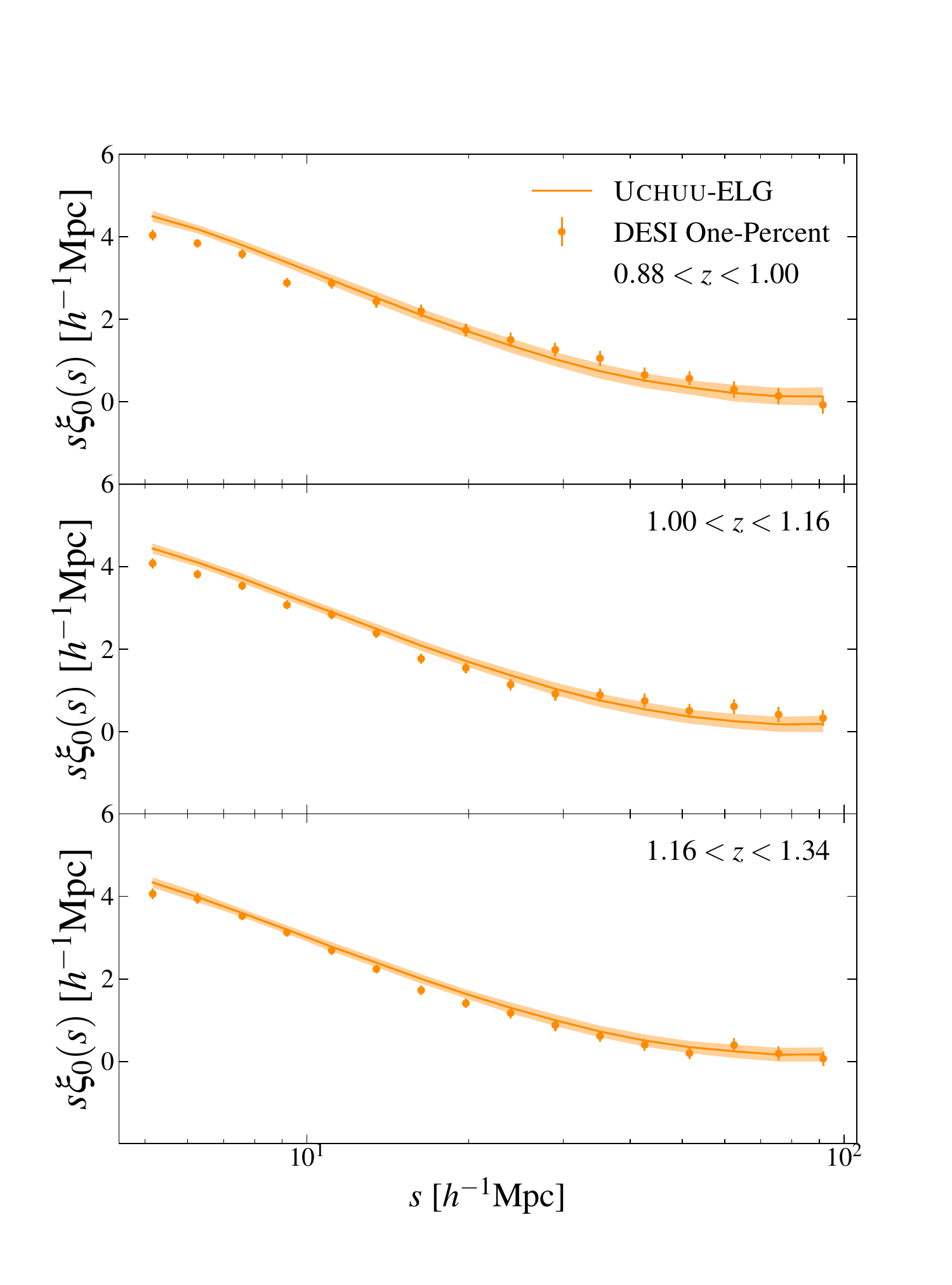}
    \includegraphics[width=0.95\columnwidth]{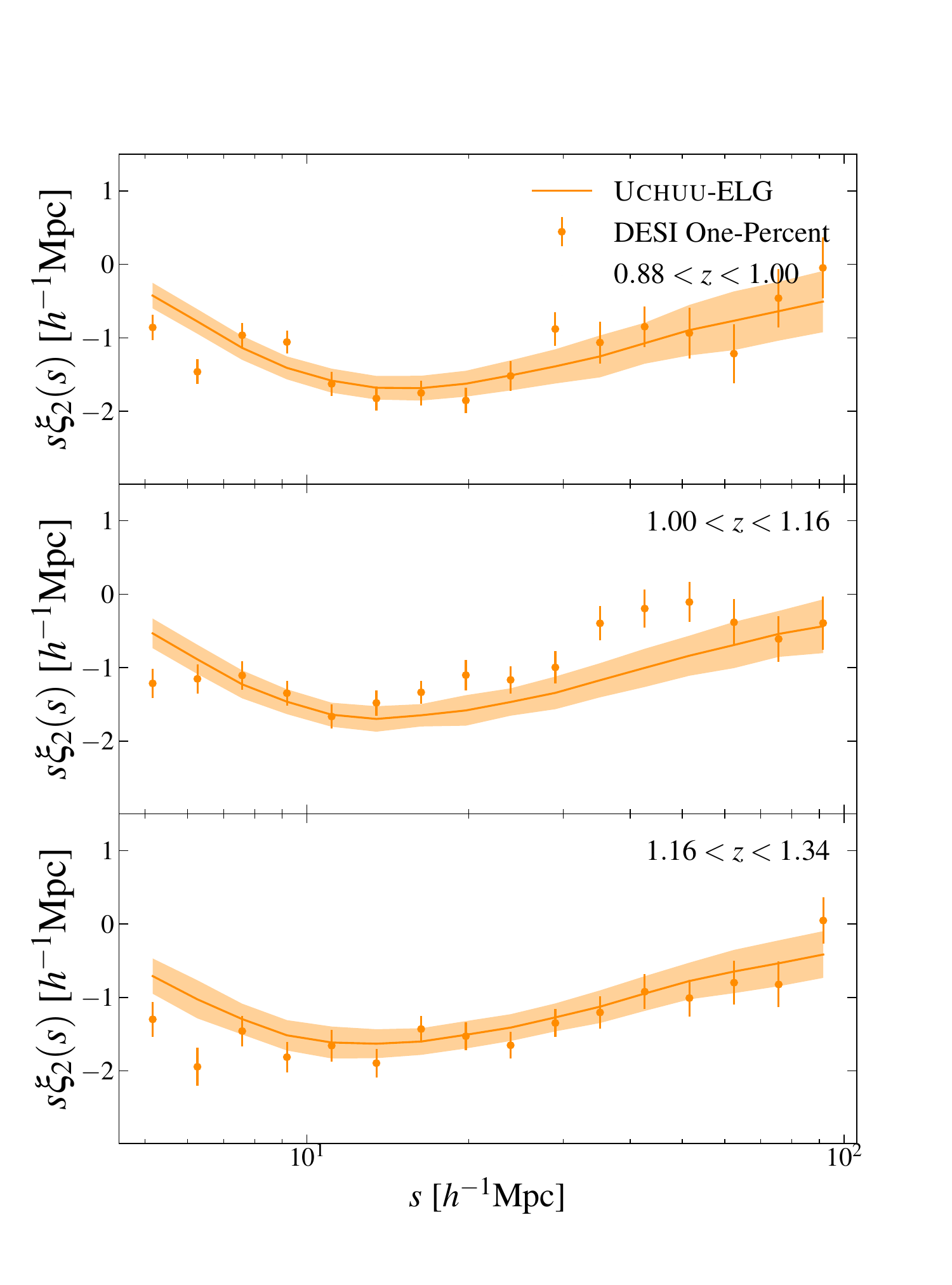}
    \caption{Measurement of the monopole and quadrupole of the redshift-space correlation function for ELGs from the DESI One-Percent parent samples in their respective redshift bins. The solid curves show the mean of the 102  \textsc{Uchuu} One-Percent mocks, where the shaded region is the error from their RMS. DESI One-Percent clustering measurements are indicated by the points with error bars, where the errors are the 1$\sigma$ scatter between the mocks. We note the agreement between data and model on scales larger than $2~\hMpc$, and the noticeable difference on the smallest scales, see text.}
    \label{fig:sv3-2pcfELG}
\end{figure*}

The correlation function of the ELGs, split into redshift bins, is shown in Figure~\ref{fig:sv3-2pcfELG}. These comparisons for the monopole exhibit similar levels of agreement between data and mocks as the combined plot shown in Figure~\ref{fig:sv3-2pcf}. The redshift-binned quadrupole, on the other hand, indicates agreement in the lowest and highest redshift bins.  However, the middle bin indicates that we observe more clustering in the data than we predict in the model. This excess is the source of the excess observed in Figure~\ref{fig:sv3-2pcf} for separations of $30 \leq s \leq 45 \hMpc$.

\begin{figure*}
    \centering
    \includegraphics[width=0.95\columnwidth]{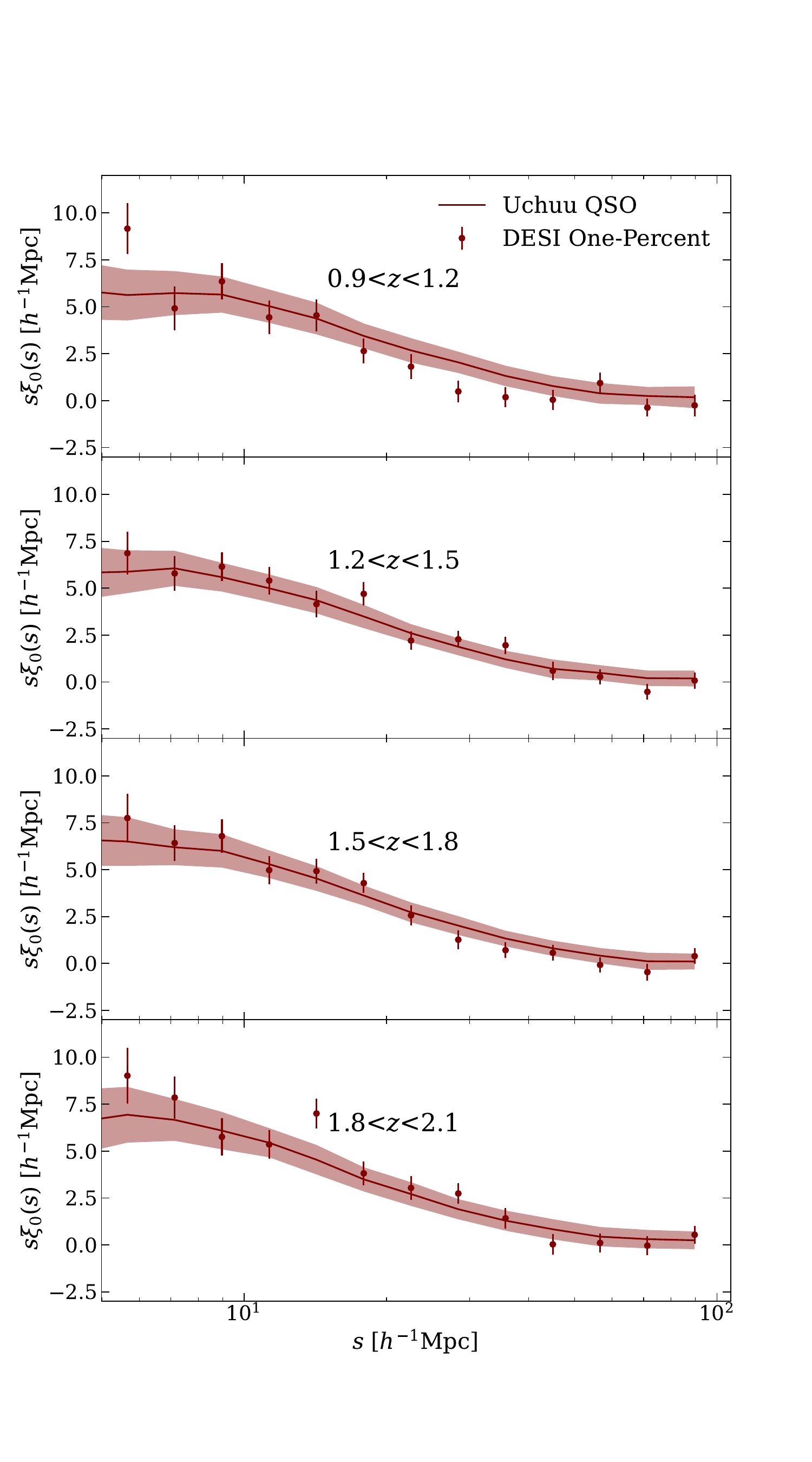}
    \includegraphics[width=0.95\columnwidth]{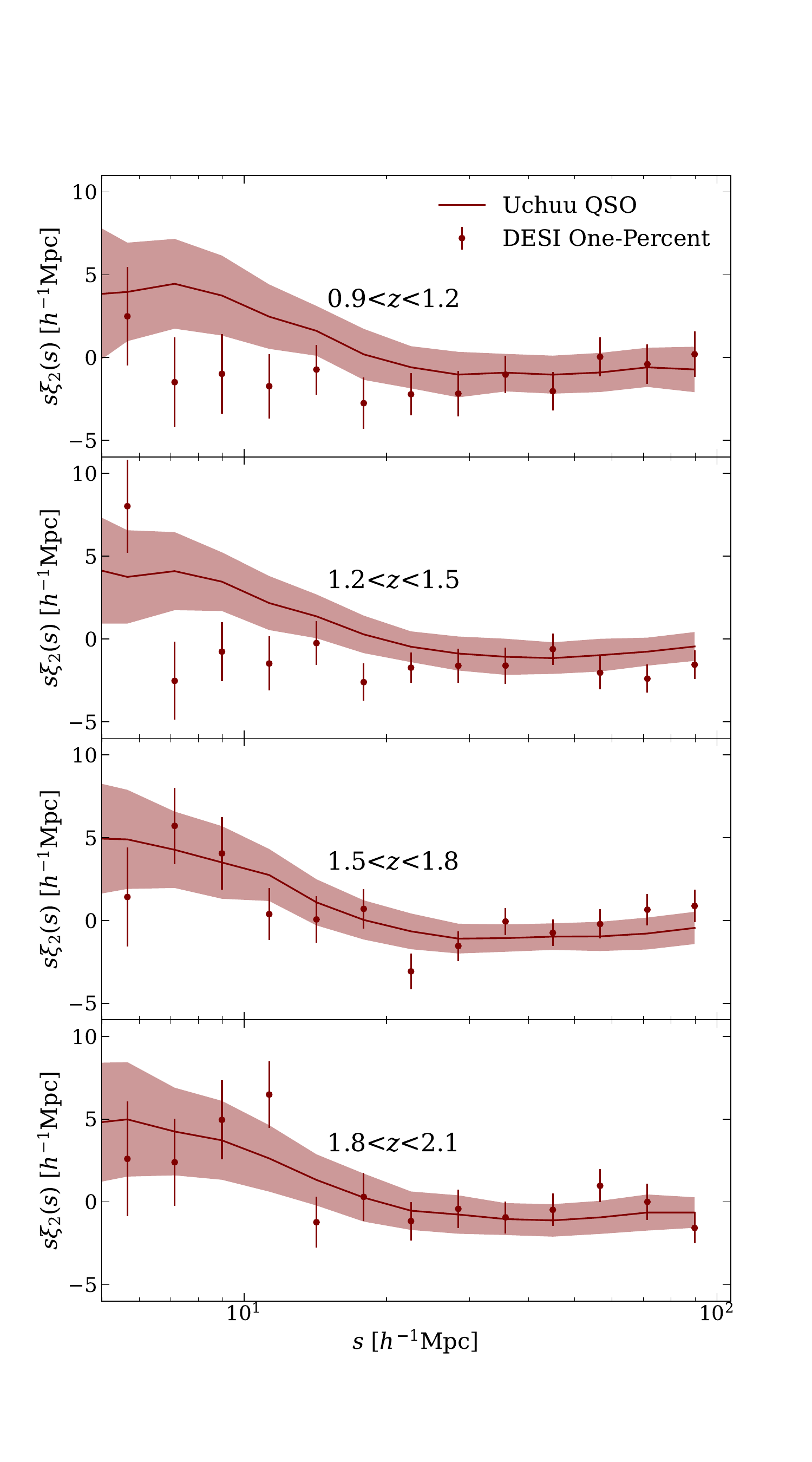}
    \caption{Clustering measurements for monopole (left panels) and quadrupole (right panels) of the DESI One-Percent QSO sample in redshift bins. The solid curves show the mean of the Uchuu mocks, and the shaded region is the
    error from their RMS. Data clustering measurements are indicated by the points with error bars, where the errors are the $1\sigma$ scatter
    between the mocks.}
    \label{fig:sv3-2pcf-qso}
\end{figure*}

For QSOs, Figure~\ref{fig:sv3-2pcf-qso} shows the correlation function in redshift bins. The large error bars on the data points reflect the relatively low statistics of the DESI QSO sample compared to that of the other tracers. For the monopole, we find agreement between the \textsc{Uchuu} mocks and the DESI One-Percent data in the respective bins, for the monopole. 
The quadrupole exhibits agreement in the two high redshift bins, but the model overpredicts the data for small separations in the two low redshift bins.  We have traced this to the lack of redshift dependence in the DESI-SV3 redshift error model. Fits to Uchuu mocks using Y1 data show that there is a strong redshift dependence in the magnitude of the redshift error on QSOs.

\begin{figure*}
    \centering
    \includegraphics[width=0.95\columnwidth]{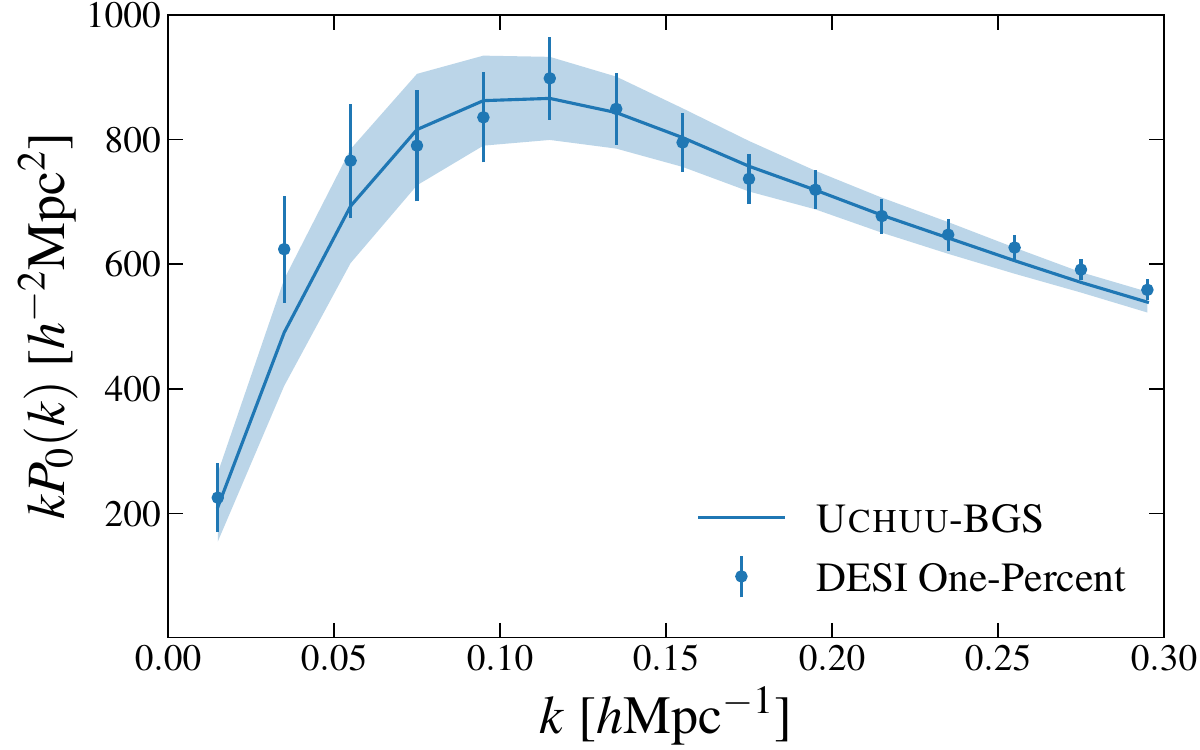}
    \includegraphics[width=0.95\columnwidth]{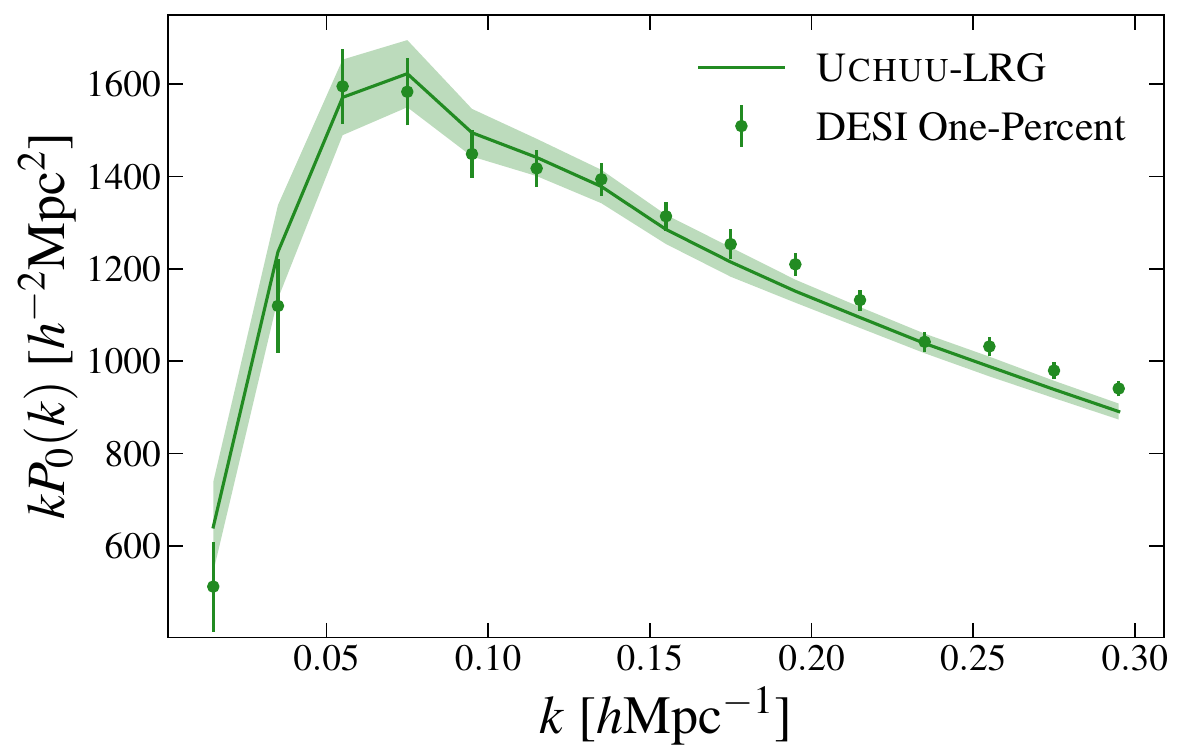}
    \includegraphics[width=0.95\columnwidth]{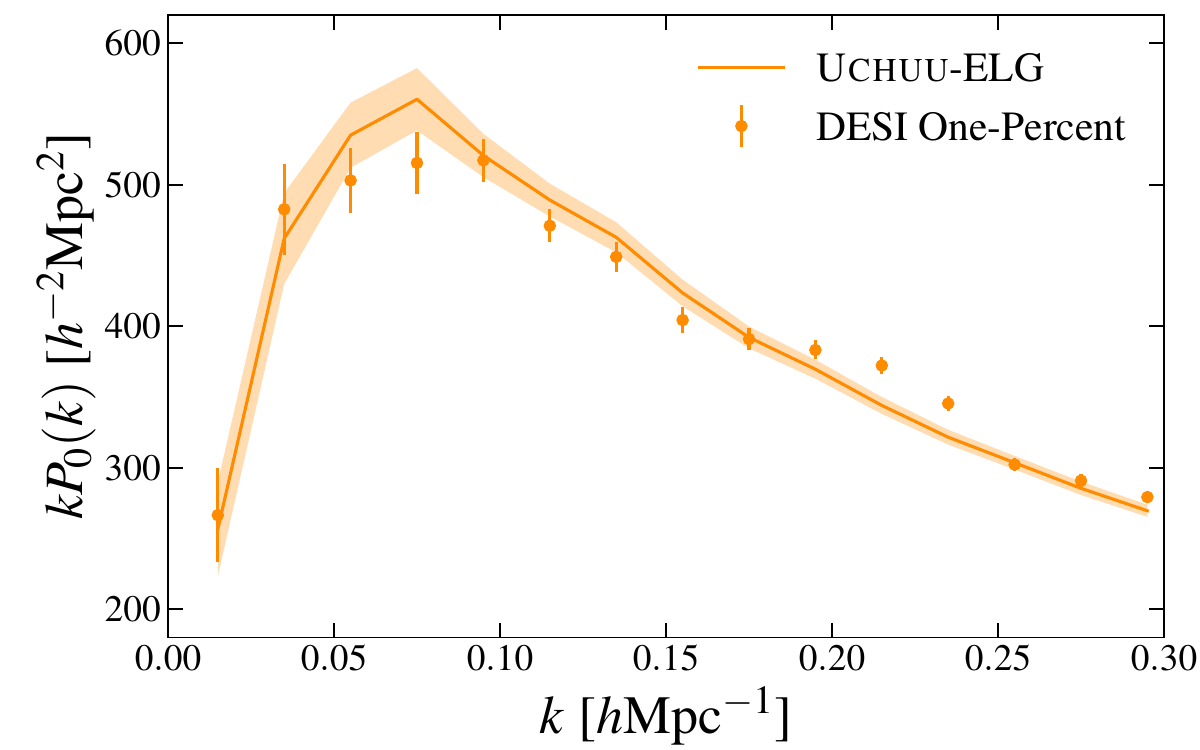}
    \includegraphics[width=0.95\columnwidth]{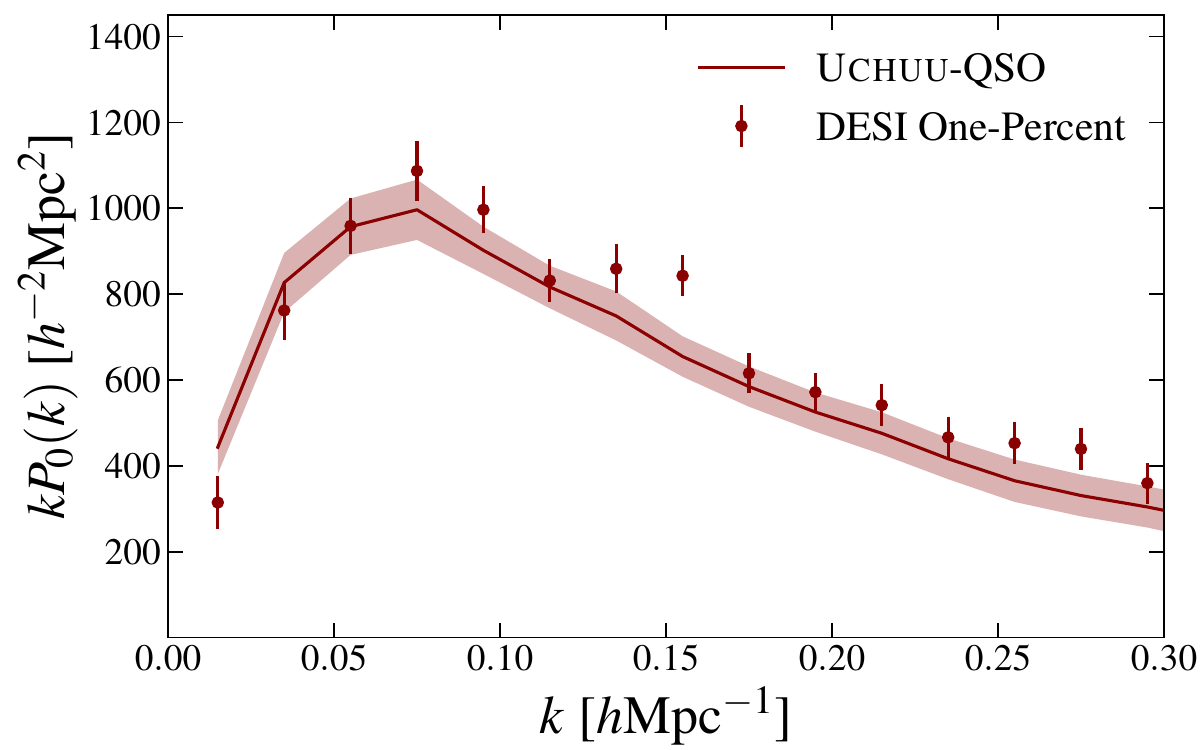}
    \caption{Power spectrum monopole and quadrupole measurements of the four DESI tracers, for the same samples as in Figure~\ref{fig:sv3-2pcf}. The BGS, LRG, ELG and QSO samples are shown in blue, green, orange and red, respectively, with the power spectrum monopole in the upper panels, and quadrupole in the lower panels. The solid lines and the points with error bars show the measurements from \textsc{Uchuu} and the One-Percent Survey data, respectively. Error bars represent the 1$\sigma$ scatter between the 102 mocks.}
    \label{fig:sv3-power-spectrum}
\end{figure*}

\subsubsection{Power spectrum}

We also measure the power spectrum monopole, $P_0(k)$, and quadrupole, $P_2(k)$, with the Python package \textsc{pypower}\footnote{\url{https://github.com/cosmodesi/pypower/}} which is based on the estimator from \citet{Hand2017}. Similarly to the correlation function measurements, incompleteness weights are applied to the survey data, and FKP weights are calculated for both the survey data and \textsc{Uchuu} lightcones from the $n(z)$ of each tracer, using the same fiducial power $P_0$ as above. To minimise the amount of aliasing from discrete Fourier sampling, we have used the piecewise cubic spline (PCS) mesh assignment scheme with a grid number $N_\mathrm{grid} = 1024$ in each dimension with interlacing~\citep{Sefusatti2016}. For each of the four tracers, Figure~\ref{fig:sv3-power-spectrum} shows the power spectrum multipoles over the wavenumber range $k \in [0.005, 0.505]~h\mathrm{Mpc}^{-1}$ in 50 uniform bins. The solid lines show the mean power spectrum over 102 \textsc{Uchuu} lightcones, and the data points show the measurements from the DESI One-Percent samples with error bars given by the standard deviation of the \textsc{Uchuu} lightcone measurements. The shaded regions correspond to the error calculated from the standard deviation of the 102 \textsc{Uchuu} lightcones.
Regarding the power spectrum monopole between $0.05 < k<0.5~h\mathrm{Mpc}^{-1}$, the residuals remain within $10\%$ for BGS, $7\%$ for LRG, and $9\%$ for ELG, and below $10\%$ for QSO up to $k<0.1~h\mathrm{Mpc}^{-1}$.

The power spectrum measurements performed here are similar to the BOSS and eBOSS power spectra~\citep{Beutler2017,GilMarin2020,mattia2021,Neveux2020}, where the local plane-parallel approximation is adopted to account for a varying line of sight~\citep{Feldman1994,Yamamoto2006}. The local line of sight is chosen to be the end-point vector to one of the galaxies in a pair, which enables fast, FFT-based evaluations to be carried out~\citep{Bianchi2015}. A minor difference here is that the normalisation factor is computed directly from the mesh field instead of relying on the angularly uniform quantity,~\( n(z) \). Besides the fact that the DESI One-Percent samples are smaller in area with stronger window effects on large scales, and smaller in size resulting in higher shot noise (which is subtracted accordingly), the power spectrum estimates are comparable to those from BOSS and eBOSS for the LRG, ELG and QSO samples. For the BGS sample, a comparison can be made with the main galaxy sample (MGS) from the SDSS survey in \citet{Tegmark2004}, \citet{Percival2007} and \citet{Ross2015}. However, it is worth noting that \citet{Tegmark2004} measures a different combination of the power spectrum multipoles with a minimum-variance quadratic estimator; focusing on the monopole only, \citet{Percival2007} has combined the MGS dataset with the 2dFGRS sample, and \citet{Ross2015} has moreover restricted MGS galaxies to those residing in high-mass haloes resulting in a larger clustering amplitude.

\subsection{Mean Halo-Occupancy of DESI One-Percent tracers}

\begin{figure*}
    \centering
    \includegraphics[width=\columnwidth]{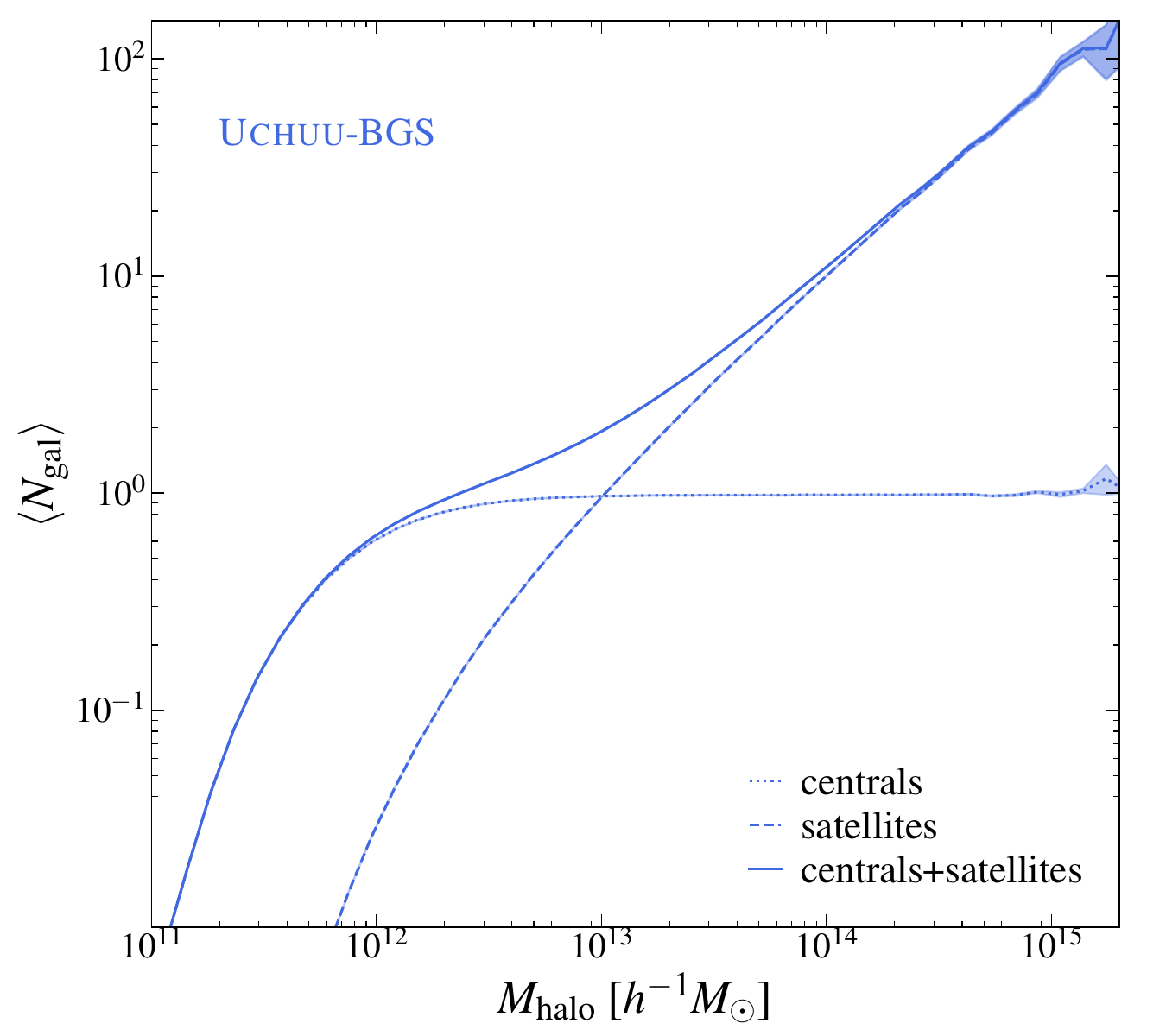}
    \includegraphics[width=\columnwidth]{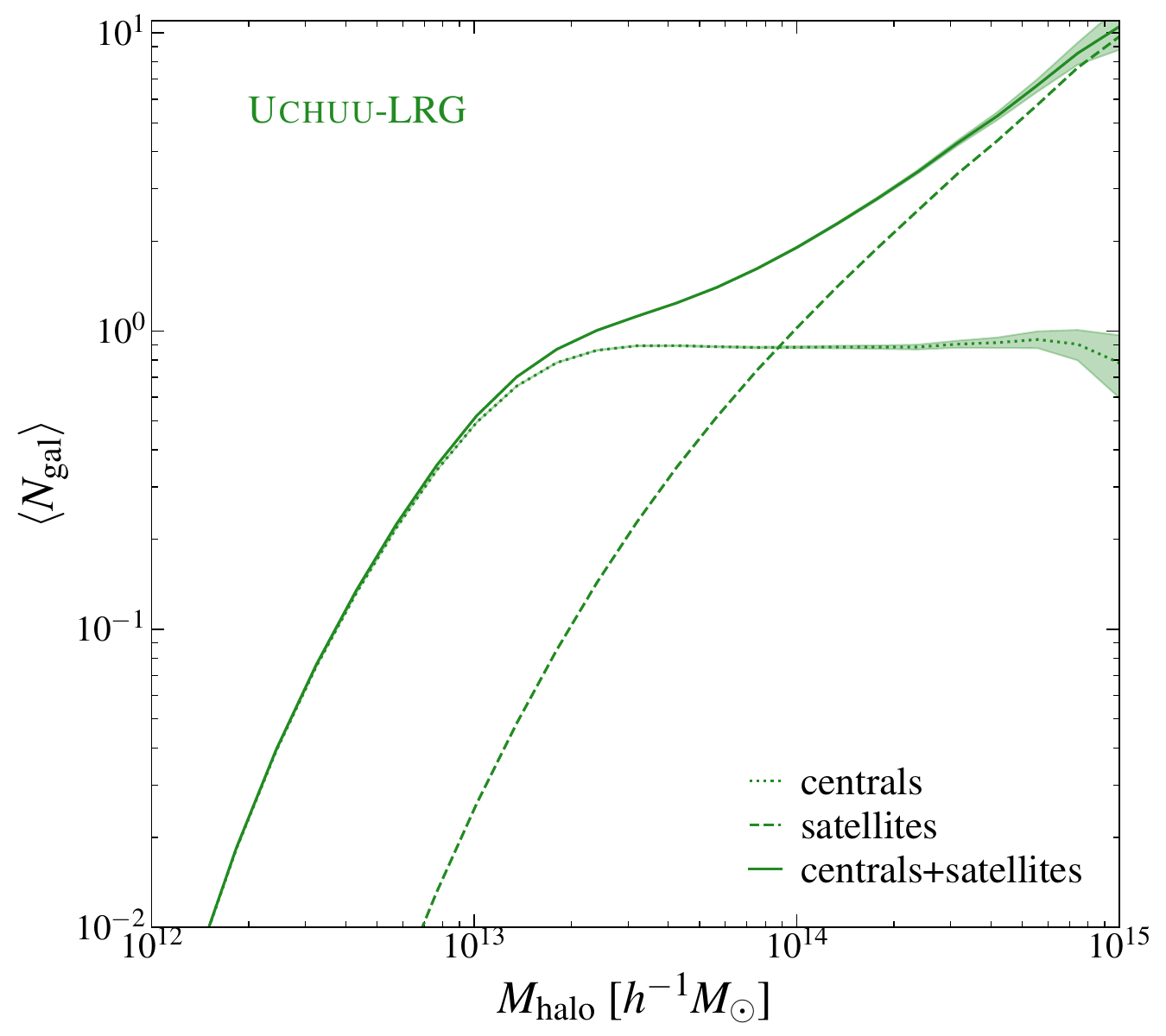}
    \includegraphics[width=\columnwidth]{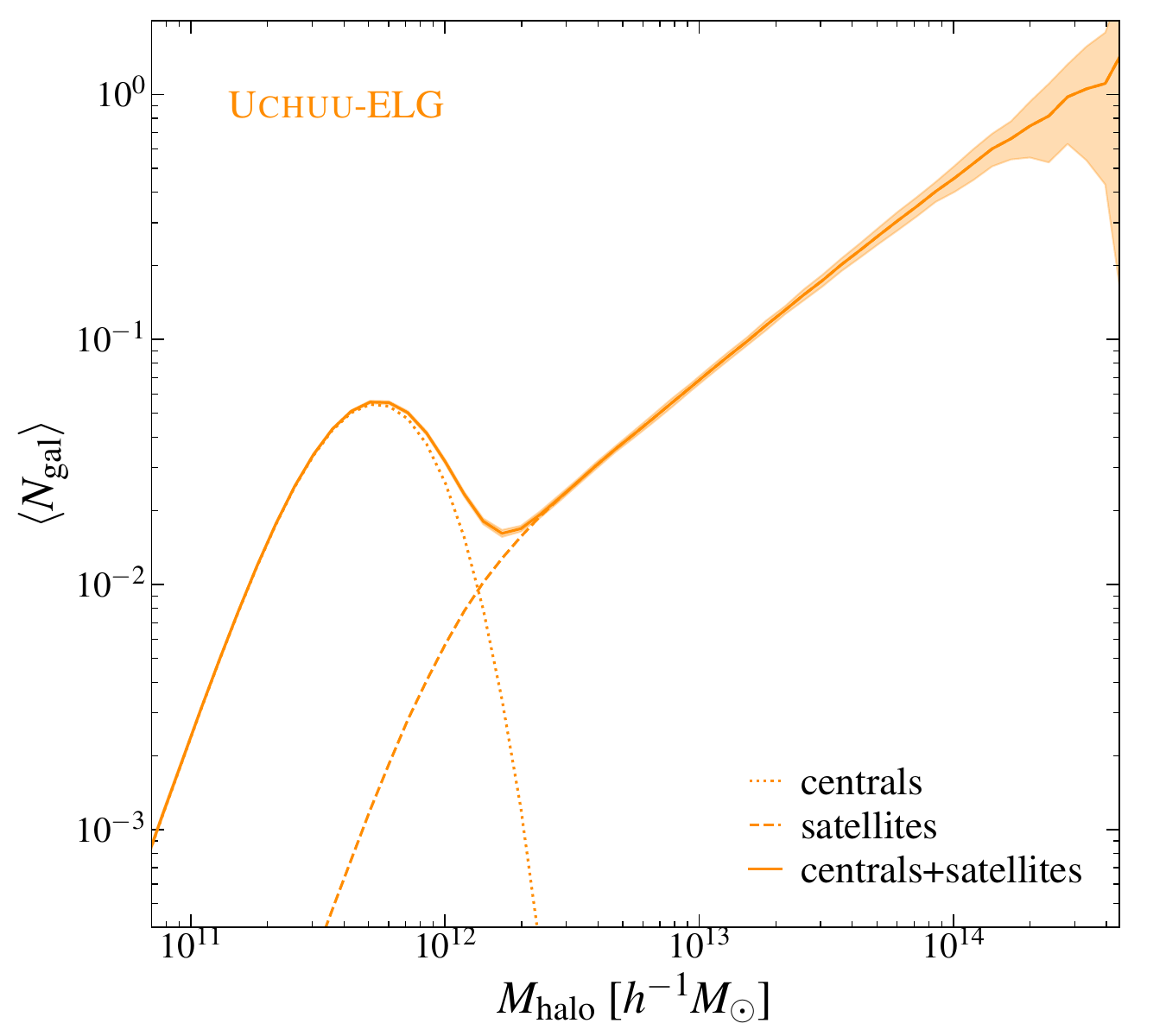}
    \includegraphics[width=\columnwidth]{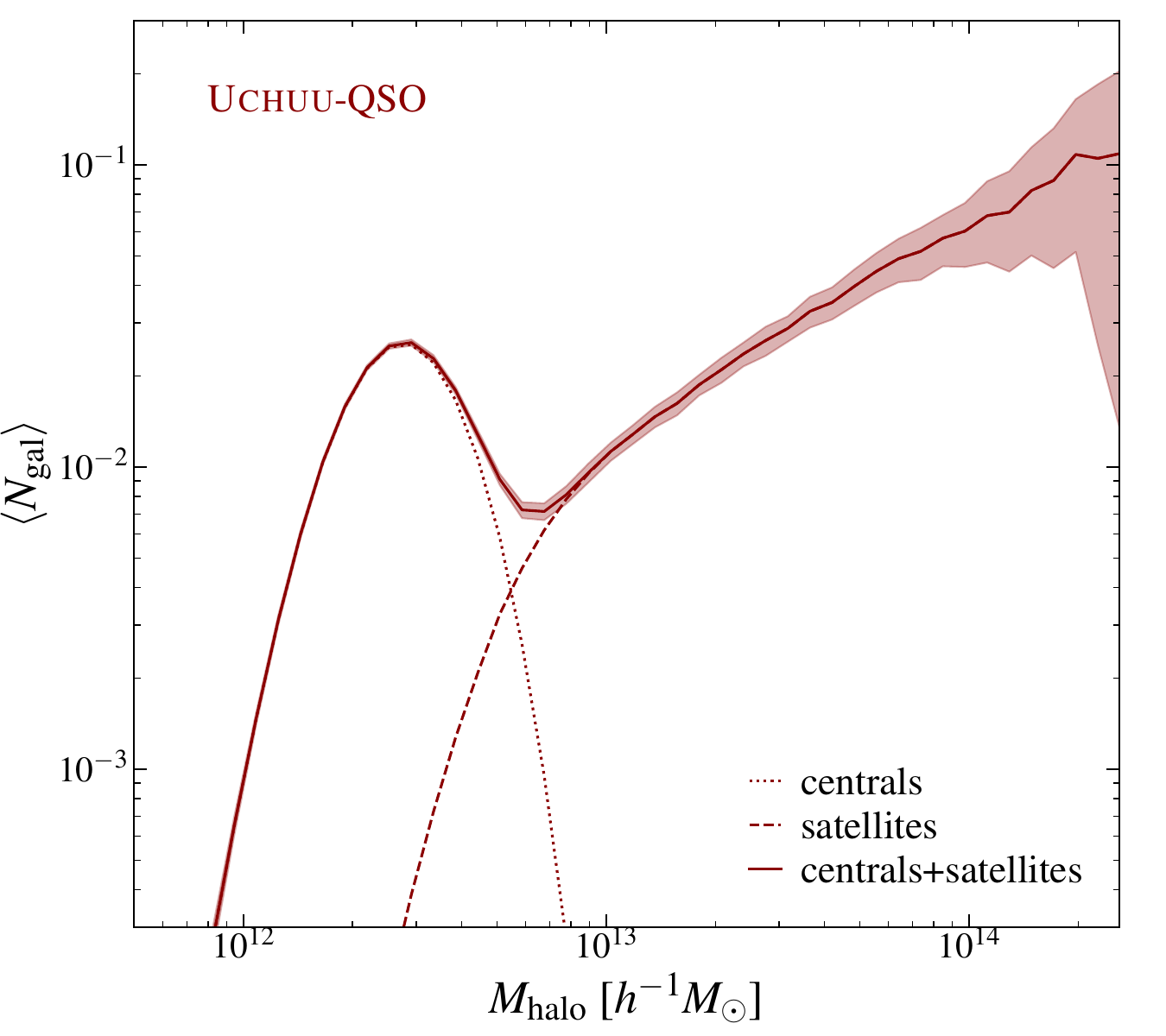}
    \caption{Mean halo occupancy of BGS (top-left panel), LRG (top-right panel), ELG (bottom-left panel), and QSO (bottom-right panel) samples, as determined from our (modified) SHAM \textsc{Uchuu} lightcones.  The mean number of galaxies of a halo with a given mass $M_\mathrm{halo}$ is denoted by $\langle N_\mathrm{gal} \rangle$.  The solid lines represent the combined centrals and satellite occupation, while the dotted and dashed lines show the mean halo occupancy for centrals and satellites, respectively. The shaded area indicates the 1$\sigma$ uncertainty of the occupation measured from the \textsc{Uchuu} lightcones. For BGS, this is a jackknife error from the full-sky mock, split into 100 jackknife regions. For the other tracers, this is the 1$\sigma$ scatter between the 102 mocks. The best-fit HOD model parameters for BGS and LRG are listed in Tables~\ref{tab:bgs-hod}~and~\ref{tab:lrg-hod}, respectively.}
    \label{fig:all-hod}
\end{figure*}

The abundance matching technique implemented with \textsc{Uchuu} provides a complete determination of the distribution and properties of all four DESI tracers within their host dark matter haloes. This allows us to estimate the mean number of galaxies or quasars within a dark matter (sub)halo of virial mass $M_\mathrm{halo}$ for each tracer sample. In Figure~\ref{fig:all-hod}, we present the mean halo occupancy as a function of halo mass for BGS BRIGHT, LRG, ELG, and QSO tracers, obtained from our independent set of \textsc{Uchuu} lightcones. The clustering signal of the same samples for the One-Percent survey is shown in Figure~\ref{fig:sv3-2pcf}. The low mass threshold of Uchuu allows us to distinguish between central galaxies/quasars residing in their host haloes and satellite galaxies/quasars that live in subhaloes. By doing so, we are able to measure the HOD separately for these two populations for all the DESI tracers. These are shown by the dotted and dashed curves in Figure~\ref{fig:all-hod}, respectively, for each tracer. 

The central and satellite HODs for the BGS sample show different behaviours. The central HOD increases smoothly as halo mass increases, with all high mass haloes containing a central BGS galaxy, while at low masses, the occupancy is zero, and there is a smooth transition in between due to scatter in the relationship between halo mass and galaxy luminosity. The satellite HOD, on the other hand, follows a power law that drops off more rapidly at low masses. For galaxy samples such as the BGS, where there is a monotonic relationship between halo mass and luminosity (with scatter), the HOD is commonly described using a 5-parameter form \citep[see][]{Zheng_2005}, where for central and satellite galaxies,
\begin{equation}
\label{eq:5param_hod}
\begin{split}
\langle N_\mathrm{cen} \rangle &= \frac{1}{2}\left[ 1 + \mathrm{erf}\left( \frac{\log M - \log M_\mathrm{min}}{\sigma_{\log M}} \right) \right] \\
\langle N_\mathrm{sat} \rangle &= \langle N_\mathrm{cen} \rangle \left( \frac{M-M_0}{M^{\prime}_{1}} \right)^\alpha.
\end{split}
\end{equation}
In this HOD parametrisation, the position and width of the central galaxy step function are set by $M_\mathrm{min}$ and $\sigma_{\log M}$. The satellite power-law slope is denoted by $\alpha$, $M^{\prime}_{1}$ the normalisation, and $M_0$ represents the low mass cutoff. 

For the LRG sample, the shape of the HOD is similar to the BGS. Since LRGs live in more massive haloes, the central HOD is shifted to higher masses compared with the BGS. However, at very high masses, the central occupancy is less than 1. As can be seen in Figure \ref{fig:lrg_stellarmass}, at very high stellar masses there seems to be an incompleteness in the observed galaxy population, so that the complete SMFs assumed in the SHAM remains above the data, resulting in the HOD dropping below 1 when the incompleteness is added. We use the 5-parameter HOD as in Eq.~\ref{eq:5param_hod} to model the LRG occupation function, which is discussed in Section~\ref{sec:lrg_hod}.

For the ELG and QSO samples, the central galaxies strongly show the influence from our Gaussian $V_{\rm mean}$ model in the HOD vs. $M_{\rm halo}$. This is expected due to the strong correspondence between $V_{\rm peak}$ and $M_{\rm halo}$. The satellite component rises more quickly at high $M_{\rm halo}$ for ELGs than QSOs. This is a result of the ELG sample having a lower best-fit $V_{\rm mean}$ than the QSO sample as shown in Table~\ref{tab:qso-shamparm} as well as the much larger sample size of ELGs. 

In upcoming sections, we will explore how the halo occupation distribution of the various tracers, which we obtained from the previously described (modified) SHAM mocks,  depends on luminosity or stellar mass as well as redshift.

\subsubsection{BGS HOD: luminosity dependency}
\label{sec:bgs_hod}

\begin{figure*}
    \centering
    \includegraphics[width=\columnwidth]{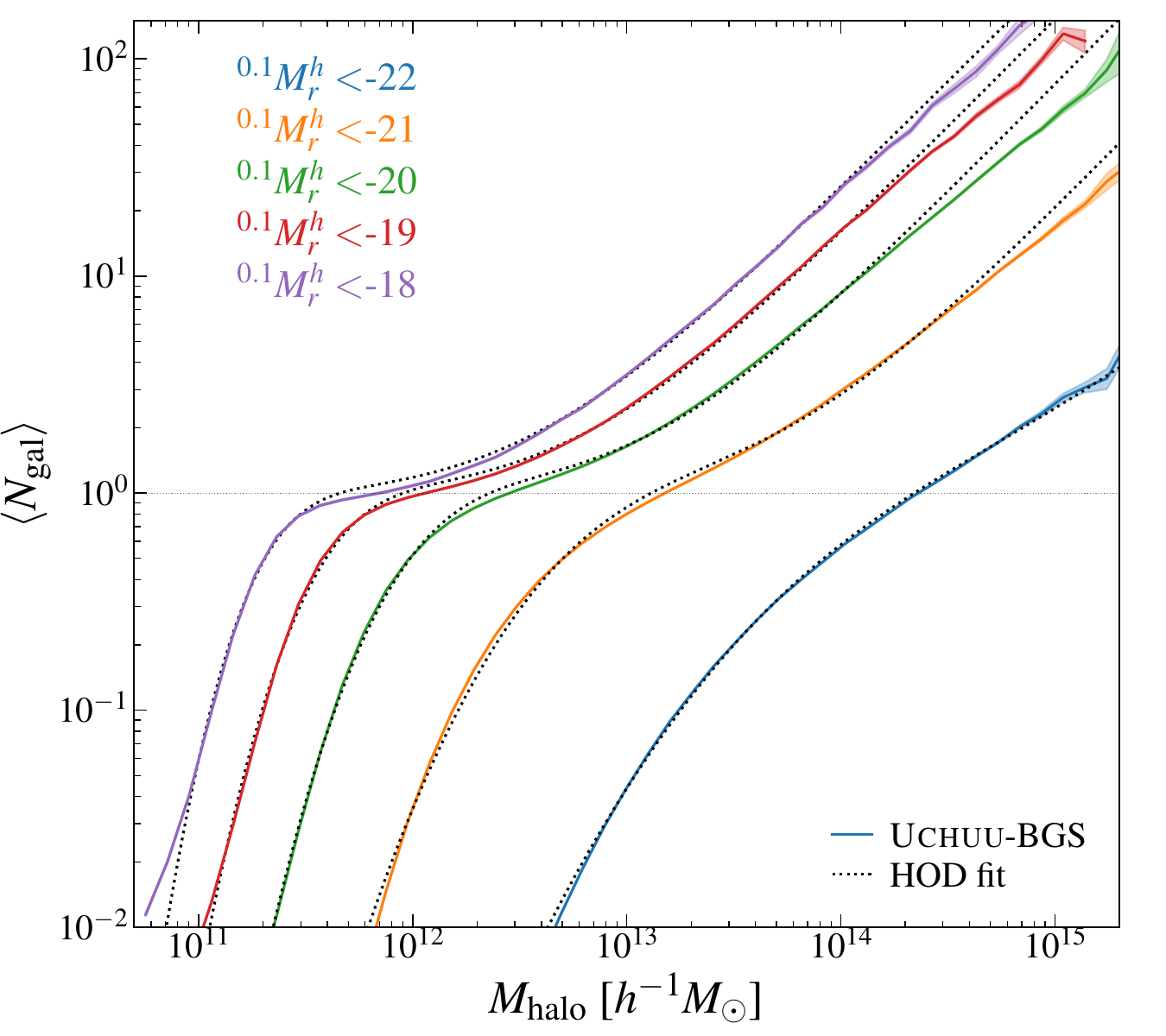}
    \includegraphics[width=\columnwidth]{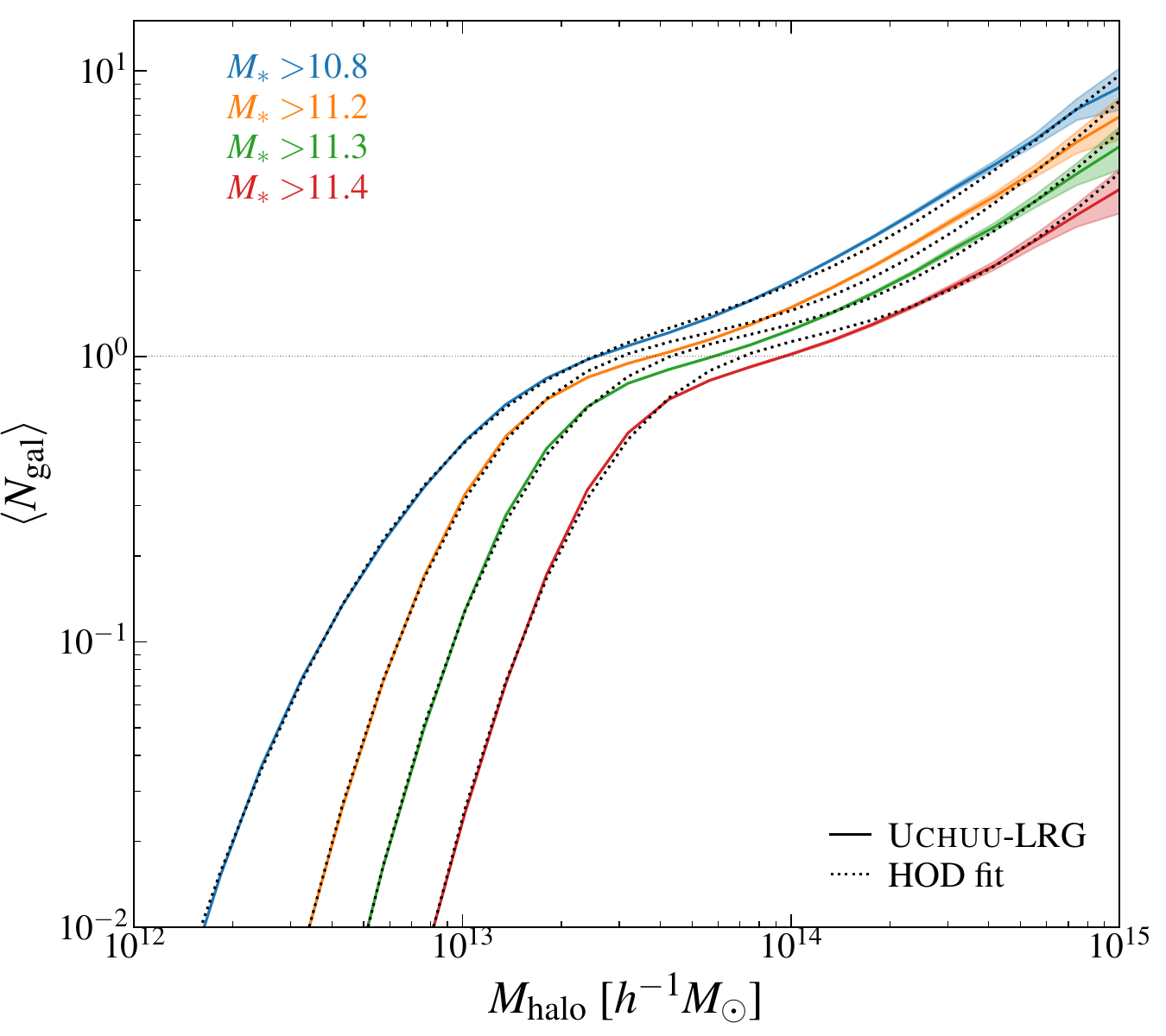}
    \caption{The same as Figure~\ref{fig:all-hod} but showing the HODs for several BGS BRIGHT luminosity-threshold (left panel) and LRG stellar mass-threshold samples (right panel), selected from our \textsc{Uchuu}-DESI lightcones. The coloured curves show the HODs measured from the full-sky mock, where the sample is indicated in the legend, and the shaded area indicates the jackknife error, using 100 jackknife regions. The best-fitting 5-parameter HOD model for each sample is shown by the black dotted curves. HOD model parameters are provided in Table~\ref{tab:bgs-hod} and Table~\ref{tab:lrg-hod} for the BGS and LRG samples, respectively.}
    \label{fig:bgslrg-hod}
\end{figure*}

We have obtained the HOD for nine different volume-limited samples, in addition to the full BGS-BRIGHT sample presented in Figure~\ref{fig:all-hod}. The first two columns of Table~\ref{tab:bgs-hod} provide the magnitude thresholds and maximum redshifts defining these samples. We display five of these HODs using coloured curves in the left panel of Figure~\ref{fig:bgslrg-hod}. Although the HOD for each sample has a shape that is very similar to the total sample shown in Figure~\ref{fig:all-hod}, the HOD shifts to higher masses for the brighter samples, because bright galaxies live in more massive haloes.

To investigate the dependence of the HOD on the absolute magnitude threshold of our galaxy samples, we fit the 5-parameter HOD form from Equation~\ref{eq:5param_hod} to the HODs obtained from the full-sky \textsc{Uchuu} BGS mock. Since the area of the full-sky mock is roughly twice as large as the combined area of the 102 smaller mocks, our HOD statistics are improved, particularly at the high-mass end. We separately fit the HOD for central and satellite galaxies using the non-linear least squares method, taking into account the uncertainty in the Uchuu HOD measurements as estimated from splitting the full sky into 100 jackknife regions. We split the area of the full sky into 100 equal area jackknife regions based on cuts in right ascension and declination, and we measure the HOD with each subvolume omitted once. We restrict the HOD fit to halo masses $M_\mathrm{halo}$ between $10^{11}$ and $3 \times 10^{15}~\hMsun$, and where the occupation number $\langle N \rangle$ is greater than $10^{-2}$. The resulting best-fitting HODs are shown as black dotted curves in Figure~\ref{fig:bgslrg-hod}.

\begin{table*}\hspace*{-1.1cm}
\centering
\setlength{\tabcolsep}{4pt}
    \begin{tabular}{ccccccccccccccc}
        \hline
        $M_r^\mathrm{th}$ & $z_\mathrm{max}$ & $N_{\rm eff}$ & $V_{\rm eff}$ & $\log n_{\rm g}^{\rm BGS}$ & $\log n_{\rm g}^{\rm Uchuu}$ & $\log M_\mathrm{min}$ & $\sigma_{\log{M}}$ & $\log{M_{0}}$ & $\log{M^{\prime}_{1}}$ & $\alpha$ & $f_{\rm sat}^{\rm HOD}$ & $f_{\rm sat}^{\rm Uchuu}$ & $b^{\rm BGS}$ & $b^{\rm Uchuu}$ \\
        \hline
        r<19.5 & 0.3 & 101909& 10.48 & -1.99 & -1.86$\pm$0.01 & 11.88 & 0.51 & 11.65 & 13.02 & 1.04 & 0.226 & 0.223 & 1.04$\pm$0.05 & 1.110$\pm$0.005  \\
        \hline
        -18.0 & 0.10 & 12159 &  0.42 & -1.52 & -1.54$\pm$0.04 & 11.31 & 0.28 & 11.49 & 12.59 & 0.99 & 0.267 & 0.26  & 0.76$\pm$0.19 & 1.027$\pm$0.018  \\
        -18.5 & 0.12 & 16764 &  0.76 & -1.66 & -1.66$\pm$0.03 & 11.44 & 0.30 & 11.59 & 12.69 & 1.00 & 0.259 & 0.25  & 0.81$\pm$0.17 & 1.065$\pm$0.016  \\
        -19.0 & 0.15 & 24089 &  1.49 & -1.73 & -1.80$\pm$0.03 & 11.59 & 0.33 & 11.69 & 12.82 & 1.00 & 0.251 & 0.246 & 1.02$\pm$0.12 & 1.110$\pm$0.012  \\
        -19.5 & 0.20 & 38522 &  3.46 & -1.92 & -1.97$\pm$0.02 & 11.76 & 0.35 & 11.79 & 12.96 & 1.01 & 0.239 & 0.237 & 0.98$\pm$0.09 & 1.144$\pm$0.009  \\
        -20.0 & 0.25 & 44867 &  6.45 & -2.14 & -2.18$\pm$0.02 & 11.99 & 0.39 & 11.96 & 13.14 & 1.01 & 0.225 & 0.225 & 1.17$\pm$0.07 & 1.206$\pm$0.007  \\
        -20.5 & 0.30 & 39591 & 10.37 & -2.38 & -2.46$\pm$0.02 & 12.30 & 0.46 & 12.10 & 13.39 & 1.02 & 0.204 & 0.209 & 1.20$\pm$0.06 & 1.289$\pm$0.006  \\
        -21.0 & 0.35 & 25511 & 14.29 & -2.74 & -2.83$\pm$0.02 & 12.73 & 0.57 & 12.21 & 13.73 & 1.02 & 0.179 & 0.188 & 1.33$\pm$0.10 & 1.414$\pm$0.010  \\
        -21.5 & 0.425& 13184 & 16.78 & -3.21 & -3.36$\pm$0.01 & 13.31 & 0.73 & 11.95 & 14.15 & 0.97 & 0.151 & 0.160 & 1.51$\pm$0.10 & 1.638$\pm$0.010  \\
        -22.0 & 0.50 &  3600 &  5.82 & -3.79 & -4.14$\pm$0.02 & 14.05 & 0.86 & ---   & 14.73 & 0.86 & 0.111 & 0.128 & 1.81$\pm$0.16 & 2.003$\pm$0.016  \\
        \hline
    \end{tabular}
\caption{HOD parameters and bias factors of the BGS parent and volume-limited samples. The first two columns show the absolute magnitude threshold and the maximum redshift used to define the volume-limited samples. All volume-limited samples have a minimum redshift $z_\mathrm{min} = 0.05$, while for the $r<19.5$ sample, the redshift range is $0.1 < z < 0.3$. The following columns give for each sample the number of galaxies, the effective volume (in $10^{6} h^{-3}\mathrm{Mpc}^{3}$), $\log_{10}$ of the galaxy number density (in $h^{3}\mathrm{Mpc}^{-3}$), the 5 parameters of the HOD model fit to its mean halo occupancy, displayed in the left panel of Figure~\ref{fig:all-hod}. Halo masses are in units of $\hMsun$. $f_\mathrm{sat}^\mathrm{HOD}$ is the satellite fraction calculated using the best-fitting 5 parameter HOD, while $f_\mathrm{sat}^\mathrm{Uchuu}$ is the satellite fraction measured directly from the \textsc{Uchuu} BGS mock. The last two columns provide the bias factors at the median redshift of each sample measured from the \textsc{Uchuu}-BGS lightcones and One-Percent data, as shown in Figure~\ref{fig:bias}. }
\label{tab:bgs-hod}
\end{table*}

The top row of Table~\ref{tab:bgs-hod} shows the best-fit HOD of the full BGS sample, as shown in Figure~\ref{fig:all-hod}, while the remaining rows list the HOD parameters for the nine volume-limited samples. We also provide the satellite fractions in Table~\ref{tab:bgs-hod} calculated from our HOD fits, and also measured directly from the \textsc{Uchuu} mocks. Since the uncertainties in the HOD measurements from the full-sky mock are very small, we also obtain very small uncertainties in the best-fit HOD parameters, and quoting them would be misleading. To provide a more realistic estimate of the HOD uncertainties, one needs to consider the uncertainties in the luminosity function and intrinsic scatter parameter used in the SHAM procedure to construct the lightcone. However, since we only have one \textsc{Uchuu} simulation, we cannot estimate these uncertainties.

Figure~\ref{fig:bgslrg-hod} compares the best-fitting BGS HODs for five of the volume-limited samples to the HODs measured from the \textsc{Uchuu} lightcone. Although the 5-parameter HOD form provides reasonable fits by eye, the small uncertainties in the HOD measurements from the mock lead to very large $\chi^2/\mathrm{dof}$ values. While the simple 5-element parametrisation approximates the HODs reasonably well, it fails to fully capture the shape predicted by the \textsc{Uchuu} BGS lightcone. Specifically, the shape of the smooth central step function does not perfectly match the shape of the error function, and the satellite power-law slope is not strictly a simple power law.

\begin{figure}
    \includegraphics[width=0.95\columnwidth]{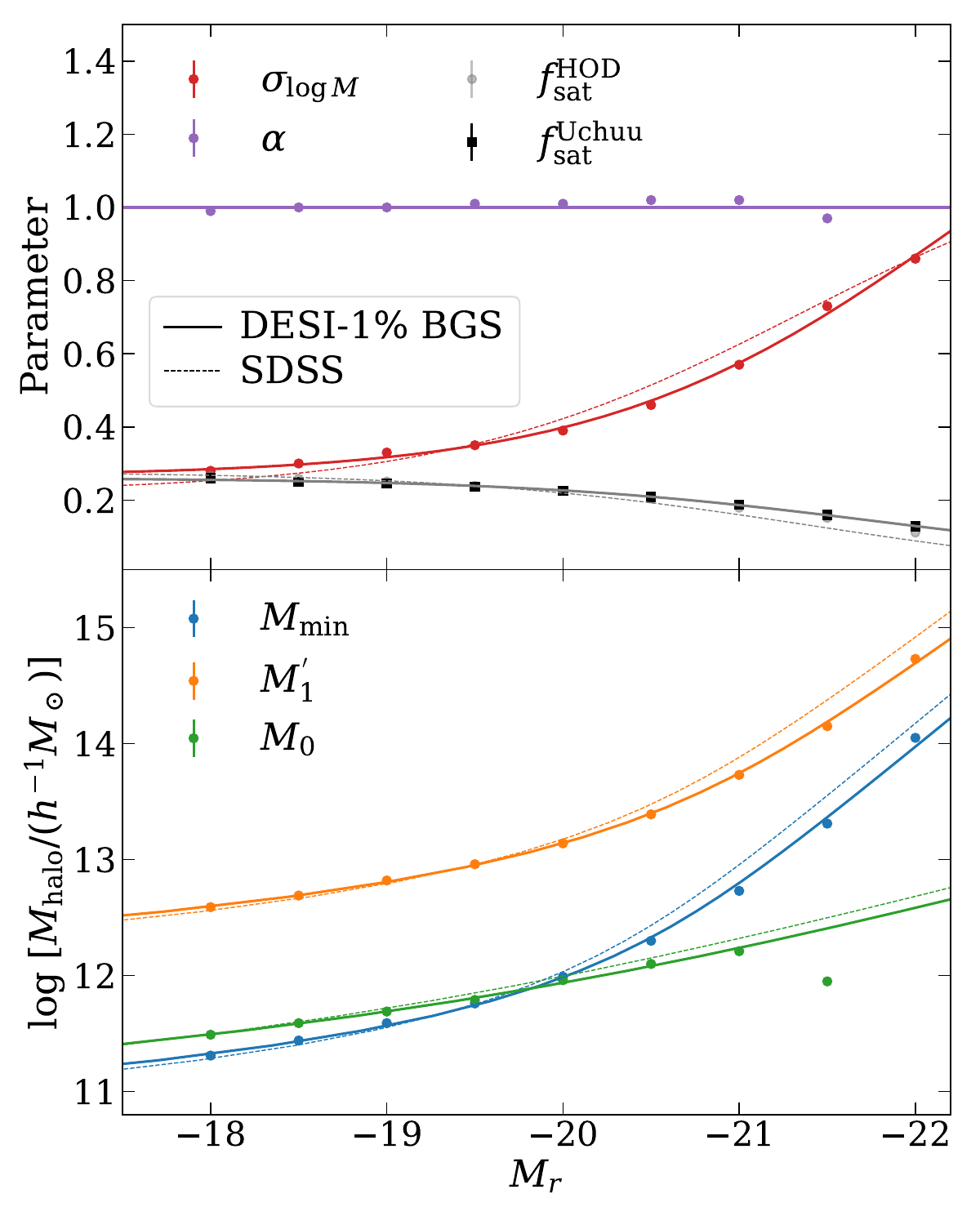}
    \caption{Best-fitting BGS HOD parameters and satellite fractions (points), as a function of magnitude threshold. The top panel shows the $\sigma_{\log M}$ (red) and $\alpha$ (purple) HOD parameters, plus the satellite fractions from our HODs (light grey) and Uchuu (dark grey). The mass parameters $M_\mathrm{min}$ (blue), $M^{\prime}_{1}$ (orange) and $M_0$ (green) are shown in the lower panel. The dashed lines indicate the SDSS results obtained from \citet{DongPaez22}, while the solid lines correspond to our approximations given by the equations included in the text. For the $M_r < -22$ sample, the $M_0$ parameter is completely unconstrained. 
    }
    \label{fig:bgs-hodparm}
\end{figure}

The trends in the best-fitting HOD parameters and satellite fractions as a function of BGS luminosity are shown in Figure~\ref{fig:bgs-hodparm} (solid symbols). These trends are consistent with previous SDSS studies such as \citet{Zehavi2011} and \citet{DongPaez22}, as indicated by the dashed line obtained from the latter. The three mass parameters increase for brighter samples, which is expected since brighter galaxies typically reside in more massive haloes. The scatter parameter $\sigma_{\log M}$, also increases for the brighter samples, as the SHAM method applies a constant intrinsic scatter to the assigned magnitudes. This constant scatter leads to a wider step function for central galaxies at brighter magnitudes, as the galaxy luminosity function falls off exponentially at the bright end. The agreement between the satellite fractions obtained from the HOD analysis and \textsc{Uchuu} is remarkable, with a decreasing trend expected as the luminosity (and hence host halo mass) of the galaxies in the sample increases.
The power-law slope $\alpha \approx 1$ for all BGS samples, but it decreases for the brightest sample. Interestingly, we do not observe an increase of $\alpha$ for the largest luminosities as reported in previous studies such as \citep[e.g.][]{Zehavi2011} for SDSS. It is worth noting that 
for the $M_r<-22$ sample, as previously mentioned, the parameter $M_0$ is unconstrained. Notably
for this sample, the low-mass cutoff in the satellite HOD is well modeled by the central HOD, and therefore, the additional $M_0$ parameter is not necessary.

We describe the relation between the mass scales in our HOD parametrisation (Equation~\ref{eq:5param_hod}) and the absolute magnitude threshold by adopting the same functional form proposed by \citet{Zehavi2011}. This is given by
\begin{equation}
    M_r = M^*_r + A -2.5 \, \log_{10} \left[ \left( \frac{x}{M_t} \right)^{\alpha_M} \exp \left( 1 - \frac{M_t}{x} \right) \right] \, ,
\end{equation}
where $x \in \{ M_\mathrm{min}, M^{\prime}_{1}, M_0 \}$.
The normalisation $A$ represents the median luminosity of central galaxies \citep[in units of magnitude, relative to $M^*_r=-20.44$,][]{Blanton2003} in haloes of the transition mass, $M_t$ (in units of $\hMsun$). $\alpha_M$ is the power-law index.
The solid lines in the bottom panel of Figure~\ref{fig:bgs-hodparm} correspond to the fits with the best-fitting parameters. For $M_\mathrm{min}$, $\{A, M_t, \alpha_M\} = \{1.450, 3.658 \times 10^{11}\hMsun, 0.320\}$. 
For the parameter $M^{\prime}_{1}$, $\{A, M_t, \alpha_M\} = \{1.168, 7.612 \times 10^{12}\hMsun, 0.366\}$. 
For $M_0$, $\{A, M_t, \alpha_M\} = \{1.880, 3.951 \times 10^{11}\hMsun, 1.0\}$.

The scatter in the relationship between halo mass and galaxy luminosity, denoted by $\sigma_{\log M}$, can be expressed through the equation
\begin{equation}
    \sigma_{\log M} = a + \frac{b-a}{1 + \exp(M_r-M^*_r+c)} \, .
\end{equation}
We also find that the same function form provides a good fit to the satellite fraction, $f_{\rm sat}^{\rm Uchuu}$, obtained from the \textsc{Uchuu} lightcones.
The solid lines in the top panel of Figure~\ref{fig:bgs-hodparm} correspond to the fits. The best-fitting parameters for $\sigma_{\log M}$ are $\{ a, b, c\} = \{ 0.265, 1.612, 1.769\}$ and for $f_{\rm sat}^{\rm Uchuu}$, we obtain $\{ a, b, c\} = \{ 0.261,  0.199, -0.742\}$. The satellite power-law slope $\alpha$ is assumed to be 1, as expected.

\subsubsection{LRG HOD: stellar mass dependency}
\label{sec:lrg_hod}

The HODs predicted by \textsc{Uchuu} for LRG samples with different stellar mass cuts are shown in the right panel of Figure~\ref{fig:bgslrg-hod}. For clarity, we show only four out of the five different samples. The stellar mass thresholds defining these samples are provided in Table~\ref{tab:lrg-hod}. Similar to the BGS sample, the HOD of the total LRG sample (shown in the top-left panel of Figure~\ref{fig:all-hod}) and the stellar mass threshold samples have very similar shapes. However, the HODs are shifted towards higher halo masses for higher stellar mass samples, as more massive galaxies tend to inhabits in more massive haloes.

To assess the dependence of the HOD on the stellar mass threshold, we used the 5-parameter HOD of Equation~\ref{eq:5param_hod} to fit the mean HOD obtained from the four \textsc{Uchuu}-LRG boxes with the SMF incompleteness already taken into account. As we did for the BGS sample, we performed separate fits for the central and satellite galaxies using the non-linear least squares method. 
For LRGs, we restrict the central (satellite) HOD fit to halo masses $M_\mathrm{halo}$ below $4 \times 10^{14}~\hMsun$ ($10^{15}~\hMsun$) and where the occupation number $\langle N \rangle$ is greater than $10^{-2}$. 
We estimated the uncertainty in the mean HOD by calculating Poisson errors, for both the galaxies and the halo, as $\sqrt{N}$, where $N$ is the number of galaxies/haloes within each host halo mass bin. 
Subsequently, we performed error propagation following the HOD formula to determine the associated uncertainty. We assumed that there were no systematic uncertainties or selection biases in the halo mass measurements or binning process.
Table~\ref{tab:lrg-hod} shows the HOD parameters that provide the best-fitting HODs, which are shown as dashed black curves in the right-hand panel of Figure~\ref{fig:bgslrg-hod}.

\begin{table*}\hspace*{-0.65cm}
\centering
\setlength{\tabcolsep}{4pt}
    \begin{tabular}{cccccccccccccc}
        \hline
        $\log{M_{\star}}$ & $N_{\rm eff}$ & $V_{\rm eff}$ & $\log n_{\rm g}^{\rm LRG}$ & $\log n_{\rm g}^{\rm Uchuu}$ & $\log{M_\mathrm{min}}$ & $\sigma_{\log{M}}$ & $\log{M_{0}}$ & $\log{M^{\prime}_{1}}$ & $\alpha$ & $f_{\rm sat}^{\rm HOD}$ & $f_{\rm sat}^{\rm Uchuu}$ & $b^{\rm LRG}$ & $b^{\rm Uchuu}$  \\
        \hline
        all  & 67466 & 91.50 & -3.286 & -3.286$\pm$0.001 & 13.04 & 0.53 & 11.95 & 14.08 & 1.03 & 0.117 & 0.119 & 1.89$\pm$0.05 & 1.847$\pm$0.005 \\
        10.8 & 66079 & 90.89 & -3.294 & -3.295$\pm$0.001 & 13.03 & 0.51 & 12.19 & 14.09 & 1.04 & 0.113 & 0.117 & 1.91$\pm$0.05 & 1.857$\pm$0.005 \\
        11.0 & 58609 & 87.19 & -3.344 & -3.347$\pm$0.001 & 13.04 & 0.44 & 12.36 & 14.15 & 1.07 & 0.103 & 0.111 & 1.96$\pm$0.05 & 1.900$\pm$0.005 \\
        11.2 & 37516 & 71.77 & -3.533 & -3.541$\pm$0.002 & 13.14 & 0.37 & 12.47 & 14.29 & 1.17 & 0.085 & 0.095 & 2.06$\pm$0.06 & 2.032$\pm$0.006 \\
        11.3 & 22006 & 51.40 & -3.784 & -3.772$\pm$0.003 & 13.29 & 0.36 & 12.72 & 14.41 & 1.21 & 0.076 & 0.081 & 2.15$\pm$0.07 & 2.189$\pm$0.007 \\
        11.4 & 11069 & 27.49 & -4.100 & -4.071$\pm$0.004 & 13.51 & 0.36 & 12.75 & 14.59 & 1.30 & 0.062 & 0.066 & 2.35$\pm$0.08 & 2.429$\pm$0.008 \\
        \hline
    \end{tabular}
\caption{HOD parameters and bias factors of the LRG samples with different stellar mass thresholds. The first column show the stellar mass (in $M_\odot$ units) threshold used in each sample. The remaining columns are the same as described in Table~\ref{tab:bgs-hod}. We provide the HOD model parameters (and $1\sigma$ errors) fitted to its mean halo occupancy, shown in the right panel of Figure~\ref{fig:bgslrg-hod}. The last two columns present the bias factors measured from DESI and \textsc{Uchuu} at the median redshift, $z=0.65$, of the samples, as shown in Figure~\ref{fig:bias}.}
\label{tab:lrg-hod}
\end{table*}

Our findings are in agreement with previous BOSS LRG studies \citep[e.g.][]{Nuza13,Tinker2017,Stoppacher2019}, showing that all halo mass HOD parameters increase with the stellar mass threshold, given that more massive galaxies inhabit more massive haloes. We also find that the power law slope, $\alpha$, increases with the stellar mass threshold, with a  value of $\alpha \approx 1$ for the complete and lowest stellar mass threshold samples. 
On the other hand, the parameter $\sigma_{\log M}$ remains constant for the 11.4, 11.3 and 11.2 stellar mass samples. However, as the stellar mass threshold decreases, the effect of SMF incompleteness enhances (see Figure~\ref{fig:lrg_stellarmass}), and a flatter central curve is obtained, which results in an increase in the value of $\sigma_{\log M}$.

As depicted in Figure~\ref{fig:bgslrg-hod}, the 5-parameter HOD approximation produces fits that agree well with the mean measurement from the boxes. Nevertheless, for host halo masses above $3 \times 10^{13}~\hMsun$, the model fails to fully capture the shape predicted by \textsc{Uchuu}. Specifically, the satellite HOD exhibits the largest difference between the model and data.

\subsubsection{ELG \& QSO HODs}
\label{sec:elgqso_hod}

Due to the substantial impact of the modified SHAM method on the resulting halo occupation distribution, and the adoption of the same procedure for selecting both ELGs and QSOs, we analyze and discuss them together.

The HOD analysis of the ELG mocks, as seen in the lower left panel of Fig~\ref{fig:all-hod}, reveals two mostly distinct components. The component of the HOD from the centrals follows a Gaussian distribution which dominates where $M_{\rm halo} < 1.25\times10^{12}~\hMsun$, while the satellites exhibit a power-law distribution which dominates at  $M_{\rm halo} > 1.25\times10^{12}~\hMsun$. The central haloes reach a peak occupation of 0.17, observed at $M_{\rm halo}=4.7\times10^{11}~\hMsun$. 

Comparing the HOD results with those from \citet{Favole2017} is challenging due to differences in the definition of halo occupancy. \citet{Favole2017} determine halo occupancy based on a denominator that includes only haloes with centrals in their mock, rather than considering all distinct haloes in the simulation. For their most inclusive sample ($L[\textsc{Oii}] > 1\times 10^{39}~\mathrm{erg}\,s^{-1}$), they observe a peak central occupancy of several percent over the mass range from $10^{12}~M_{\odot}$ to $3.2\times10^{13}~M_{\odot}$.

Similar to the ELG HOD, the QSO HOD also exhibits a Gaussian distribution for the centrals, characterized by $M_{\rm halo} < 6.8\times10^{12}~\hMsun$, and a power-law behavior for the satellites, which dominates at $M_{\rm halo} > 6.8\times10^{12}~\hMsun$. The mean halo mass for the central quasars is $M_{\rm halo} = 2.5\times10^{12}~\hMsun$. This slightly deviates from the reported mean halo mass values reported in \citet{rodrigueztorres17} for eBOSS QSOs, ranging from $3.2\times10^{12}~\hMsun$ to
 $6.6\times10^{12}~\hMsun$. This is expected since the reported $V_\mathrm{mean}$ values in their study are higher compared to the values presented in this paper. However, despite these differences in numerical values, the shapes of the HOD in both DESI and eBOSS studies exhibit a similar pattern.

\subsection{Large-scale bias of all four tracers}

\begin{figure*}
    \centering
    
    \includegraphics[width=0.95\columnwidth]{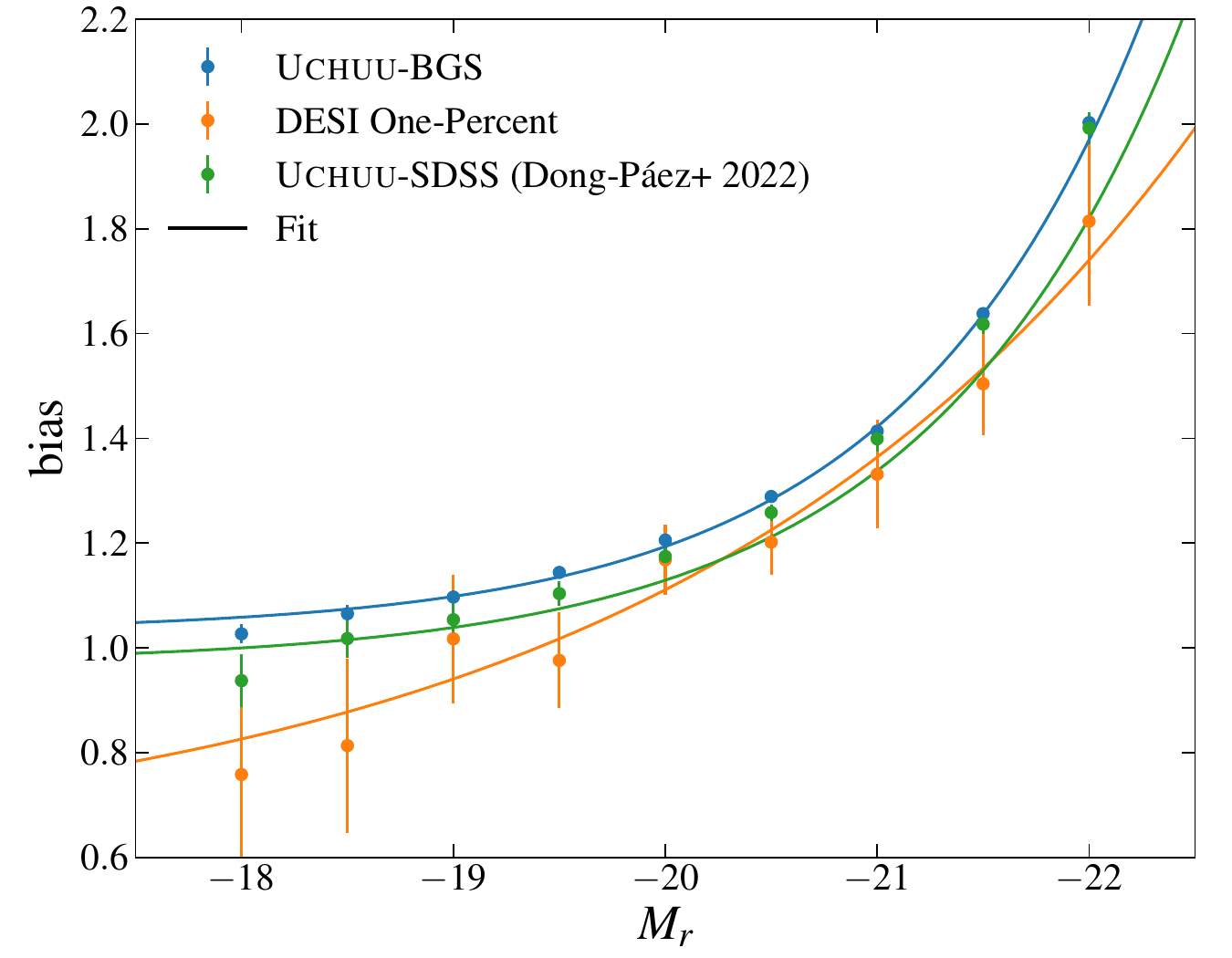}
    \includegraphics[width=0.95\columnwidth]{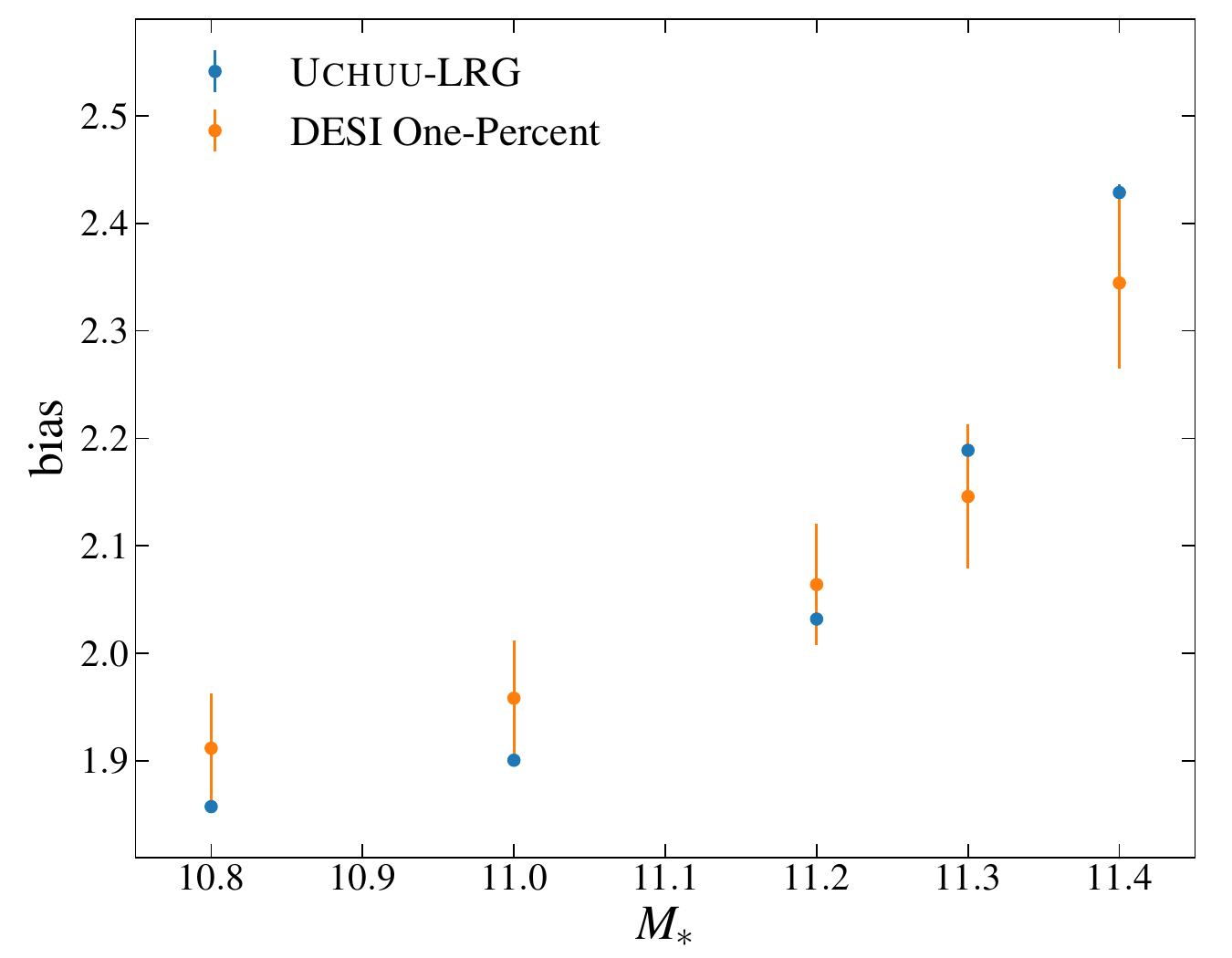}
    \includegraphics[width=0.95\columnwidth]{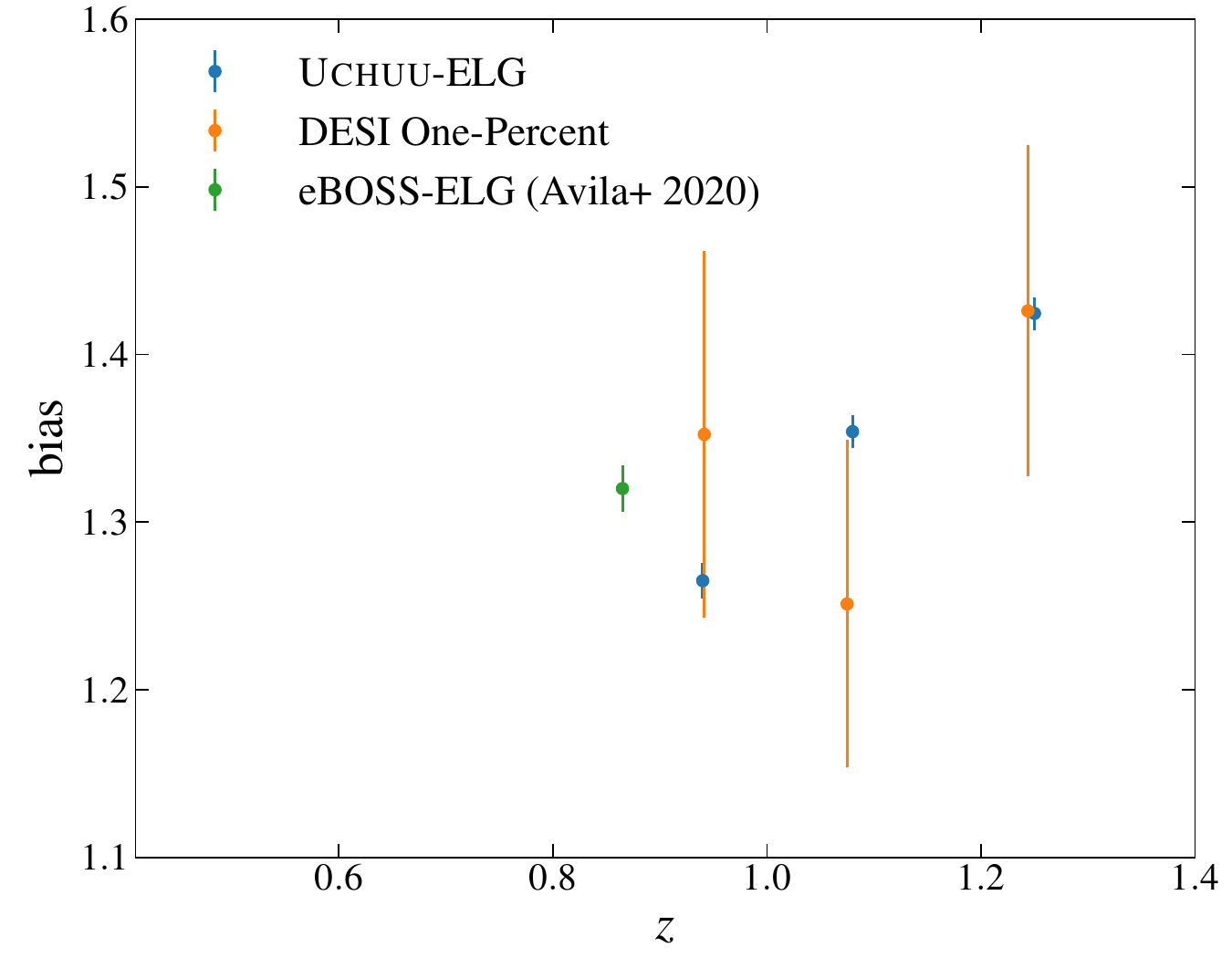}
    \includegraphics[width=0.95\columnwidth]{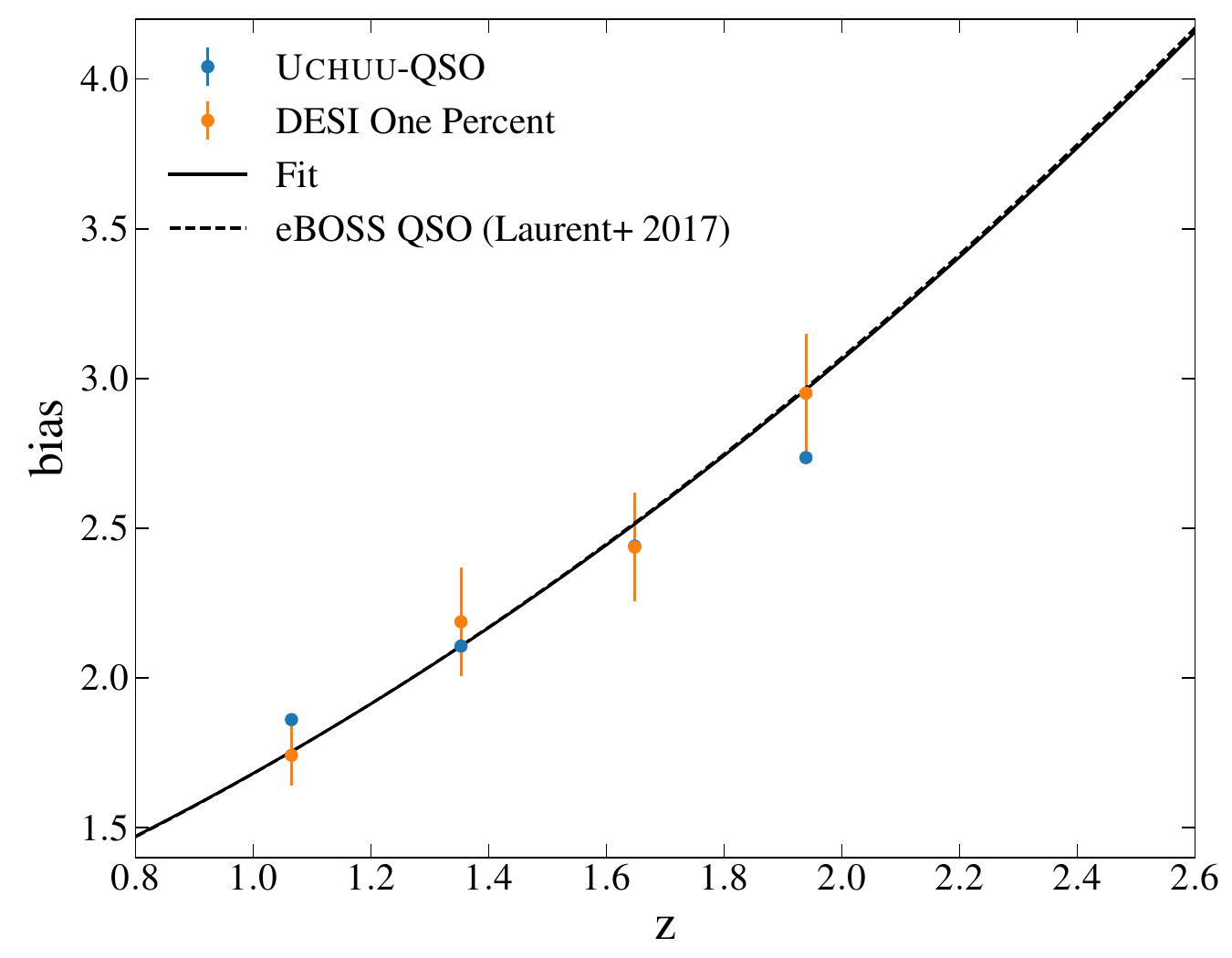}
    
    \caption{\textit{Top left}: Bias of the BGS magnitude threshold volume-limited samples, at the median redshift of each sample. Blue points indicate the mean bias measured from the 102 \textsc{Uchuu}-BGS lightcones, where the small error bars are the error on the mean. Orange points show the bias measured from the BGS One-Percent Survey clustering measurements, where the errors are the 1$\sigma$ scatter between the lightcones. The green points show bias measurements from the \textsc{Uchuu} mock, but cut to the SDSS volume limited samples, with redshift limits provided in Table~2 of \citet{Zehavi2011}. The lines show our fits to these bias measurements as a function of magnitude using Equation~\ref{eq:bias_bgs}. \textit{Top right}: The same as the top left panel, but showing the bias measurements for the LRG stellar mass threshold samples.   \textit{Bottom left}: QSO bias measurements as a function of redshift. The black dashed lines show the parametric fit from \citet{laurent2017clustering}, and the black solid line shows a similar fit to the DESI One-Percent data.}
    \label{fig:bias}
\end{figure*}

The large-scale bias, $b$, for each of the four DESI tracers was measured from the DESI One-Percent Survey and compared to their prediction obtained from the \textsc{Uchuu} lightcones in the Planck cosmology. The results are presented in Figure~\ref{fig:bias}. We measure the linear bias by fitting 
\begin{equation}
    \xi_0(s) = b^2 \left(1 + \frac{2}{3}\beta + \frac{1}{5}\beta^2 \right) \xi_\mathrm{lin}(s)
\end{equation}
to our correlation function monopole measurements, $\xi_0(s)$, over a given range of separations. $\xi_\mathrm{lin}(s)$ is from the linear power spectrum at the redshift of our galaxy sample, and $\beta=\Omega_\mathrm{m}^{0.6}/b$ \citep[see][]{Kaiser1987,Hamilton1998}. 
The details and description of results are provided below for each tracer.  

The linear bias for the BGS magnitude threshold samples is measured over the separation range $8 < s < 20~\hMpc$ for most samples, except for the three brightest thresholds, where we use $15 < s < 40~\hMpc$, at the median redshift of each volume-limited sample (blue symbols in the top-left panel of Figure~\ref{fig:bias}). As expected, the brightest galaxies have the highest bias, since they live in the most massive haloes, which are more strongly clustered. The bias values measured from the DESI One-Percent data (orange symbols) tend to be smaller than the predictions from the \textsc{Uchuu} mock. However, this can be explained by cosmic variance, since the volume of the One-Percent Survey is small and the uncertainties are large. We also show the agreement with bias factors measured by \citet{DongPaez22} for the SDSS, who also used the \textsc{Uchuu} simulation (green symbols).
Note that we are fitting a linear model to clustering measurements on relatively small quasi-linear scales. On very large scales where linear theory is valid, our One-Percent Survey clustering measurements are dominated by noise due to the small survey volume. To obtain precise bias values, we had to restrict our fits to small scales. We have checked that the ratio $\xi_0(s)/\xi_\mathrm{lin}(s)$ is flat over the scales used when fitting the bias. For the brightest samples, this ratio deviates from being flat at larger scales than the faint samples, which is why we use a different fitting range. Even on these small scales, the clustering of the faint samples are affected by cosmic variance, which is reflected in the large bias errors. Using larger datasets in the future will allow us accurately measure the bias on truly linear scales.

Following \citet{Zehavi2011}, the bias as a function of magnitude can be modelled as
\begin{equation}
    b(<M_r) = B_0 + B_1 \times 10^{B_2(M^*_r-M_r)/2.5}
\label{eq:bias_bgs}
\end{equation}
where $M^*_r = -20.44$. We fit this to the bias measurements from the \textsc{Uchuu}-BGS mock, and the One-Percent BGS data, taking into account the uncertainties shown in Figure~\ref{fig:bias}. For the \textsc{Uchuu}-BGS mock, we measure $\{B_0,B_1,B_2\} = \{1.03\pm0.01, 0.24\pm0.01, 0.95\pm0.04\}$, while for the DESI BGS, our best-fitting values are $\{B_0,B_1,B_2\} = \{0.59\pm0.43, 0.62\pm0.45, 0.43\pm0.29\}$. Since the uncertainties in the BGS bias measurements are large, the uncertainties in these parameters are also large. For \textsc{Uchuu}-SDSS $\{B_0,B_1,B_2\} = \{0.97\pm0.03, 0.27\pm0.02, 0.91\pm0.08\}$. We find reasonable agreement between these fits, although for the One-Percent data, the parameter $B_0$ is smaller than is measured in the mock. We caution against a direct comparison to the bias measurements of \citet{Zehavi2011}, since these are done at $z=0$, and in a different cosmology. However, the \textsc{Uchuu}-SDSS mock matches well the clustering from SDSS \citep{DongPaez22}.

The top-right panel of Figure~\ref{fig:bias} shows the large-scale bias factors of the LRGs, measured between $15$ and $35~\hMpc$, for the different stellar mass threshold samples. The results are as expected: the higher the stellar mass threshold, the larger the bias, since more massive galaxies reside in more massive haloes which are, as mentioned above for BGS galaxies, more strongly clustered. The model and data agree within the uncertainties. The bias factors are similar to those measured for BOSS LRGs.

The bias of ELGs vs. redshift is shown in the bottom-left panel of Figure~\ref{fig:bias}. The bias was calculated for each box defined in Table~\ref{tab:elg-shamparm} and for the data cut into the same redshift range covered by the mocks. As in \cite{Avila:2020rmp} we use a separation range of $20~\hMpc$ to $55~\hMpc$ to facilitate comparison with their results. We find agreement between the bias measured from the data and the bias measured from the mocks. Additionally, we have overplotted the single bias measurement of \citet{Avila:2020rmp} at the median redshift of their eBOSS ELG sample as a comparison. Our bias measurements are consistent with these measurements.

Finally, the large-scale bias of the QSOs is shown in the bottom-right panel of Figure~\ref{fig:bias}, as a function of redshift, with bias factor measured in the separation range $10 < s < 85~\hMpc$. The increasing bias measurements with redshift are consistent with those obtained by \citet{KrolewskiQSObias} using the DESI two-month data. We compare our results to those obtained by the eBOSS QSO survey, as presented by the $b(z)$ parametrisation in \citet{laurent2017clustering}. We adopted the same $b(z)$ form to fit our DESI bias estimates, i.e.,
\begin{equation}
     b(z) = \alpha [(1 + z)^{2} - 6.565] + \beta,
\end{equation}
 where $z$ is the median redshift value from each QSO sample and $\alpha$ and $\beta$ are parametric constants. We obtained the following values from the DESI QSO $b(z)$ fit:  $\alpha = 0.278 \pm 0.011$, and $\beta = 2.383 \pm 0.033$, including higher redshift bias estimates from \citet{KrolewskiQSObias}. This DESI QSO bias parametrisation is shown as a solid line in Figure~\ref{fig:bias}. The One-Percent and eBOSS data points are consistent with each other. The fitted One-Percent $\alpha$ and $\beta$ values are also in agreement with the eBOSS results $\alpha = 0.278 \pm 0.018$, and $\beta = 2.393 \pm 0.042$.

\section{Summary}
\label{sec:concl}

This paper presents a detailed overview of the process employed to generate simulated lightcones for the DESI One-Percent Survey. These lightcones are constructed within the framework of the flat-LCDM Planck cosmology model using \textsc{Uchuu}, a 2.1 trillion particle $N$-body simulation specifically designed for the DESI survey. \textsc{Uchuu} enables precise numerical resolution of dark matter haloes and subhaloes across a large volume, encompassing diverse scales ranging from galaxy clusters to dwarf galaxies.
The \textsc{Uchuu} haloes are populated with DESI galaxies and quasars using the SHAM method. In this approach, the peak maximum circular velocity serves as a proxy for (sub)halo mass. The construction of BGS and LRG lightcones follows the conventional SHAM technique. However, for ELG and QSO tracers, a modified SHAM method is employed, incorporating additional parameters. This methodology takes into consideration the redshift evolution of the tracers and their clustering dependence on fundamental properties like luminosities and stellar masses. Moreover, the \textsc{Uchuu} lightcones provide covariance errors for the clustering measurements of all four tracer samples in the DESI One-Percent Survey.
We carry out a thorough comparison of the measured clustering signals for each galaxy and quasar sample in the DESI One-Percent Survey with the corresponding predictions from \textsc{Uchuu}. Additionally, we determine the
halo occupancy and large-scale bias factors for all four DESI targets.

Our main results are summarized as follows:

\begin{enumerate}
    \item We measure the redshift-space two-point correlation function monopole and quadrupole over the scales from $0.3~\hMpc$ to $100~\hMpc$ for BGS and from $5~\hMpc$ to $100~\hMpc$ for LRG, ELG, and QSO. Additionally, we measure the power spectrum monopole from $0.02 ~h\mathrm{Mpc}^{-1}< k < 0.3~h\mathrm{Mpc}^{-1}$.
    Overall, we find consistency between the One-Percent Survey measurements and the theoretical predictions based on the Planck cosmology using the \textsc{Uchuu} lightcones. We find some differences, which can be attributed to incompleteness at the massive end of the stellar mass function (for LRGs), a simplistic model of the galaxy-halo connection (for ELGs and QSOs), and cosmic variance (as observed for the BGS on large scales).
    \item The clustering measurements for the entire BGS One-Percent sample agree with \textsc{Uchuu} on scales below $30~\hMpc$. The luminosity-dependent clustering for the BGS sample is also studied. A fair agreement is observed between \textsc{Uchuu} and BGS data for bright samples, but an offset is seen for fainter samples due to cosmic variance in the small volume of the One-Percent data, which may also explain the deviations above $30~\hMpc$. It is worth noting that the SHAM modeling should improve once we incorporate an intrinsic scatter dependency with galaxy luminosity.  
    \item For the LRG full sample, agreement between data and \textsc{Uchuu} is observed, but a systematically underprediction is found for \textsc{Uchuu} compared to DESI at $~6-8~\hMpc$. This difference is consistent with the incompleteness of the stellar mass function at the massive end, after studying the dependence of clustering on stellar mass and redshift.
    \item For the ELGs, there is agreement between \textsc{Uchuu} and DESI on large scales. However, \textsc{Uchuu} in tandem with our simplified SHAM modeling for ELGs underpredicts clustering at the smallest scales. This is an area that can be improved in future modeling efforts.    
    \item For QSOs, agreement is found between \textsc{Uchuu} and the DESI One-Percent data in the respective redshift bins.
    \item For the BGS and LRG galaxy samples where there is a monotonic relationship between halo mass and luminosity/stellar mass (with scatter), the HOD is reasonably described adopting the commonly used 5-parameter HOD form. A similar shape is seen in the HODs of BGS galaxies and LRGs. Since LRGs reside in more massive haloes, the halo occupancy of central haloes is shifted towards higher masses compared to the BGS. We also study the HOD dependence with luminosity and stellar mass for the BGS and LRGs, respectively. For the former, we provide approximations of the measured HOD 5-parameters as a function of absolute magnitude, which agree perfectly well with those obtained from cutting the \textsc{Uchuu} mock to the same volume limited samples as the SDSS survey. 
    \item The ELG halo occupation consists of a Gaussian component for centrals with low halo masses ($M_{\rm halo} < 1.25\times10^{12}~\hMsun$ and a peak occupation of 0.17 at $M_{\rm halo} = 4.7\times10^{11}~\hMsun$), and a power law component for satellites with higher halo masses ($M_{\rm halo} > 1.25\times10^{12}~\hMsun$). The QSO HOD exhibits a similar form: a Gaussian component for central haloes ($M_{\rm halo} < 6.8\times10^{12}~\hMsun$ with a peak occupation of 0.03 at $M_{\rm halo} = 2.9\times10^{12}~\hMsun$), and a power-law component for the satellites ($M_{\rm halo} > 6.8\times10^{12}~\hMsun$).
    \item The linear bias factors were measured for all four tracers from the DESI One-Percent Survey and compared to predictions based on the \textsc{Uchuu} lightcones in the Planck cosmology. The bias values measured from the DESI One-Percent BGS data tend to be smaller than the \textsc{Uchuu} predictions, possibly due to cosmic variance and the limited volume of the survey. Additionally, we measured the bias as a function of absolute magnitude threshold for the various BGS volume-limited samples, and there is agreement with that observed with that obtained from the SDSS samples. The dependence of bias with stellar mass threshold is obtained for LRGs, which agrees with BOSS LRGs. For the ELG and QSO samples, the bias factors measured from the data and mocks agree. The QSO bias measurements increase with redshift, and a parametrisation for this is provided. We compare the bias of the ELG and QSO samples to previous eBOSS results, to which they also agree.
\end{enumerate} 

The findings presented in this study play a crucial role in refining and optimizing essential components of cosmology models. They provide valuable insights into improving the construction of simulated lightcones, enhancing the galaxy-halo connection schemes, and advancing our understanding of clustering signals.  Through careful analysis and interpretation of the results from the One-Percent Survey, we can significantly enhance the accuracy and reliability of the final survey's cosmology interpretation. In conclusion, the lessons learned from the current size of DESI data are pivotal in ensuring the success of the final survey. As the next step, we encourage readers to explore forthcoming papers that will be based on the first year of DESI data and improved \textsc{Uchuu} lightcones, especially regarding issues caused by cosmic variance, which are expected to improve with the much larger Year-1 dataset.

\section*{Acknowledgements}

The authors are grateful to Arnaud de Mattia and Lado Samushia for their valuable comments on the manuscript. 
FP thanks Tomoaki Ishiyama and Anatoly Klypin for their contributions and insightful discussions regarding the Uchuu simulation and various aspects of the galaxy-halo connection.
FP and JE acknowledge support from the Spanish MICINN funding grant PGC2018-101931-B-I00. 
RK, JL and RV acknowledge support of the U.S. Department of Energy (DOE) in funding grant DE-SC0010129 for the work in this paper.
AS, SC and PN acknowledge STFC funding ST/T000244/1 and ST/X001075/1.
SA acknowledge support of the Department of Atomic Energy, Government of India, under project no. 12-R\&D-TFR-5.02-0200. SA was partially supported by the European Research Council through the COSFORM Research Grant (\#670193) and STFC consolidated grant no. RA5496.
Computational resources for JL, RV, and RK were provided by SMU's Center for Research Computing.
The \textsc{Uchuu} simulation was carried out on the Aterui II supercomputer at CfCA-NAOJ. We thank IAA-CSIC, CESGA, and RedIRIS in Spain for hosting the Uchuu data releases in the \textsc{Skies \& Universes} site for cosmological simulations. The analysis done in this paper have made use of NERSC at LBNL and $skun6$@IAA-CSIC computer facility managed by IAA-CSIC in Spain (MICINN EU-Feder grant EQC2018-004366-P).

This material is based upon work supported by the DOE, Office of Science, Office of High-Energy Physics, under Contract No. DE–AC02–05CH11231, and by the National Energy Research Scientific Computing Center, a DOE Office of Science User Facility under the same contract. Additional support for DESI was provided by the U.S. National Science Foundation (NSF), Division of Astronomical Sciences under Contract No. AST-0950945 to the NSF’s National Optical-Infrared Astronomy Research Laboratory; the Science and Technology Facilities Council of the United Kingdom; the Gordon and Betty Moore Foundation; the Heising-Simons Foundation; the French Alternative Energies and Atomic Energy Commission (CEA); the National Council of Science and Technology of Mexico (CONACYT); the Ministry of Science and Innovation of Spain (MICINN), and by the DESI Member Institutions: \url{https://www.desi.lbl.gov/collaborating-institutions}. Any opinions, findings, and conclusions or recommendations expressed in this material are those of the author(s) and do not necessarily reflect the views of the U. S. National Science Foundation, the U. S. Department of Energy, or any of the listed funding agencies.

The authors are honored to be permitted to conduct scientific research on Iolkam Du’ag (Kitt Peak), a mountain with particular significance to the Tohono O’odham Nation.

\section*{Data Availability}

The Uchuu-DESI lightcones are
available at \url{https://data.desi.lbl.gov}.

The data points and Python scripts for reproducing all figures in this publication are available at \url{https://doi.org/10.5281/zenodo.8006825}.



\bibliographystyle{aa}

\bibliography{desi-eda-uchuu}

\begin{appendix}
\section{Author Affiliations}
\label{sec:affil}

\begin{hangparas}{.5cm}{1}

$^{1}${Instituto de Astrof\'{i}sica de Andaluc\'{i}a (CSIC), Glorieta de la Astronom\'{i}a, s/n, E-18008 Granada, Spain}

$^{2}${Institute for Computational Cosmology, Department of Physics, Durham University, South Road, Durham DH1 3LE, UK}

$^{3}${Department of Physics, Southern Methodist University, 3215 Daniel Avenue, Dallas, TX 75275, USA}

$^{4}${Institute of Space Sciences, ICE-CSIC, Campus UAB, Carrer de Can Magrans s/n, 08913 Bellaterra, Barcelona, Spain}

$^{5}${Institute for Astronomy, University of Edinburgh, Royal Observatory, Blackford Hill, Edinburgh EH9 3HJ, UK}

$^{6}${Tata Institute of Fundamental Research, Homi Bhabha Road, Mumbai 400005, India}

$^{7}${Dipartimento di Fisica ``Aldo Pontremoli'', Universit\`a degli Studi di Milano, Via Celoria 16, I-20133 Milano, Italy}

$^{8}${Department of Physics \& Astronomy and Pittsburgh Particle Physics, Astrophysics, and Cosmology Center (PITT PACC), University of Pittsburgh, 3941 O'Hara Street, Pittsburgh, PA 15260, USA}

$^{9}${Department of Physics and Astronomy, University of California, Irvine, 92697, USA}

$^{10}${Centre for Extragalactic Astronomy, Department of Physics, Durham University, South Road, Durham, DH1 3LE, UK}

$^{11}${Lawrence Berkeley National Laboratory, 1 Cyclotron Road, Berkeley, CA 94720, USA}

$^{12}${Physics Dept., Boston University, 590 Commonwealth Avenue, Boston, MA 02215, USA}

$^{13}${Department of Physics \& Astronomy, University College London, Gower Street, London, WC1E 6BT, UK}

$^{14}${Department of Physics and Astronomy, The University of Utah, 115 South 1400 East, Salt Lake City, UT 84112, USA}

$^{15}${Instituto de F\'{\i}sica, Universidad Nacional Aut\'{o}noma de M\'{e}xico,  Cd. de M\'{e}xico  C.P. 04510,  M\'{e}xico}

$^{16}${Kavli Institute for Particle Astrophysics and Cosmology, Stanford University, Menlo Park, CA 94305, USA}

$^{17}${SLAC National Accelerator Laboratory, Menlo Park, CA 94305, USA}

$^{18}${Departamento de F\'isica, Universidad de los Andes, Cra. 1 No. 18A-10, Edificio Ip, CP 111711, Bogot\'a, Colombia}

$^{19}${Observatorio Astron\'omico, Universidad de los Andes, Cra. 1 No. 18A-10, Edificio H, CP 111711 Bogot\'a, Colombia}

$^{20}${Department of Astrophysical Sciences, Princeton University, Princeton NJ 08544, USA}

$^{21}${Center for Cosmology and AstroParticle Physics, The Ohio State University, 191 West Woodruff Avenue, Columbus, OH 43210, USA}

$^{22}${Department of Physics, The Ohio State University, 191 West Woodruff Avenue, Columbus, OH 43210, USA}

$^{23}${The Ohio State University, Columbus, 43210 OH, USA}

$^{24}${Department of Physics, The University of Texas at Dallas, Richardson, TX 75080, USA}

$^{25}${Departament de F\'{i}sica, Serra H\'{u}nter, Universitat Aut\`{o}noma de Barcelona, 08193 Bellaterra (Barcelona), Spain}

$^{26}${Institut de F\'{i}sica d’Altes Energies (IFAE), The Barcelona Institute of Science and Technology, Campus UAB, 08193 Bellaterra Barcelona, Spain}

$^{27}${NSF NOIRLab, 950 N. Cherry Ave., Tucson, AZ 85719, USA}

$^{28}${Instituci\'{o} Catalana de Recerca i Estudis Avan\c{c}ats, Passeig de Llu\'{\i}s Companys, 23, 08010 Barcelona, Spain}

$^{29}${Department of Physics and Astronomy, Siena College, 515 Loudon Road, Loudonville, NY 12211, USA}

$^{30}${Department of Physics and Astronomy, University of Sussex, Brighton BN1 9QH, U.K}

$^{31}${National Astronomical Observatories, Chinese Academy of Sciences, A20 Datun Rd., Chaoyang District, Beijing, 100012, P.R. China}

$^{32}${Department of Physics and Astronomy, University of Waterloo, 200 University Ave W, Waterloo, ON N2L 3G1, Canada}

$^{33}${Perimeter Institute for Theoretical Physics, 31 Caroline St. North, Waterloo, ON N2L 2Y5, Canada}

$^{34}${Waterloo Centre for Astrophysics, University of Waterloo, 200 University Ave W, Waterloo, ON N2L 3G1, Canada}

$^{35}${Space Sciences Laboratory, University of California, Berkeley, 7 Gauss Way, Berkeley, CA  94720, USA}

$^{36}${University of California, Berkeley, 110 Sproul Hall \#5800 Berkeley, CA 94720, USA}

$^{37}${Department of Physics, Kansas State University, 116 Cardwell Hall, Manhattan, KS 66506, USA}

$^{38}${Department of Physics and Astronomy, Sejong University, Seoul, 143-747, Korea}

$^{39}${CIEMAT, Avenida Complutense 40, E-28040 Madrid, Spain}

$^{40}${Department of Physics, University of Michigan, Ann Arbor, MI 48109, USA}

$^{41}${University of Michigan, Ann Arbor, MI 48109, USA}

\end{hangparas}

\end{appendix}

\label{lastpage}
\end{document}